\documentclass[11pt]{article}
\usepackage[T1]{fontenc}
\usepackage[utf8]{inputenc}
\usepackage[backend=biber,sorting=none,sortcites,citestyle=numeric-comp,bibstyle=phys,doi=false,eprint=true,url=false,biblabel=brackets,isbn=true,articletitle=true]{biblatex}   
\usepackage{authblk}

\usepackage{microtype}
\usepackage{mathpazo} 
\usepackage{eulervm}  

\usepackage{amsmath, amssymb, amstext, amsfonts}
\usepackage{mathtools}
\usepackage{dsfont}
\usepackage{graphicx}
\usepackage[dvipsnames]{xcolor}
\usepackage[margin=1in]{geometry}
\usepackage{float}
\usepackage{ragged2e}
\usepackage[caption=false, justification=justified]{subfig}
\DeclareCaptionJustification{justified}{\justifying}
\usepackage{placeins}
\usepackage{amsthm}
\usepackage[title,titletoc]{appendix}
\usepackage[colorlinks=true, linkcolor = BlueViolet, citecolor = BlueViolet, urlcolor = BlueViolet]{hyperref}
\usepackage{enumitem}

\usepackage{subdepth} 

\newtheorem{theorem}{Theorem}[section]
\newtheorem{definition}[theorem]{Definition}

\newtheorem{proposition}[theorem]{Proposition}

\DeclareMathOperator{\Tr}{Tr}
\DeclareMathOperator{\sinc}{sinc}

\DeclareMathOperator{\symdiff}{\triangle}
\DeclareMathOperator{\sgn}{sgn}

\newcommand{\eshell}{\Sigma}
\newcommand{\diff}[1]{\ensuremath{\text{d} #1}}
\newcommand{\Eshell}[1]{\eshell_{#1}}

\newcommand{\ptau}{p}

\newcommand{\uh}{\hat{U}_H}
\newcommand{\uhd}[1]{\hat{U}_{H #1}}
\newcommand{\uc}{\hat{U}_C}
\newcommand{\ucp}{\hat{U}_{C'}}
\newcommand{\ucb}{\hat{U}_{C''}}
\newcommand{\ucj}[1]{\hat{U}_{C,#1}}
\newcommand{\twi}{w}

\newcommand{\uerr}{\hat{U}_\Delta}
\newcommand{\phierr}{\phi_\Delta}

\newcommand{\proj}{\hat{\Pi}}

\newcommand{\idop}{\hat{\mathds{1}}}
\newcommand{\dptG}[2]{\Gamma^{(#1)}_{#2}}
\newcommand{\zprec}{\eta}
\newcommand{\meanz}{z}

\newcommand{\uone}{\hat{U}_1}
\newcommand{\utwo}{\hat{U}_2}
\newcommand{\aot}{\hat{A}_{12}}
\newcommand{\ato}{\hat{A}_{21}}

\newcommand{\eps}{\overline{\epsilon}}
\newcommand{\vareps}{\varepsilon}

\newcommand{\htime}{{\rm H}}

\newcommand{\edit}[1]{{#1}} 
\newcommand{\editb}[1]{{#1}}

\newcommand{\newedit}[1]{{#1}} 
\newcommand{\newadd}[1]{{#1}} 

\begin{document}

\title{Dynamical quantum ergodicity from energy level statistics}
\author[1]{Amit Vikram}
\author[1]{Victor Galitski}
\affil[1]{Joint Quantum Institute and Department of Physics, University of Maryland, College Park, MD 20742, USA}

\date{}
\maketitle

\begin{abstract}
   Ergodic theory provides a rigorous mathematical description of \newedit{chaos in} classical dynamical systems, including a formal definition of the ergodic hierarchy. \newedit{How ergodic dynamics is reflected in the energy levels and eigenstates of a quantum system is the central question of quantum chaos, but a rigorous quantum notion of ergodicity remains elusive.} Closely related to \newedit{the classical ergodic} hierarchy is a less-known notion of cyclic approximate periodic transformations [see, {\em e.g.}, I.~Cornfield, S.~Fomin, and Y.~Sinai, Ergodic Theory (Springer-Verlag New York, 1982)], which maps any ``ergodic'' dynamical system to a cyclic permutation on a circle and arguably represents the most elementary \newedit{form} of ergodicity. This paper shows that cyclic ergodicity generalizes to quantum dynamical systems,
\newedit{and provides a rigorous observable-independent} definition of quantum ergodicity.
 It implies the ability to construct an orthonormal basis, where quantum dynamics transports any initial basis vector to \newedit{have a sufficiently large overlap with each of the} other basis vectors \newedit{in a cyclic sequence}.
 It is proven that the basis, maximizing the overlap over all such quantum cyclic permutations, is obtained via the discrete Fourier transform of the energy eigenstates, \newedit{with overlaps given by specific measures of spectral rigidity}. This relates quantum cyclic ergodicity to energy level statistics. \newedit{The level statistics of Wigner-Dyson random matrices, usually associated with quantum chaos on empirical grounds, is derived as a special case of this general relation}. \newedit{To demonstrate generality,} we prove that irrational flows on a 2D torus are classical and quantum cyclic ergodic, \newedit{with spectral rigidity distinct from Wigner-Dyson}. Finally, we motivate a quantum ergodic hierarchy of operators \newedit{and discuss connections to eigenstate thermalization}.
This work provides a general framework for transplanting some rigorous \edit{concepts} of ergodic theory to quantum dynamical systems.
\end{abstract}

\newpage 

{
\hypersetup{linkcolor=black} 
\setcounter{tocdepth}{2}
\tableofcontents
 }
 \newpage
 
\section{Introduction}
\label{sec:intro}

\subsection{Background and motivation}
\label{sec:motivation}

\begin{figure}[!h]
    \centering
    \includegraphics[width=0.6\columnwidth]{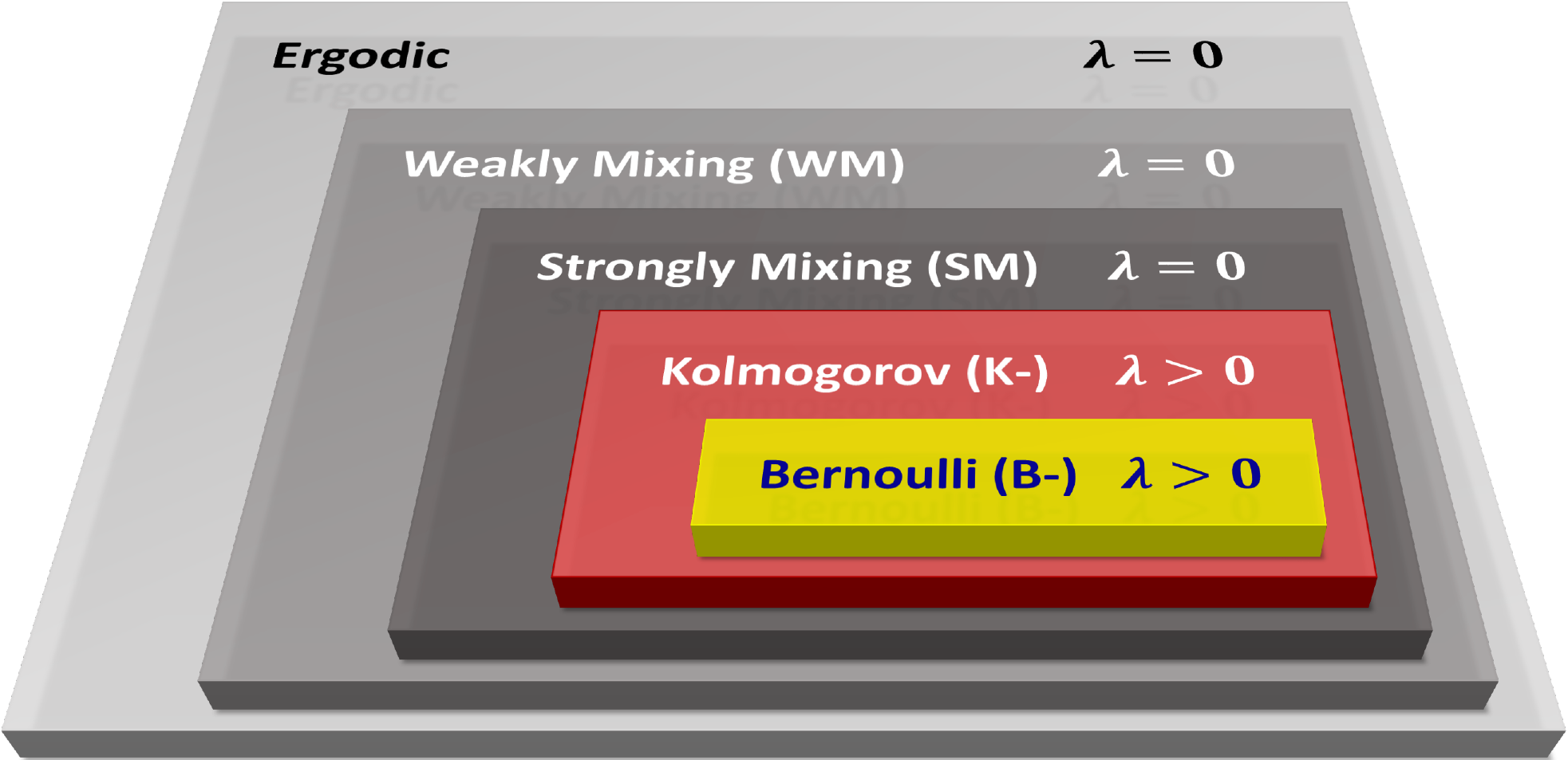}
    \caption{\newedit{Venn diagram of} the classical Ergodic Hierarchy~\cite{PlatoErgodic, Ott}. \newedit{Each successive subset denotes a level of higher guaranteed randomness within the preceding level}. $\lambda$ indicates the maximal Lyapunov exponent, whose nonzero value is a defining signature of chaos~\cite{PlatoErgodic, Ott}.
(Reused from Ref.~\cite{EfimPolygon})}
    \label{fig:ErgodicHierarchy}
\end{figure}

Ergodic theory~\cite{Sinai1976, SinaiCornfield, HalmosErgodic} concerns itself with a study of the statistical properties of classical dynamical systems, centered around a mathematically precise classification of classical dynamics into different levels of randomness called the ergodic hierarchy~\cite{PlatoErgodic, Ott} (see Fig.~\ref{fig:ErgodicHierarchy}).  These levels, such as ergodic, mixing, K-mixing and others~\cite{HalmosErgodic, Sinai1976, SinaiCornfield, PlatoErgodic, Ott} (in order of increasing randomness, discussed in more detail in Sec.~\ref{sec:cl_ErgodicHierarchy}), can be used to motivate different elements of classical statistical mechanics~\cite{PlatoErgodic, Ott}: ergodicity justifies the use of the microcanonical ensemble, and mixing the approach to thermal equilibrium, while K-mixing is responsible for chaotic dynamics.

\edit{In quantum mechanics, on the other hand, it has remained unclear if any pertinent ``ergodic'' dynamical properties can even be defined~\cite{Haake}}. Instead, our present understanding of quantum statistical mechanics is founded on a much less precise, but empirically successful, connection to the statistical properties of random matrices~\cite{Haake, Mehta}. Direct contact with the thermalization of observables is made through a comparison of the energy eigenstates (or eigenvectors) \newedit{in a ``physical'' basis} of a system with random eigenvectors, via the Eigenstate Thermalization Hypothesis (ETH)~\cite{LLStatMech, deutsch1991eth, srednicki1994eth, srednicki1999eth, rigol2008eth, DAlessio2016, deutsch2018eth, subETH} and related approaches~\cite{tumulka_CT, CanonicalTypicalityPSW, NormalTypicality, ReimannRealistic, ShortEqb, Borgonovi2016, WuErgMix, Zelditch, Anantharaman}.
\newedit{While observables that satisfy ETH show thermalization behaviors resembling ergodicity and mixing~\cite{DAlessio2016, deutsch2018eth}, every system also has observables that do not satisfy ETH, and it is not well understood from first principles which of these are to be regarded as the ``physical'' observables --- except as determined empirically~\cite{DAlessio2016, Borgonovi2016, deutsch2018eth}}.



Such observable-dependent ambiguities are avoided in the comparison of the statistics of energy eigenvalues (i.e. level statistics) of a system with those of random matrices, at the apparent cost of direct dynamical relevance to thermalization. This approach is based on the observation that on quantization, typical classically non-ergodic systems show highly fluctuating energy spectra with Poisson (locally uncorrelated) level statistics~\cite{BerryTabor}, while typical classically chaotic systems show rigid spectra with the local level statistics of Wigner-Dyson random matrices~\cite{DKchaos, CGV, BerryStadium, BGS} (\newedit{within the eigenspaces of conserved quantities associated with all classical symmetries preserved on quantization; emergent quantum symmetries and conserved quantities must also be accounted for, e.g., in certain quantizations of billiards on arithmetic domains and cat maps~\cite{Haake}}). A semiclassical ``periodic orbit'' argument for Wigner-Dyson level statistics soon followed~\cite{HOdA, BerrySpectralRigidity} (with further developments in, e.g., Refs.~\cite{argaman, SieberRichter, HaakePO, HaakePO2, RichterReview}), assuming \newedit{semiclassical contributions from a ``uniform'' distribution of isolated periodic orbits in the classical phase space of} a K-mixing system~\cite{Haake}. \newedit{However, Wigner-Dyson statistics has also been seen numerically~\cite{GiraudWDwithoutPO, Lozej2021, CasatiWang} on quantization of merely mixing~\cite{Lozej2021, CasatiWang} and merely ergodic~\cite{GiraudWDwithoutPO, CasatiWang} systems without K-mixing/chaos, and even without periodic orbits~\cite{GiraudWDwithoutPO, CasatiWang, CasatiProsenTriangleMap}. These observations highlight the need for a theoretical understanding of spectral rigidity with regard to ergodicity, without relying on K-mixing or periodic orbits (as also anticipated in Ref.~\cite{CasatiProsenTriangleMap}).}
\vphantom{\cite{BlackHoleRandomMatrix}, \cite{ShenkerThouless}, \cite{KosProsen2018}, \cite{ChanScrambling}, \cite{ChanExtended}, \cite{BertiniProsen}, \cite{ChalkerSum}, \cite{SSS}, \cite{ExpRamp1}, \cite{ExpRamp2}, \cite{ExpRamp3}, \cite{ExpRamp4}, \cite{LiaoCircuits}, \cite{WinerSpinGlass}} 

Similar trends of spectral rigidity \newedit{being associated with ``chaotic'' behavior} have been observed analytically \newedit{(often using basis-dependent analogues of periodic orbit theory, for ensembles of systems)} and numerically in fully quantum many-body systems~\cites{BlackHoleRandomMatrix, ShenkerThouless, KosProsen2018, ChanScrambling, ChanExtended, BertiniProsen, ChalkerSum, SSS, ExpRamp1, ExpRamp2, ExpRamp3, ExpRamp4, LiaoCircuits, WinerSpinGlass, DAlessio2016},
 where judgements of the ``chaoticity'' of a system have been largely based on intuition.
\newadd{It remains unclear exactly what kind of ergodic dynamics is represented by spectral rigidity \textit{especially} in such fully quantum systems, in addition to those with a classical limit}.
\newedit{While} correlation functions of \textit{local} observables have been rigorously characterized in a manner similar to the ergodic hierarchy in the specific case of dual-unitary quantum circuits~\cite{ProsenErgodic, ClaeysErgodicCircuits, ArulCircuit}, \newedit{no direct link to level statistics is known even in this case}. \newedit{Further, the only known dynamical signature of spectral rigidity --- a (gradually weakening) \textit{suppression}} \edit{of extremely small (vanishing in the thermodynamic limit) quantum fluctuations representing recurrences at late times, called the ``ramp''~\cite{BlackHoleRandomMatrix}, appearing in autocorrelation} functions~\cite{SSS, ThoulessRelaxation} and quantum simulation protocols designed specifically to measure spectral rigidity~\cite{SFFmeas, pSFF} \newedit{--- is a subleading effect showing no transparent connection to the ergodic hierarchy}. \newadd{The fundamental open problem indicated here, of understanding the precise relationship between ergodicity and the basis-independent energy level statistics of a general quantum dynamical system, forms the central motivation for this work.}

\subsection{Summary of results}
\label{sec:summary}

\newedit{In this work, we develop an approach that provides a first general answer to the above problem within the fully quantum regime, by introducing suitable ergodic properties (independent of ``physical'' bases or observables) in the Hilbert space of a quantum dynamical system and deriving their formal connection to energy level statistics. Qualitatively, our approach is based on the following observations: 1. Cyclic permutations of a discrete set of states are the only invertible discrete processes (in other words, permutations of states) whose repeated action is ergodic, i.e., every initial state evolves into every other state in the set; 2. A quantum cyclic permutation of an orthonormal basis of states is a unitary operator whose eigenvalues are regularly spaced (roots of unity); 3. Spectral rigidity, as usually studied in quantum chaos, is essentially a measure of how close a given energy spectrum is to a regularly spaced spectrum. Thus, by defining a quantum version of ergodicity in terms of ``closeness'' to cyclic permutation dynamics in a finite-dimensional Hilbert space (further justified by a similar approach to classical ergodicity in the mathematical literature~\cite{SinaiCornfield, KatokStepin1, KatokStepin2}), we show that spectral rigidity is most directly a measure of this form of ergodicity in any quantum dynamical system.
In physical terms, the suppression of recurrences of initial states due to spectral rigidity, indicated by the ``ramp'' in quantum fluctuations, is what allows certain initial states to ``cyclically'' \textit{visit} other states under unitary dynamics (over the time scale of the ramp) due to the conservation of probability. This ``cyclic'' form of ergodicity encoded in the energy level statistics is, remarkably, quite distinct from the familiar quantum thermalization process of the \textit{spreading} of an initial state over a ``physical'' basis (which occurs before most of the ramp~\cite{ShenkerThouless}) associated with random eigenstates.}

\newedit{Formally, a rigorous treatment of classical ergodicity in terms of closeness to cyclic permutations of discretized cells in a continuum phase space (or ``cyclic approximate periodic transformations'') was developed in Refs.~\cite{KatokStepin1, KatokStepin2}} (see also Refs.~\cite{Sinai1976, SinaiCornfield, KatokSinaiStepin, Nadkarni} for reviews and related results). \newedit{Central to this treatment are a discrete counterpart to ergodicity and a discrete prerequisite for strong mixing (see Sec.~\ref{sec:cl_cyclic} for the precise relationship), which we respectively call cyclic ergodicity and aperiodicity. An abstract ``quantization'' of these properties, where the discretized phase space cells correspond to pure states in an orthonormal basis (qualitatively representing the smallest Planck-sized phase space cells allowed by quantum mechanics), yields the desired quantum notions of ergodicity. These \textit{extrapolate} the classical notions to fully quantum dynamics, rather than relying on any specific quantization techniques developed for various classical systems. Such an abstract approach is justified, and perhaps even necessitated, by the established difficulty~\cite{Zelditch, Anantharaman} of relating energy levels to ergodicity in a ``physical'' basis (see Sec.~\ref{sec:motivation}). Correspondingly,} our quantum definitions are in the context of a general autonomous (i.e. time-independent) unitary quantum dynamical system and do not rely on a classical limit.

\begin{figure}[!hbt]
    \centering
    \includegraphics[width=0.6\columnwidth]{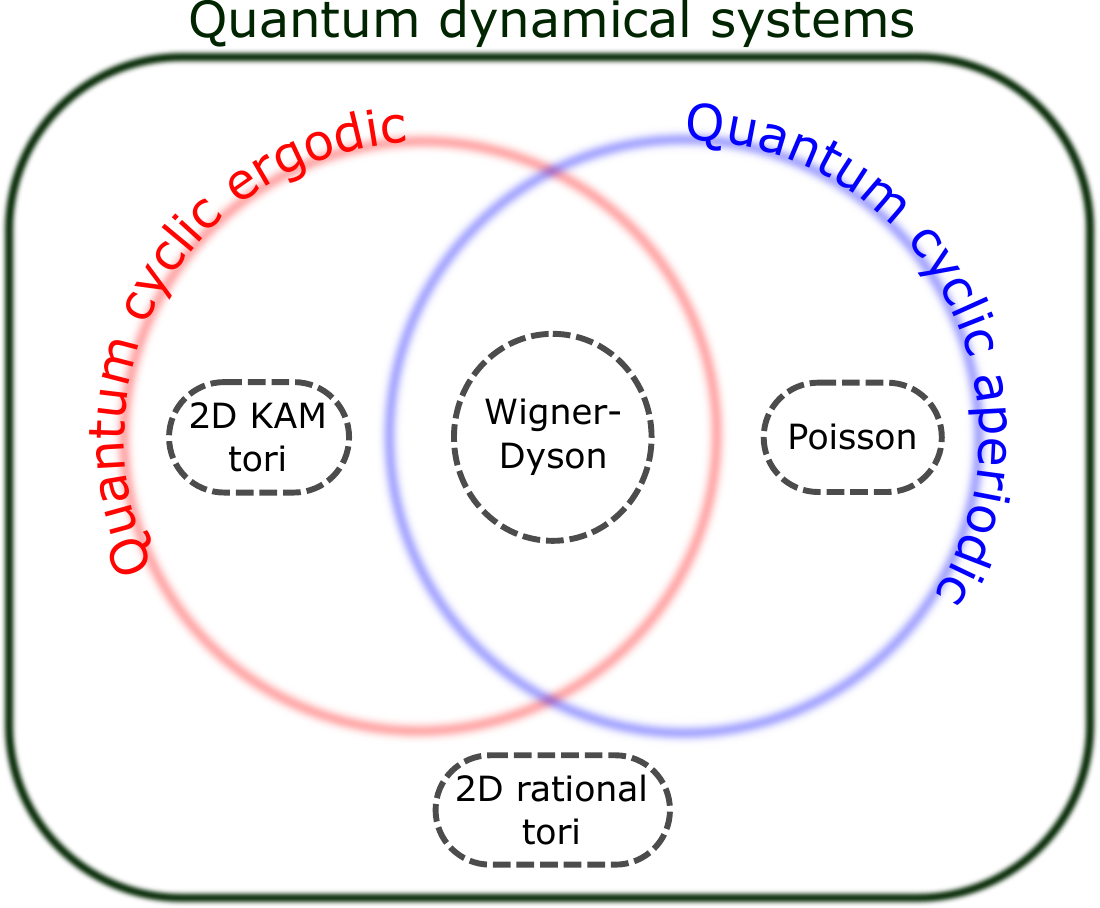}
    \caption{Depiction of the set of quantum dynamical systems in terms of the observable-independent ergodic properties proposed in this work, namely quantum cyclic ergodicity and aperiodicity. One distinct type of energy level statistics exemplifying each possibility for the presence or absence of these properties is shown (with dashed boundaries). The location of each such example in this diagram is analytically justified in the course of the main text.
    }
    \label{fig:QuantumErgodicProperties}
\end{figure}

\newadd{Some key physical takeaways from our approach are:
\begin{enumerate}
\item Classical cyclic ergodicity and aperiodicity generalize to quantum mechanics in a surprisingly direct way (unlike the continuum ergodic/chaotic properties of Fig.~\ref{fig:ErgodicHierarchy}) as fundamental quantum dynamical properties in the Hilbert space, including in systems without a classical limit. This allows a general observable-independent definition of quantum cyclic ergodicity and aperiodicity.
\item Quantum cyclic ergodicity requires the \textit{existence} of an orthonormal basis where every time-evolving basis state ``visits'' every other (fixed) basis state successively in a cyclic sequence; quantum cyclic aperiodicity requires the \textit{existence} of an orthonormal basis where no time-evolving basis state ``visits'' its original self; both requirements apply within a specified range of times. Here, a state ``visits'' another if their overlap exceeds that between two generic randomly chosen states.
\item Spectral rigidity directly characterizes the presence of quantum cyclic ergodicity and aperiodicity in a system, rather than conventional ergodicity, mixing or K-mixing as believed in the ``quantum chaos conjecture''~\cite{CGV, BGS}.
This clarifies the precise dynamical role played by level statistics in relation to the ergodic hierarchy, beyond a generic ``signature of quantum chaos''~\cite{Haake}.
\end{enumerate}
Thus, this work provides a system-independent framework that addresses the general connection between ergodic (quantum) dynamics and energy level statistics, in a way that captures the observable-independence of the latter.}
We further provide both analytical and numerical evidence for the applicability of this framework to different physically relevant types of energy level statistics: Wigner-Dyson (seen near-universally in intuitively ``quantum chaotic'' systems) and Poisson (seen near-universally in ``non-ergodic'' systems) \newedit{as associated with random matrix theory~\cite{Haake}}, and \newedit{also the non-random-matrix-like} spectral rigidity of quantized Kolmogorov-Arnold-Moser (KAM) tori (classically ergodic, non-mixing systems with no periodic orbits) and \newedit{fluctuating spectra in} rational tori (classically non-ergodic collections of periodic orbits) which occur as subsets of classically integrable systems~\cite{Ott}.  A depiction of the resulting logical relationships is shown in Fig.~\ref{fig:QuantumErgodicProperties}.

\subsection{Structure of this paper}
\label{sec:paperstructure}

\edit{The rest of this paper is organized as follows. The basic theory of observable-independent quantum ergodic properties and their connections to energy level statistics is set up in Secs.~\ref{sec:ErgodicReview}, \ref{sec:quantumcyclic}, and \ref{sec:levelstatistics}. Sec.~\ref{sec:ErgodicReview} reviews some necessary aspects of classical ergodic theory, including the use of cyclic permutations to ``discretize'' a dynamical system~\cite{KatokStepin1, KatokStepin2, SinaiCornfield, Sinai1976}, and defines cyclic ergodicity and cyclic aperiodicity \newedit{(Definitions \ref{def:cl_cyc_erg}, \ref{def:cl_cyc_apd})} as discrete, primitive properties related to ergodicity and mixing that can be extended to quantum mechanics. Sec.~\ref{sec:quantumcyclic} motivates and defines their analogues in quantum mechanics \newedit{(Definitions~\ref{def:qcycerg}, \ref{def:qcycapd}, \ref{def:QuantumErgodicity})}. Sec.~\ref{sec:levelstatistics} shows that quantum cyclic ergodicity and aperiodicity in a system are quantitatively determined, \newedit{through discrete Fourier transforms of the energy eigenstates\footnote{We emphasize that the fact that the connection to level statistics is realized in specific bases, unambiguously specified relative to the energy eigenbasis, does not reduce the basis-independence of the definitions. This is in the same sense in which the energy eigenvalues are basis-independent, even though a unitary time evolution operator is only diagonal in the energy eigenbasis.} (Eq.~\eqref{eq:DFTbasis})}, by specific measures of energy level statistics: namely, mode fluctuations~\cite{Lozej2021, Aurich1, Aurich2} \newedit{(Eq.~\eqref{eq:dft_cyc_ergodicity})} and the spectral form factor (SFF)~\cite{Haake} \newedit{(Eq.~\eqref{eq:dft_cyc_aperiodicity})}. Remarkably, Wigner-Dyson spectral rigidity emerges naturally at this stage, as a direct consequence of some simple dynamical conditions \newedit{(Proposition ~\ref{prop:GaussWD}); this allows an intuitive visualization of random matrix statistics in terms of ergodic Hilbert space dynamics (Fig.~\ref{fig:qcycpermutationExact})}.}

\edit{The subsequent sections are concerned with demonstrating applications to different
situations of physical interest. Sec.~\ref{sec:TypicalCyclic} considers
``typical'' systems, \newedit{relating quantum cyclic ergodicity to the ``ramp'' in the SFF via the conservation of probability (Sec.~\ref{sec:errorboundSFF})}, and provides detailed analytical \newedit{(Secs.~\ref{sec:errorboundSFF}, \ref{sec:GaussWD})} and numerical \newedit{(Fig.~\ref{fig:RMT_ergodicity})} evidence that quantum systems with Wigner-Dyson spectra are \newedit{(quantum cyclic)} ergodic and aperiodic, while those with Poisson spectra are aperiodic but not ergodic \newedit{in the appropriate subspaces} --- covering the \newedit{forms of level statistics associated with random matrices}~\cite{Haake, Mehta}. Sec.~\ref{sec:KAMtori} considers 2D linear flows on tori, for which we are able to explicitly construct cyclic permutations, and analytically prove both classical and quantum cyclic ergodicity and non-aperiodicity \newedit{(Theorem~\ref{thm:KAMcycerg})} as well as higher-than-Wigner-Dyson spectral rigidity \newedit{(Eq.~\eqref{eq:KAMtorus_spectralrigidity})} for every 2D KAM torus (where the linear flow has an irrational frequency ratio), with an analytical argument that the non-ergodic and non-aperiodic rational tori (with a rational frequency ratio) have comparable, but not identical, spectral rigidity to Poisson. This covers some physically interesting systems with atypical level statistics, and suggests a wider applicability of cyclic ergodicity and aperiodicity in characterizing quantum dynamics than random matrix theory, while establishing the possibility of analytical proofs of spectral rigidity for individual systems in this framework (which has been considered a mathematically and conceptually challenging problem~\cite{Zelditch, Anantharaman}).
Finally, Sec.~\ref{sec:discussion} discusses some insights about the connection between the above \newedit{eigenvalue-based} dynamical properties and thermalization \newedit{as determined by eigenstates} that may be gained from cyclic permutations, in a largely semi-qualitative manner that may motivate future rigorous work.}

\section{A short review of classical ergodic theory}
\label{sec:ErgodicReview}

In classical ergodic theory~\cite{HalmosErgodic, Sinai1976, SinaiCornfield, PlatoErgodic}, one is concerned with dynamics on a phase space (or a smaller region of interest, \edit{such as an energy shell of a Hamiltonian system}) $\mathcal{P}$, with an operator $\mathcal{T}^t:\mathcal{P}\to \mathcal{P}$ that evolves points in the space by the time $t$ (which may be a continuous or discrete variable, corresponding to flows or maps). The main questions of interest are which regions of phase space are explored over time by an initial point, and how rapidly a typical point explores these regions. These questions are conveniently posed when there is a measure $\mu(A) \geq 0$ defined for subsets $A \subseteq \mathcal{P}$ that is preserved by time evolution, $\mu(\mathcal{T}^t A) = \mu(A)$ (in Hamiltonian dynamics, this measure is given by the phase space volume $\int_A \diff^n q \diff^n p$). An important feature of such systems is guaranteed by the \textit{Poincar\'{e} recurrence theorem}~\cite{Sinai1976, HalmosErgodic, SinaiCornfield}: for any $A \subseteq \mathcal{P}$ such that $\mu(A) > 0$, almost every point in $A$ eventually returns to $A$, each within some (long) finite time (i.e. with the exceptions forming a set of measure zero). Given such a measure, how well an initial point explores the phase space is generally expressed through correlation functions of various sets, the behavior of which is classified into the ergodic hierarchy~\cite{Ott, PlatoErgodic}. In what follows, we normalize $\mu(\mathcal{P}) = 1$.

\subsection{The classical ergodic hierarchy}
\label{sec:cl_ErgodicHierarchy}

We first ask whether almost all initial points explore every region of nonzero measure in $\mathcal{P}$. If so, the dynamics is said to be \textit{ergodic} in $\mathcal{P}$. If not, $\mathcal{P}$ can be decomposed into (say) $M$ subsets that are invariant under $\mathcal{T}$, i.e. $\mathcal{P} = \bigcup_{j=1}^{M}\mathcal{P}_j$ (each with a measure induced by $\mu$), such that the dynamics is ergodic within each $\mathcal{P}_j$.
\edit{Ergodicity in $\mathcal{P}$ can be shown to be equivalent~\cite{HalmosErgodic,Sinai1976, SinaiCornfield, PlatoErgodic, Ott} to time-averaged correlations (e.g. presence of the system in a region $B \subseteq \mathcal{P}$) becoming independent of the initial condition (e.g. starting in the region $A \subseteq \mathcal{P}$) with only measure zero exceptions:}
\begin{equation}
    \lim_{T\to\infty} \frac{1}{2T}\int_{-T}^{T}\diff t\ \mu\left[(\mathcal{T}^t A)\cap B\right] = \mu(A)\mu(B),\ \forall\ A,B \subseteq \mathcal{P}.
    \label{eq:cl_ergodic}
\end{equation}
\edit{These measure zero exceptions could be isolated periodic orbits or other closed surfaces of lower dimension than $\mathcal{P}$}. Here, we use $\diff t$ either as a continuous integration measure or that corresponding to a discrete sum, depending on the domain of $t$.

\textit{Mixing} is a property in which time evolution eventually becomes uncorrelated with initial conditions, and represents how rapidly typical points explore a phase space region $\mathcal{P}$ on which time evolution is ergodic. \edit{Generically, a system is mixing if any ensemble $A$ with sufficiently many `neighboring' initial states (i.e. of nonzero measure) in $\mathcal{P}$ eventually spreads out equally according to certain criteria over every region of $\mathcal{P}$ (with deviations from this behavior allowed for a vanishing fraction of times in the case of \textit{weak} mixing).} The simplest such criterion is expressed in terms of two element correlation functions \edit{eventually becoming uncorrelated}~\cite{HalmosErgodic,Sinai1976, SinaiCornfield, PlatoErgodic, Ott}:
\begin{equation}
    \lim_{t\to\infty}\ \mu\left[(\mathcal{T}^t A)\cap B\right] = \mu(A)\mu(B),\ \forall\ A,B \subseteq \mathcal{P},
    \label{eq:cl_mixing}
\end{equation}
and is conventionally merely called mixing (\edit{weak mixing corresponds to the limit converging except at a vanishing fraction of times,} instead of exactly for all $t\to \infty$~\cite{HalmosErgodic, PlatoErgodic, Ott}). This can be extended to higher order correlation functions~\cite{SinaiCornfield, rokhlin1967}, and the dynamics is said to be \textit{K-mixing} when all higher order correlation functions become uncorrelated in the above sense~\cite{SinaiCornfield, Ott, PlatoErgodic}, \edit{which leads to chaotic behavior, e.g. nonzero Lyapunov exponents~\cite{Ott, PlatoErgodic}}. These criteria form a hierarchy in the sense that K-mixing implies mixing, which implies ergodicity~\cite{PlatoErgodic, Ott}. Additional levels of randomness may also be considered~\cite{PlatoErgodic, Ott}; see Fig.~\ref{fig:ErgodicHierarchy} for a depiction of the hierarchy of Ref.~\cite{PlatoErgodic}.

It is interesting to note that if one defines a unitary operator $U_{\mathcal{T}}$ induced by $\mathcal{T}$ \edit{on phase space functions $f(x \in \mathcal{P})$ via $U_{\mathcal{T}}^{t}f(x) \equiv f(\mathcal{T}^t x)$} (Koopman and von Neumann's Hilbert space representation of classical mechanics~\cite{HalmosErgodic, Sinai1976, SinaiCornfield, KatokStepin2, KatokSinaiStepin, Nadkarni}), some of these properties can be translated to those of the eigenvalues and eigenfunctions of $U_{\mathcal{T}}$, whose direct extensions to quantum mechanics have been previously considered~\cite{CastagninoErgodic1}. For instance, ergodicity translates to non-degenerate eigenvalues with eigenfunctions of uniform magnitude, and weak mixing to a continuous spectrum with no non-constant eigenfunction, of $U_{\mathcal{T}}$~\cite{HalmosErgodic}. For a discrete and finite quantum spectrum that corresponds to phase spaces or energy shells of finite measure by Weyl's law~\cite{Haake}, the eigenvalues are almost always non-degenerate (i.e. are non-degenerate or can be made so by infinitesimal perturbations) and the spectrum is necessarily discrete, prompting us to seek alternate avenues in which the above properties are at best emergent in the classical limit.

\subsection{Discretizing ergodicity with cyclic permutations}
\label{sec:cl_cyclic}

We eventually want to understand how quantum mechanics with its discrete set of energy levels can lead to ergodic and mixing behaviors, defined classically for continuous systems. A useful bridge between the continuum and discrete descriptions is offered by the technique of discretizing an arbitrary dynamical system in terms of cyclic permutations, which have been studied \edit{as ``cyclic approximate periodic transformations''} in Refs.~\cite{KatokStepin1, KatokStepin2, Sinai1976, SinaiCornfield, KatokSinaiStepin, Nadkarni}. Here, we discuss and adapt the elements of this framework that are most relevant for our purposes, following Ref.~\cite{KatokStepin2}, \edit{leading to the definition of cyclic ergodicity and cyclic aperiodicity respectively as a discrete version of ergodicity and a discrete prerequisite for strong mixing. These ideas are then illustrated with simple examples, which are directly used later in the study of quantized KAM tori in Sec.~\ref{sec:KAMtori}}.

\edit{A simple motivation for studying cyclic permutations is as follows. All invertible maps on a discrete set of states act as permutations of these states. Every permutation can be decomposed into a set of cyclic permutations, each acting on a separate subset of states. Among these, the only \textit{ergodic} permutations --- under whose repeated action every discrete state visits every other state --- are cyclic permutations of \textit{all} states with no further decomposition into subsets. This suggests the strategy of probing the ergodicity of an invertible autonomous dynamical system by comparing its dynamics to the repeated action of a cyclic permutation on some discretized states, in the continuum limit.}


To this end, let $\mathtt{C}= \lbrace C_k\rbrace_{k=0}^{n-1}$ be a decomposition of the phase space $\mathcal{P}$ into a large number of $\mu$-disjoint (i.e. with measure zero intersection) closed sets of identical measure, $\mu(C_k) = 1/n$; we will implicitly consider $\mathtt{C}$ to be part of a well-defined sequence of such decompositions, with $n\to\infty$ through a subset of the positive integers. Introduce a time evolution operator $\mathcal{T}_C$ on $\mathcal{P}$ that \edit{effects a cyclic permutation of the $C_k$}, i.e. $\mathcal{T}_C C_k = C_{k+1}$ (with $(n-1)+1 \equiv 0$ i.e. the addition is modulo $n$). \edit{For some choice of discrete time step $t_0$, we would like to see if successive actions of $\mathcal{T}^{t_0}$ on the $C_k$ closely resemble the cyclic permutation effected by $\mathcal{T}_C$}. Thus, as a measure of how well $\mathcal{T}_C$ approximates $\mathcal{T}^{t_0}$ \edit{over a single time step}, we define the single-step error of the permutation (differing from that in Ref.~\cite{KatokStepin2} by the factor of $1/2$):
\begin{equation}
    \eps_{C}(t_0) = \frac{1}{2}\sum_{k=0}^{n-1}\ \mu\left[(\mathcal{T}^{t_0}C_k) \symdiff C_{k+1}\right],
    \label{eq:cl_cyc_err}
\end{equation}
where $A \symdiff B = (A \cup B) - (A\cap B)$. \edit{The error measures the fraction of $\mathcal{T}^{t_0} C_k$ that lies outside $C_{k+1}$, averaged over all initial sets $C_k$.} 

\begin{figure}[!hbt]
\centering
\subfloat[][Partitioning of phase space into $\lbrace C_k\rbrace_{k=0}^{10-1}$.]{
    \centering
    \includegraphics[width=0.4\columnwidth]{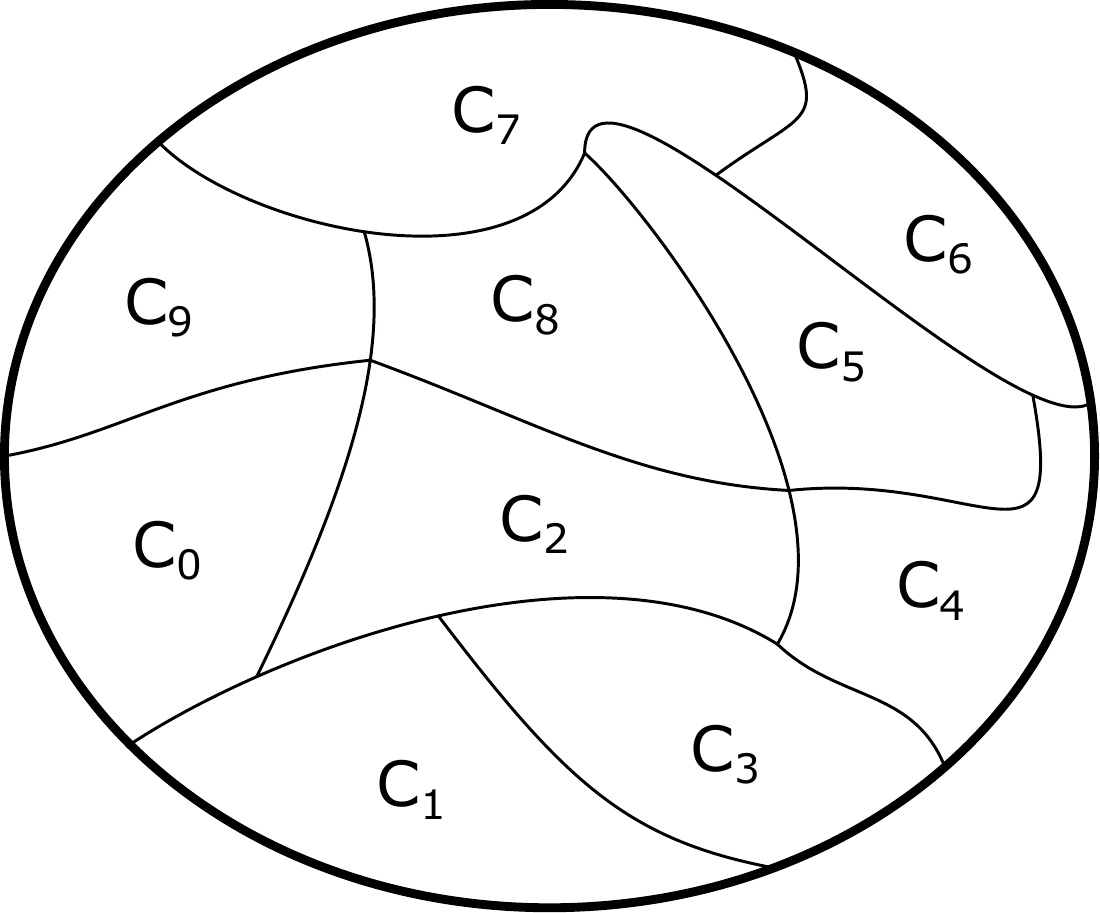}
    \label{fig:phasespacepartition}
}
\qquad \qquad
\subfloat[][Contribution to error from $(\mathcal{T}^{t_0} C_0) \symdiff C_1$ (shaded region).]{\includegraphics[width=0.4\columnwidth]{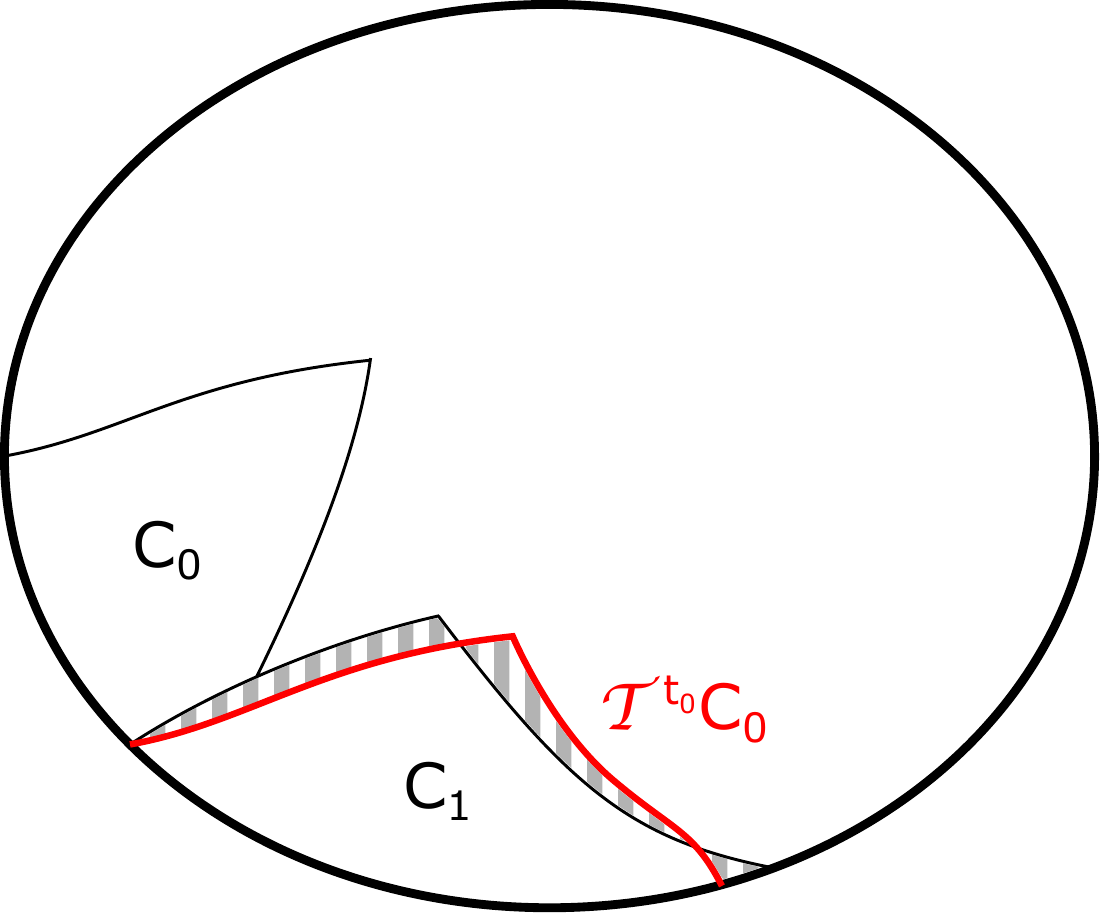}
    \label{fig:phasespaceerror}}
    \caption{Schematic depiction of an $(n=10)$-element cyclic permutation for some phase space $\mathcal{P}$ (interior of ellipse).}
    \label{fig:cycpermutation1}
\end{figure}

We will often directly call $\mathtt{C}$ the cyclic permutation in place of $\mathcal{T}_C$ for brevity, as any $\mathcal{T}_C$ that cyclically permutes the elements of $\mathtt{C}$ has the same error. We also note that \edit{we can choose $t_0$ to} depend on $n$, in particular \edit{with $t_0 \to 0$ as $n\to\infty$} for a flow with a continuous time variable, \edit{so that the discrete steps $\mathcal{T}^{\ptau t_0}$ with $\ptau \in \mathbb{Z}$ approach the continuous time flow $\mathcal{T}^t$ with $t \in \mathbb{R}$ as the decomposition becomes more refined with increasing $n$}. \edit{As a simple example, the rotation $\mathcal{T}^t\theta_0 = \theta_0 + \omega t$ (modulo $2\pi$, which will be left implicit in what follows) on a 1D circle with angular coordinate $\theta \in [0,2\pi)$ is approximated by the $n$-element cyclic permutation $\mathtt{C}$ comprising of equal intervals of the circle $C_k = \lbrace \theta \in [2\pi k/n, 2\pi (k+1)/n]\rbrace$, with \textit{zero} error if $t_0(n) = 2\pi/(n\omega)$. In contrast, higher dimensional ergodic systems typically take infinitely long to explore all of $\mathcal{P}$, and we expect $t_0 n\to \infty$ in addition to $t_0\to 0$ in such cases}. A schematic of a generic cyclic permutation is depicted in Fig.~\ref{fig:cycpermutation1}.

The single-step error $\eps_C(t_0)$ can serve as a probe of ergodicity. \edit{Refs.~\cite{KatokStepin1, KatokStepin2} showed that for a given dynamical system, the \textit{existence} of any cyclic permutation with error $\eps_C(t_0) < 2/n$ guarantees ergodicity as long as all nonzero measure sets can be constructed as unions of the $C_k$ in the $n\to \infty$ limit, while $\eps_C(t_0) < 1/n$ with $t_0 n \to \infty$ rules out (strong) mixing; a version of their proof is recounted in Appendix~\ref{app:cl_erg_errors} for the interested reader. However, these results allow room for cyclic permutations with a larger single-step error to lead to ergodicity or prohibit mixing. Here, we define \textit{cyclic ergodicity} (a notion implicit in their proofs) and \textit{cyclic aperiodicity} (based on an observation in Ref.~\cite{KatokStepin2}) as more fundamental discrete dynamical properties based on cyclic permutations that can be used to partly characterize the ergodic and mixing nature of dynamical systems as described below.}

\subsubsection{Classical cyclic ergodicity and aperiodicity}
\label{sec:KAMtoriCycErgApd}
\begin{figure}[!hbt]
\centering
\subfloat[][Cyclic ergodicity.]{\includegraphics[width=0.4\columnwidth]{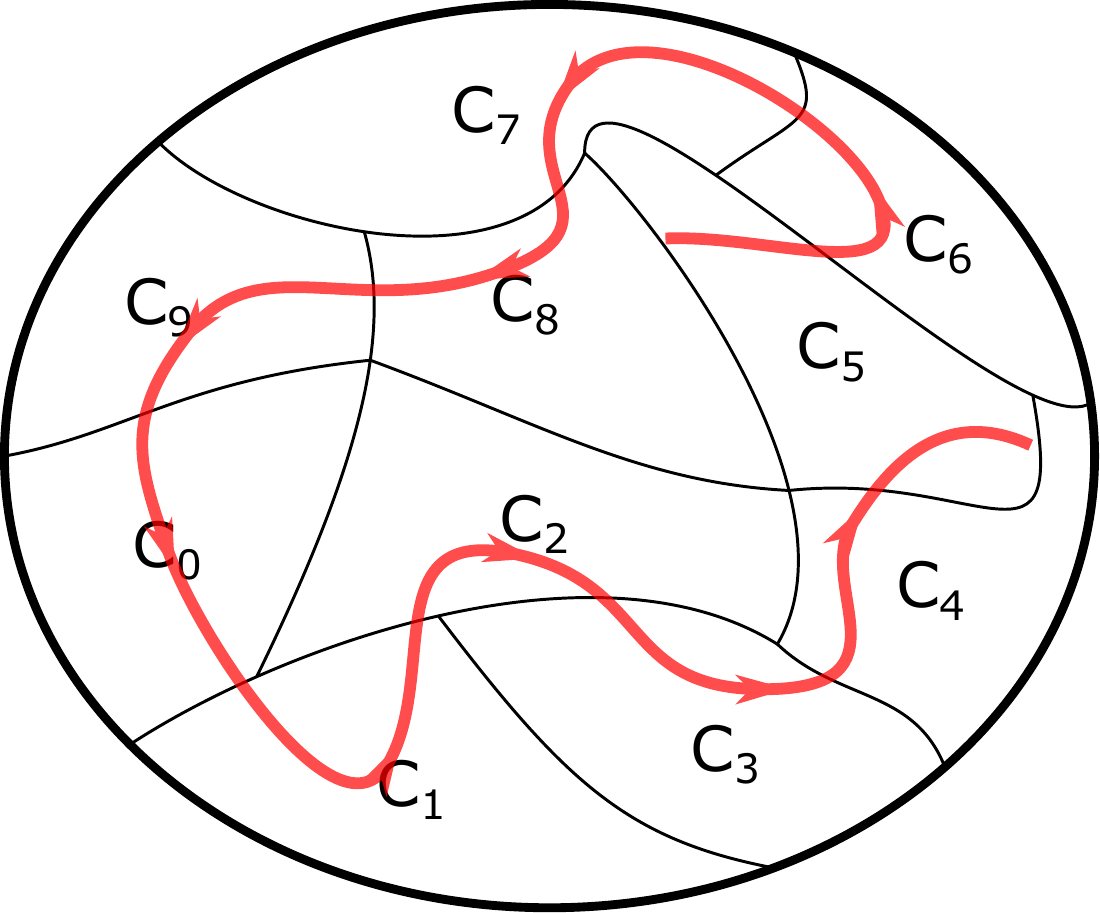}
    \label{fig:phasespaceergodicity}}
\qquad \qquad
\subfloat[][Non-ergodicity.]{\includegraphics[width=0.4\columnwidth]{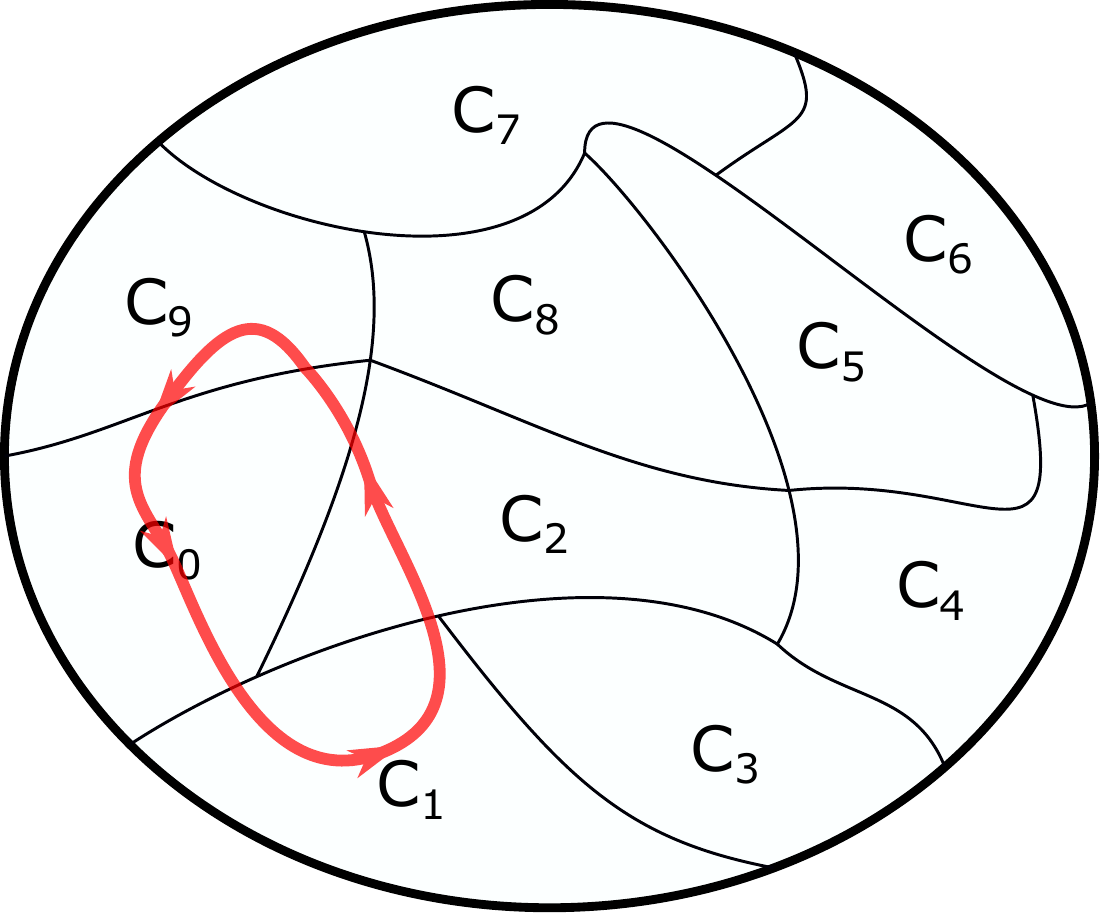}
    \label{fig:phasespacenonergodicity}}

    \caption{Schematic depiction of cyclic ergodicity and non-ergodicity for the cyclic permutation of Fig.~\ref{fig:cycpermutation1}. The trajectory may be thought of as the future and past history of the center of $C_0$, up to $\pm 5 t_0$ (arrows indicate the forward flow of time).}
    \label{fig:cycpermutation2}
\end{figure}

\edit{Cyclic ergodicity is a discrete counterpart to ergodicity in a continuous system, that directly formalizes the notion of ``(almost) all points visiting the neighborhood of every other point'' in $\mathcal{P}$; 
 the additional cyclic ordering of the ``points'' ensures that the discretized dynamical system is invertible (and first-order/``memoryless''). In terms of the decomposition $\mathtt{C}$, we want every $C_k$ to visit (i.e. have nonzero intersection with) every other $C_{k'}$ at least once during its future and past time evolution in steps of $t_0$. The cyclic structure (in the ordering of the $C_{k'}$s for all $C_k$s) can also be motivated as follows: if a given initial set $C_j$ intersects almost all of $C_k$ after one step of $\mathcal{T}^{t_0}$ and then $C_\ell$ after two steps, $C_k$ must intersect almost all of $C_\ell$ after one step. Formally, we have the following definition \editb{that considers a $C_k$ to visit another if their \textit{fractional} overlap (overlap divided by $1/n$, the measure of each $C_k$) is larger than a given precision $\zprec(n)$ as $n\to\infty$, such as $\zprec(n) = n^{-1}$} (depicted schematically in Fig.~\ref{fig:cycpermutation2}):}
\begin{definition}[\textbf{Classical cyclic ergodicity}]
A cyclic permutation $\mathtt{C}$ shows cyclic ergodicity \editb{with a given precision $\zprec(n) > 0$} iff any element $C_j \in \mathtt{C}$ sequentially intersects a sufficiently large fraction of every other $C_k \in \mathtt{C}$ at least once under (future and past) time evolution:
\begin{equation}
    n \mu\left[(\mathcal{T}^{\ptau t_0} C_j)\cap C_{j+\ptau}\right] > \zprec(n) \text{ as } n\to \infty,\ \forall\ j \text{ and }\lvert \ptau\rvert \leq \frac{n}{2},
    \label{eq:cl_cyc_erg_def}
\end{equation}
where $\ptau$ represents the number of integer steps of time evolution in units of $t_0$.
\label{def:cl_cyc_erg}
\end{definition}
\edit{For a given dynamical system, \textit{if} there exists a sequence \editb{(implicitly labeled by $n$)} of cyclic permutations $\mathtt{C}$ such that
\begin{enumerate}
\item $\mathtt{C}$ shows cyclic ergodicity \editb{for at least one choice of $\zprec(n)$}, and
\item $\mathtt{C}$ satisfies the \textit{generating property}: every nonzero measure set in $\mathcal{P}$ contains at least one $C_k$ in the $n\to \infty$ limit,
\end{enumerate}
then it follows rigorously that every nonzero measure set visits every other, and $\mathcal{T}^{t_0}$ is therefore ergodic in $\mathcal{P}$ (see Refs.~\cite{KatokStepin2, SinaiCornfield} and Appendix~\ref{app:cl_erg_errors}; we also recall that generally $t_0 \to 0$ for a continuous flow). It is also convenient to call the system cyclic ergodic in $\mathcal{P}$ \editb{for a given $\zprec(n)$}, without reference to the generating property, if it admits a sequence of cyclic ergodic $\mathtt{C}$ \newedit{\editb{with precision} $\zprec(n)$ (mainly to anticipate its quantum counterpart in Sec.~\ref{sec:quantumcyclic})}.}




\edit{For strong mixing (Eq.~\eqref{eq:cl_mixing}), we require any initial region e.g. $C_k$ to become spread out over all of $\mathcal{P}$ as $t\to \infty$. This requires that each $C_k$ \textit{on average} intersects no more than a vanishing fraction of itself in the $n\to \infty$ limit (so that $\mathcal{T}^t C_k$ is not preferentially distributed in $C_k$ for almost all $C_k$), for any $t$ with $t\to \infty$. Correspondingly, we call a system cyclic aperiodic if every sequence of cyclic permutations satisfies cyclic aperiodicity (a necessary condition for strong mixing):}

\begin{definition}[\textbf{Classical cyclic aperiodicity}]
A cyclic permutation $\mathtt{C}$ shows cyclic aperiodicity iff $\mathcal{T}^t C_k$ never returns to  intersect a non-vanishing fraction of $C_k$ on average, at all times later than $\overline{t}(n)$:
\begin{equation}
    \sum_{j=0}^{n-1} \mu\left[(\mathcal{T}^{t} C_j)\cap C_{j}\right] \to 0\ \text{as}\ n\to \infty,\ \forall\ t  > \overline{t}(n),
    \label{eq:cl_cyc_apd_def}
\end{equation}
for every $\overline{t}(n) \to \infty$ as $n\to \infty$.
\label{def:cl_cyc_apd}
\end{definition}

\edit{In light of these definitions, Refs.~\cite{KatokStepin1, KatokStepin2} effectively show that $\vareps_C(t_0) < 2/n$ for a cyclic permutation implies cyclic ergodicity, while $\vareps_C(t_0) < 1/n$ with $t_0 n \to \infty$ implies a violation of cyclic aperiodicity (note that the reverse implication in both cases is not always true), as the error generated in each step is not sufficient to lead to zero overlap of $\mathcal{T}^{\ptau t_0} C_j$ with $C_{j+p}$ by $p = n/2$ (thereby maintaining cyclic ergodicity) or $p=n$ (thereby maintaining cyclic ergodicity and violating cyclic aperiodicity) respectively; see Appendix~\ref{app:cl_erg_errors}. Thus, the existence of a sequence of cyclic permutations $\mathtt{C}$ with $\vareps_C(t_0) < 2/n$ or $\vareps_C(t_0) < 1/n$ for a dynamical system (satisfying the generating property, with $t_0 n \to \infty$) respectively implies that it is ergodic, or ergodic and not strongly mixing.}

\subsubsection{Examples}
\label{sec:cl_cyc_examples}
\edit{\textit{Example 1:} A simple illustration of these ideas is afforded by the example of rotations on a circle. A continuous rotation $\mathcal{T}^t\theta_0 = \theta_0+\omega t$ is ergodic (any initial point covers the entire circle) and not-mixing (the angular width of any initial region is preserved); as discussed after Eq.~\eqref{eq:cl_cyc_err}, the $n$-element cyclic permutation $\mathtt{C}$ with
\begin{equation}
    C_k = [2\pi k/n, 2\pi (k+1)/n] \text{ and } t_0 = 2\pi/(\omega n)
    \label{eq:example1_ck}
\end{equation}
approximates this system with $\eps_C(t_0) = 0$. $\mathtt{C}$ then shows cyclic ergodicity, and violates cyclic aperiodicity as $t\to \infty$ due to its periodicity, thereby implying ergodic and non-mixing behavior. It is also worth considering an additional static degree of freedom, e.g. a cylinder $\mathrm{Cyl} = \lbrace (\theta, z)\rbrace$ with $z \in [0,1]$ and $\mathcal{T}^t z = z$, in which case $\mathcal{T}$ is not ergodic on $\mathrm{Cyl}$. A physical example of this type is the 1D harmonic oscillator of frequency $\omega$, with conserved amplitude $A$ (with, say, $z = \tanh A$) and phase $\theta(t)$. Here, the cyclic permutation $\overline{\mathtt{C}}$ comprised of ``lengthwise strips'' $\overline{C}_k = C_k \times [0,1]$ (where $C_k$ is given by Eq.~\eqref{eq:example1_ck}) is cyclic ergodic \newedit{for any given $0 < \zprec(n) \leq 1$}; however, arbitrary nonzero measure sets in $\mathrm{Cyl}$ (particularly those that do not span the entire range of $z$ for all $\theta$, e.g. sets with $z \in [0,1/2]$) do not contain any $\overline{C}_k$. Thus, $\overline{\mathtt{C}}$ does not satisfy the generating property, remaining consistent with the non-ergodicity of $\mathcal{T}$ on $\mathrm{Cyl}$ in spite of its cyclic ergodicity.}

\edit{\textit{Example 2:} Somewhat more nontrivial is the discrete rotation on the circle,
\begin{equation}
\mathcal{T}^t \theta_0 = \theta_0 + 2\pi\alpha t,\ t \in \mathbb{Z},
\end{equation}
in steps of the angle $2\pi\alpha$ with $0 \leq \alpha < 1$, which is readily seen to be ergodic only for irrational $\alpha$ and decomposes into infinitely many periodic orbits for rational $\alpha$ (and is mixing in neither case). Here, the construction of cyclic permutations relies on the approximation of $\alpha$ by a rational number~\cite{IwanikRotation}. If $\alpha = (m+\delta_n)/n$ where $m$, $n$ are coprime integers and $\lvert \delta_n\rvert < 1/2$, we can construct an $n$-element cyclic permutation given by the intervals
\begin{align}
    C_k = [2\pi k m/n,\ &2\pi (km+1)/n] \text{ and } t_0 = 1, \nonumber \\
    &\text{ with error } \eps_C(t_0 = 1) = \lvert \delta_n\rvert.
    \label{eq:cl_irr_rotation_cyclic}
\end{align}
For all irrational $\alpha$, we can find an infinite sequence of coprime $m, n$ with $n\to \infty$ satisfying
\begin{equation}
    \lvert \delta_n\rvert < \frac{c}{n},\text{ for any constant } c \in \left[\frac{1}{\sqrt{5}},1\right),
    \label{eq:irrationalDiophantine}
\end{equation}
by Dirichlet's and Hurwitz' theorems on Diophantine approximations~\cite{DirichletHurwitzDiophantine}. The error then satisfies $\eps_C(t_0) < 1/n$ in any such sequence, establishing ergodicity (as the $C_k$ can be used to construct any finite interval as $n\to\infty$) as well as non-aperiodicity by the bounds~\cite{KatokStepin2, SinaiCornfield} discussed in Sec.~\ref{sec:KAMtoriCycErgApd} (see also Appendix~\ref{app:cl_erg_errors}), with the latter implying the absence of mixing (as $t_0 n \to \infty$). This leaves the case of rational $\alpha = p/q$ with coprime $p$ and $q$, for which it is useful to consider the regions $B_k = [2\pi k p/q, 2\pi (kp+1)/q]$; each point within any such $B_k$ lies on a different periodic orbit and therefore cannot visit another point in the same $B_k$. Split any such $B_{k}$ into two nonzero measure regions $B_{k1}$ and $B_{k2}$ which consequently never visit each other. Any cyclic permutation satisfying the generating property must possess some elements $C_{k1} \subseteq B_{k1}$ and $C_{k2} \subseteq B_{k2}$ as $n\to\infty$, which cannot visit each other by the preceding discussion; thus, no cyclic permutation satisfying the generating property can be cyclic ergodic for rational $\alpha$.}





\edit{In summary, this subsection discussed the connection between certain} properties of discretized classical dynamics and levels of the ergodic hierarchy. \edit{In particular, the \textit{existence} of at least one cyclic permutation satisfying cyclic ergodicity guarantees ergodicity (among regions of the phase space containing a discretized element), while mixing requires that every cyclic permutation satisfies aperiodicity (in the infinite time limit). We note that it is not generally known how to establish such properties for classical cyclic permutations except in cases with a sufficiently small single-step error~\cite{SinaiCornfield, KatokStepin1, KatokStepin2, IwanikRotation} (including the examples discussed above), but we will see that this difficulty is largely mitigated at a formal level after quantization.}

\section{Dynamical quantum ergodicity and cyclic permutations}
\label{sec:quantumcyclic}

This section aims to find parallels to the discretized classical ergodic properties of the previous section in quantum mechanics, \edit{which we propose as a starting point for a precise study of quantum ergodicity. In Sec.~\ref{sec:qm_cyclic}, we define quantum cyclic ergodicity and aperiodicity for cyclic permutations of orthonormal pure states in the Hilbert space, and subsequently the ergodicity and aperiodicity of a quantum system in any subspace of its Hilbert space --- which is such that the energy levels of the system encode all the relevant properties. In Sec.~\ref{sec:quantumerrorbounds}, we briefly discuss the formal quantum analogues of the classical error bounds (see Sec.~\ref{sec:cl_cyclic}) which can be used to constrain the time evolution of a cyclic permutation based entirely on the single-step error in certain cases; a more general characterization of quantum cyclic permutations requires a quantitative study of their connection to energy level statistics, which is taken up in Sec.~\ref{sec:levelstatistics}.}

\edit{We consider a general autonomous quantum system with unitary time evolution.} Let $\uh(t) = \uh^t(1)$ be the unitary time evolution operator, with $D$ (possibly nonunique) eigenstates $\left\lbrace\lvert E_n\rangle\right\rbrace_{n=0}^{D-1}$ and $D$ (correspondingly, possibly degenerate) eigenvalues $e^{-iE_n t}$:
\begin{equation}
    \uh(t)\lvert E_n\rangle = e^{-i E_n t}\lvert E_n\rangle.
    \label{eq:TheUnitary}
\end{equation}
The time variable $t$ can be chosen to be continuous or discrete, with $E_n$ respectively corresponding to the eigenvalues of a Hamiltonian or eigenphases of a Floquet map. Without loss of generality, we will use terminology associated with Hamiltonians in what follows.

\edit{It is worth noting that such an autonomous unitary evolution is itself never ergodic in the Hilbert space (even after restricting to normalized states), but always has $D$ conserved quantities $\lvert \langle E_n\vert \psi(t)\rangle\rvert^2$ for a time-evolving state $\lvert \psi(t)\rangle$. Moreover, all systems with rationally incommensurate/generic energy levels (including the vast majority of those considered both ``quantum chaotic'' and ``integrable/non-ergodic'') have no further conserved quantities, and consequently have the exact same number of ergodic subsets in the Hilbert space~\cite{DAlessio2016, deutsch2018eth}. Thus, a different conceptual basis is necessary to define and understand quantum ergodicity in an observable-independent manner, while maintaining a connection to some meaningful notion of ergodicity. The main takeaway from this section is that suitably defined quantum cyclic permutations of pure states are a promising candidate for this purpose, allowing a natural quantization of cyclic ergodicity and aperiodicity.}

\edit{Before embarking on a detailed discussion of quantum cyclic permutations (which may be defined in systems with or without a classical limit), we mention a semiclassical motivation for considering orthonormal pure state cyclic permutations. Weyl's law~\cite{Haake} (generally used for semiclassical calculations of the density of states) effectively assigns to each phase space region $A$ a number of orthonormal pure states $P(A) \propto \mu(A)$ in the semiclassical regime; see also Refs.~\cite{StechelMeasure,StechelHeller} for a related ``quantum measure algebra''. With $A=C_k$ in an $n$-element classical cyclic permutation $\mathtt{C}$, we have $\mu(C_k) = (1/n) \to 0$ classically as $n\to \infty$, suggesting that it is natural to associate the smallest number $P(C_k)=1$ of pure states with each $C_k$, i.e. to consider pure state cyclic permutations in the fully quantum description to represent the classical $n\to \infty$ limit. The invertibility of the cyclic permutation in the discretized classical system translates to the unitarity of the associated quantum cyclic permutation of an orthonormal basis.}


\subsection{Pure state cyclic permutations for quantum dynamics}
\label{sec:qm_cyclic}


We work in an invariant subspace $\Eshell{d} \subseteq \mathcal{H}$ (an `energy subspace') spanned by any subset of suitably relabeled eigenstates $\left\lbrace\lvert E_n\rangle\right\rbrace_{n=0}^{d-1}$; \edit{$\uh(t)$ will henceforth refer to the restriction of the time evolution operator to $\Eshell{d}$ unless specified otherwise. In practice, $\Eshell{d}$ may be chosen depending on convenience to be e.g. in most cases, an energy shell of a physical system spanned by all levels with energies in a range $[E, E+\Delta E]$ \newedit{(which is most likely to show ``ergodicity'' for any width $\Delta E$ less than the energy scale to which spectral rigidity extends as discussed in Sec.~\ref{sec:modefluctuations})}, or the restriction of such a shell to a subspace with fixed values of conserved quantities \newedit{showing spectral rigidity in systems with additional symmetries}. The main question of physical interest is whether a physical system is ergodic within such a (restricted) energy shell. However, the formal considerations of this section apply quite generally to any energy subspace.}

We seek pure state cyclic permutations that approximate $\uh(t)$ within this energy subspace. To this end, let $\mathcal{C} = \lbrace\lvert C_k\rangle\rbrace_{k=0}^{d-1}$ be an orthonormal basis spanning $\Eshell{d}$ with the unitary cycling operator $\uc\lvert C_k\rangle = \lvert C_{k+1}\rangle$. The eigenvalues of $\uc$ are necessarily distinct $d$-th roots of unity, $\lbrace\exp(-2\pi i n/d)\rbrace_{n=0}^{d-1}$. It is convenient to introduce the $\ptau$-step persistence amplitudes of $U_H(\ptau t_0)$ relative to the action of $\uc^\ptau$,
\begin{equation}
    z_k(\ptau, t_0) = \left\lvert \langle C_{k+\ptau}\rvert \uh(\ptau t_0)\lvert C_k\rangle\right\rvert,
    \label{eq:q_persistence_def}
\end{equation}
for some choice of $t_0$; these satisfy $z_k(\ptau, t_0) \in [0,1]$, \newedit{and represent the overlap amplitude between the time evolved $\lvert C_k\rangle$ and the original $\lvert C_{k+p}\rangle$}.
Then, we say that $\uc$ approximates $\uh(t_0)$ with $\ptau$-step error
\begin{equation}
    \vareps_C(\ptau, t_0) = 1-\left(\min_{k \in \mathbb{Z}_d} z_k(\ptau, t_0)\right)^2,
    \label{eq:q_err_def}
\end{equation}
where $\mathbb{Z}_{n} = \lbrace 0,\ldots,n-1\rbrace$. A pure state approximation scheme for unitaries has been constructed in Ref.~\cite{Nadkarni}, in analogy with certain classical non-cyclic transformations (indirectly related to classical cyclic permutations~\cite{ChaconApproximation}), to formalize results on e.g. the (non-)degeneracy of the classical unitary $U_{\mathcal{T}}$ in classical ergodic theory~\cite{Sinai1976, HalmosErgodic, ChaconApproximation, KatokStepin2}.
As we will see in Sec.~\ref{sec:DFTsAreOptimal}, the construction of pure state cyclic permutations as above allows us to go much further, and tackle non-trivial measures of the level statistics of $\uh(t)$ that can e.g. distinguish between Wigner-Dyson and Poisson statistics, \edit{seen respectively in typical ``quantum chaotic'' and ``non-ergodic'' systems}.

\begin{figure*}[!hbt]
\centering
\subfloat[][Cyclic ergodicity and aperiodicity.]{\includegraphics[width=0.4\textwidth]{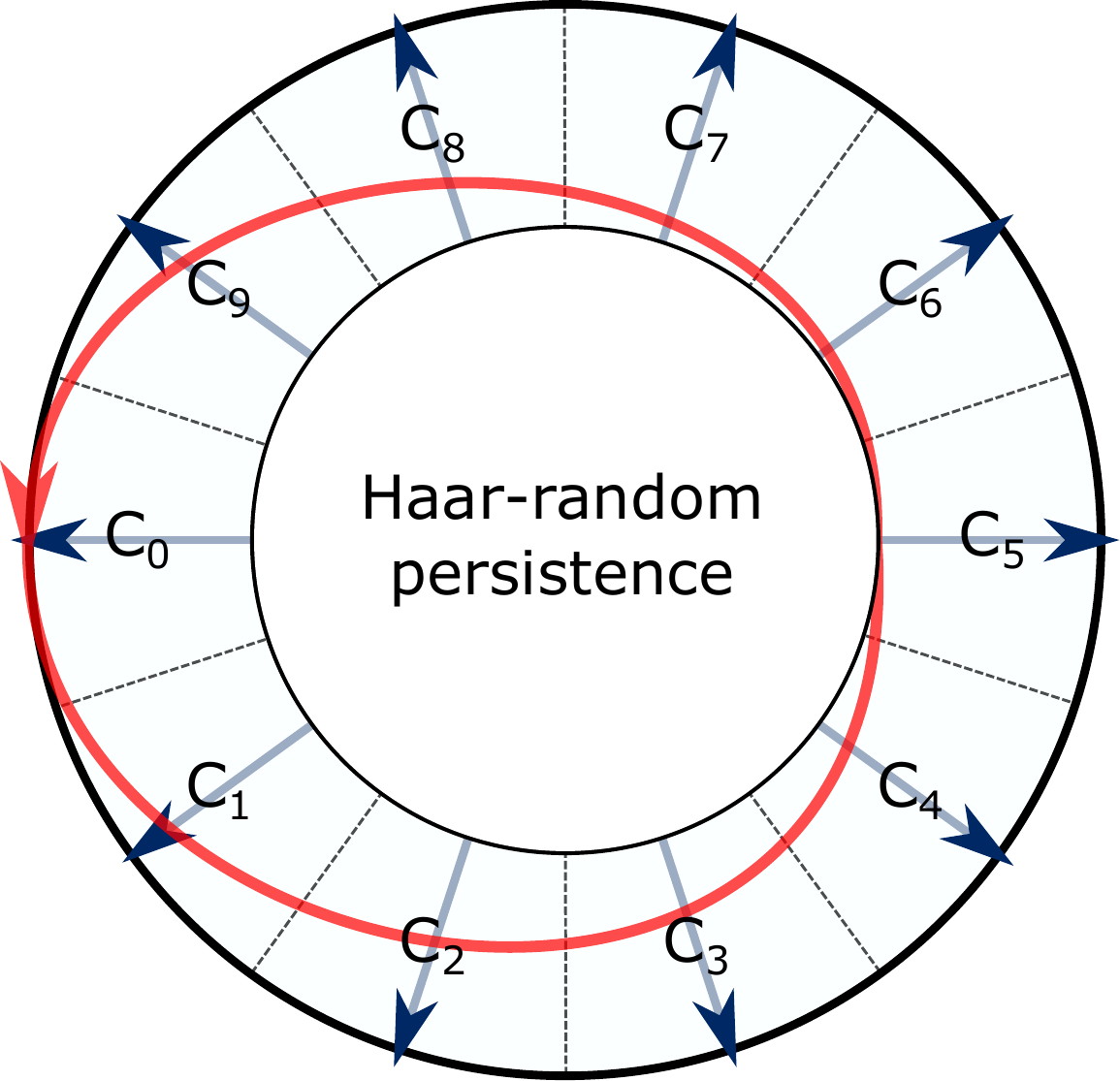} \label{fig:quantumergodicity}}
\qquad \qquad
\subfloat[][Non-ergodicity.]{\includegraphics[width=0.4\textwidth]{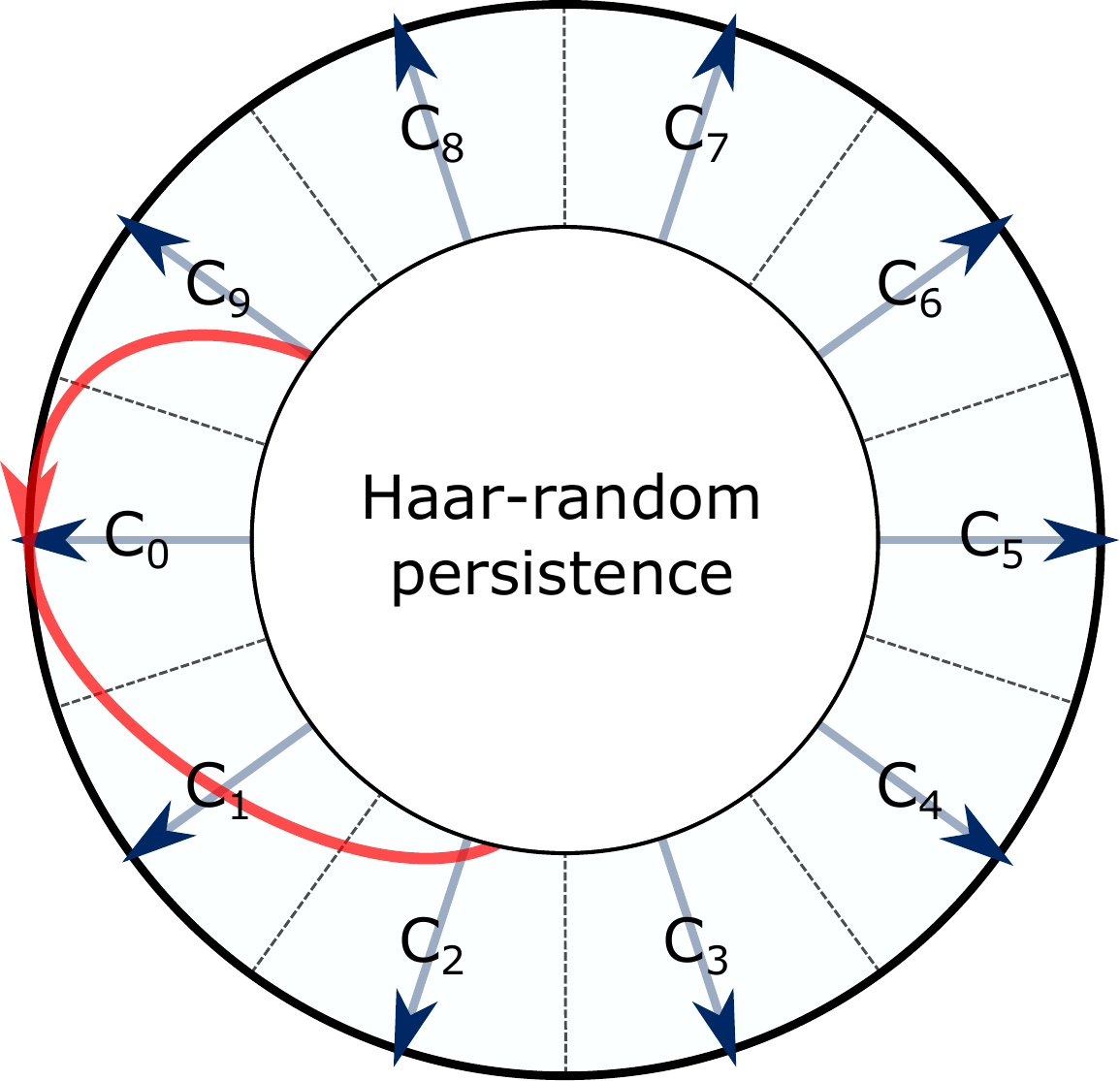} \label{fig:quantumnonergodicity}}

    \caption{Schematic depiction of (a) cyclic ergodicity and aperiodicity, and (b) non-ergodicity, for a $(d=10)$-element quantum cyclic permutation in a polar representation $(r,\theta) \in [0,1]\times [0,2\pi)$ of the corresponding $(d=10)$-dimensional Hilbert space $\eshell_d$; the angular direction is parametrized as $\theta = 2\pi p/d$, and the radial coordinate is $r = f\left(\left\lvert \langle C_0\rvert (\uc^\dagger)^p \lvert \psi\rangle\right\rvert\right)$ for any vector $\lvert \psi\rangle \in \eshell_d$, where $f:[0,1]\to [0,1]$ is some monotonic function and $\uc^p$ is extrapolated to non-integer $p$ in some convenient manner (e.g. connecting the $\lvert C_k\rangle$ along some smooth path). The basis vectors $\lvert C_k\rangle$ are depicted by arrows representing the corresponding axes. Each of (\ref{fig:quantumergodicity}) and (\ref{fig:quantumnonergodicity}) may be loosely regarded as a ``quantization'' of the respective classical versions in Fig.~\ref{fig:cycpermutation2}. The trajectory indicates the persistence amplitude $z_0(\ptau,t_0)$ of the initial state $\lvert C_0\rangle$ ($1$ at the outermost boundary, $0$ at the center) in the radial direction up to $\lvert \ptau \rvert = 5$ (visible trajectory) and beyond ("Haar-random persistence"). The region of "Haar-random persistence" refers to $z_0(\ptau, t_0) \lesssim O(d^{-1/2})$, which includes all typical Haar random states by canonical typicality~\cite{tumulka_CT, CanonicalTypicalityPSW}. Consequently, this region has by far the largest (Haar) volume in $\eshell_d$, while the depicted outer regions of non-random overlap with the $\lvert C_k\rangle$ together form a relatively tiny fraction of the space.}
    \label{fig:qcycpermutation}
\end{figure*}

In analogy with the definitions for classical cyclic permutations (Eqs.~\eqref{eq:cl_cyc_erg_def} and \eqref{eq:cl_cyc_apd_def}), we can define cyclic ergodicity and cyclic aperiodicity for these pure state quantum cyclic permutations as below (see Fig.~\ref{fig:qcycpermutation} for a schematic depiction, and Fig.~\ref{fig:qcycpermutationExact} in Sec.~\ref{sec:modefluctuations} for examples with exact numerical data).
\begin{definition}[\textbf{Quantum cyclic ergodicity}]
A pure state quantum cyclic permutation $\mathcal{C}$ shows cyclic ergodicity \editb{with precision $\zprec(d)$} iff
    \begin{equation}
            \left\lvert \langle C_{k+\ptau}\rvert \uh(\ptau t_0)\lvert C_k\rangle\right\rvert > \zprec(d),\ \forall\ k \text{ and } \lvert \ptau\rvert \leq d/2.
            \label{eq:q_cyclic_ergodicity}
        \end{equation}
        \label{def:qcycerg}
\end{definition}
Eq.~\eqref{eq:q_cyclic_ergodicity} states that ($\mathcal{C}$ shows cyclic ergodicity iff) any initial state $\lvert C_k\rangle \in \mathcal{C}$ ``visits'' all the other elements of $\mathcal{C}$ sequentially with \editb{sufficiently large overlap \newedit{(i.e. $z_k(\ptau, t_0) > \zprec(d)$)}}, at least once in its future and past evolution. Similarly, in place of a vanishing \textit{average} self-intersecting fraction for classical cyclic aperiodicity in Eq.~\eqref{eq:cl_cyc_apd_def}, we define quantum cyclic aperiodicity in terms of \editb{a sufficiently small} mean overlap amplitude of a pure state in $\mathcal{C}$ with itself, \editb{with an additional restriction to a time interval $(t_1, t_2)$ with $0 < t_1 < t_2$:}

\begin{definition}[\textbf{Quantum cyclic aperiodicity}]
A pure state quantum cyclic permutation $\mathcal{C}$ shows cyclic aperiodicity \editb{in a time interval $(t_1,t_2)$ with precision $\zprec(d)$} iff
    \begin{equation}
            \frac{1}{d}\sum_k \left\lvert\langle C_{k}\rvert \uh(t)\lvert C_k\rangle\right\rvert \leq \zprec(d),\ \forall\ t:\ \lvert t\rvert \in (t_1,t_2).
            \label{eq:q_cyclic_aperiodicity}
    \end{equation}
    \label{def:qcycapd}
\end{definition}


\edit{\editb{The question of interest is now the choice of precision $\zprec(d)$ and the time interval $(t_1,t_2)$ that are most useful in physical situations}. An important consideration for the former is suggested by canonical typicality~\cite{tumulka_CT, CanonicalTypicalityPSW}, which here refers to the observation that a typical ``Haar random''~\cite{Mehta} pure state $\lvert \psi\rangle$ in $\Eshell{d}$ (\newedit{by which we mean a normalized state $\langle\psi\vert\psi\rangle = 1$ that is randomly chosen with respect to the Haar/``uniform'' distribution, invariant under unitary transformations, in the space of such states}) has what we call a ``random'' overlap with every pure state in a given orthonormal basis $\lbrace \lvert \phi_k\rangle\rbrace_{k=0}^{d-1} \in \Eshell{d}$, namely $\lvert \langle \phi_k \vert \psi\rangle\rvert \in O(d^{-1/2})$, in a fairly strong sense (\newedit{i.e. $\sum_{j=1}^{n} \lvert \langle \phi_{k_j} \vert \psi\rangle\rvert^2 \approx n/d$ for large $n$}, appearing equally distributed~\cite{tumulka_CT, CanonicalTypicalityPSW} over large collections of the $\lvert \phi_k\rangle$). Thus, the action of $\uh(t_0)$ on any generic choice of cyclic permutation $\mathcal{C} \subset \Eshell{d}$, such that $\uh(t_0)\lvert C_k\rangle$ looks Haar random, is overwhelmingly likely to have $p$-step persistence amplitudes of $z_k(p; t_0) \in O(d^{-1/2})$ for almost all $p$ in \textit{any} system. \newedit{Consequently,} the simplest nontrivial definition of cyclic ergodicity for a quantum cyclic permutation (i.e. one not automatically satisfied by a generic cyclic permutation in every system) would be} \newedit{one that requires a larger overlap than this ``trivial'' random value to consider a given state as having ``visited'' another state.} \editb{Thus, the most physically useful choices of $\zprec(d)$ should formally satisfy $\zprec(d) \in O(d^{\varepsilon-1/2})$ for any $\varepsilon>0$ to account for all ``non-trivial'' overlaps and $\zprec(d) \notin O(d^{-1/2})$ to exclude trivial overlaps as $d\to\infty$, while taking values physically regarded as ``somewhat larger than'' $d^{-1/2}$ for finite $d$. For any such choice, we use the asymptotic notation:
\begin{align}
f(d) > \zprec(d) \implies f \gg O(d^{-1/2}), \label{eq:ggnotation} \\
f(d) \leq \zprec(d) \implies f \lesssim O(d^{-1/2}). \label{eq:lesssimnotation}
\end{align}
}

\editb{Now we consider the physically appropriate time interval $(t_1,t_2)$ (which stands} \edit{in contrast to allowing/requiring infinitely long times in classical aperiodicity}). The quantum recurrence theorem~\cite{QuantumRecurrences} (Poincar\'{e} recurrence for the flow $\varphi_n(t) = \varphi_n(0) -E_n t$ of phases $\varphi_n = \arg \langle E_n\vert \psi(t)\rangle$ in the energy eigenbasis) guarantees that aperiodicity will eventually be violated for every cyclic permutation after some extremely long time (possibly exponentially large in $d$; see e.g.~\cite{BrownSusskind2} for a related discussion of recurrences), \edit{and we also do not want to rule out recurrences for small times $t\sim O(1)t_0$ when the state is still ``near'' the initial state. For this reason}, \editb{$t_1/t_0 \notin O(1)$ and $d < t_2/t_0 \in O(d)$ are physically appropriate choices, in which case we will write
\begin{equation}
t_0 \ll \lvert t\rvert = O(t_0 d), \label{eq:timeintervalnotation}
\end{equation}
in place of $\lvert t\rvert \in (t_1,t_2)$ in Eq.~\eqref{eq:q_cyclic_aperiodicity}. To summarize, Definitions~\ref{def:qcycerg} and \ref{def:qcycapd} rigorously apply for any finite $d$, while our choice of precision $\zprec(d)$ and aperiodicity time interval $(t_1,t_2)$ suitable for physics are informed by (and mathematically rigorous in) the $d\to\infty$ limit.}

\edit{\newedit{We emphasize} that it is cyclic ergodicity and aperiodicity --- among classically relevant dynamical properties --- that have \textit{direct} quantum dynamical counterparts as given by Definitions~\ref{def:qcycerg} and \ref{def:qcycapd}, \newedit{allowing an observable-independent definition of quantum ergodicity in Definition~\ref{def:QuantumErgodicity} below. While} ``continuum'' (or conventional) ergodicity and mixing \newedit{are \textit{classically} fundamental and observable-independent, any attempted quantization appears to acquire a strong dependence on the choice of ``physical'' observables~\cite{DAlessio2016, deutsch2018eth, Borgonovi2016} due to the necessary involvement of eigenstates} (this is briefly discussed further in Sec.~\ref{sec:q_erg_mix})}. Consequently, we will often drop the ``cyclic'' qualifier for quantum cyclic ergodicity and aperiodicity, \newedit{and treat them as the fundamental quantum ergodic properties of interest} in the remainder of this paper.

 

It is now convenient to define when a quantum system is ergodic or aperiodic in an energy subspace $\Eshell{d}$. \edit{For ergodicity, we require the \textit{existence} of an ergodic cyclic permutation in $\Eshell{d}$, with the range of the time step $t_0$ restricted by $t_0 d < T$, for $T$ small enough to avoid \newedit{the time scale at which quantum recurrences may unavoidably occur in any system}. For aperiodicity, we note that a cyclic permutation composed of the energy eigenstates, e.g. $\lvert C_k\rangle = \lvert E_k\rangle$, violates aperiodicity (and ergodicity)  with the maximum possible value ($1$) for the mean overlap in Eq.~\eqref{eq:q_cyclic_aperiodicity};
we are therefore able to construct non-aperiodic cyclic permutations \newedit{(those that violate Eq.~\eqref{eq:q_cyclic_aperiodicity})} for any quantum system and cannot require all cyclic permutations to be aperiodic (as a prerequisite for mixing), a situation which has no classical analogue in the absence of superpositions. On the other hand, the aperiodicity of a cyclic permutation $\mathcal{C}$ is governed by the following inequality:
\begin{equation}
    \frac{1}{d}\sum_k \left\lvert\langle C_{k}\rvert \uh(t)\lvert C_k\rangle\right\rvert \geq \frac{1}{d}\left\lvert \Tr\left[\uh(t)\right]\right\rvert,
    \label{eq:apd_trace_bound}
\end{equation}
which follows from writing the trace in the $\lbrace \lvert C_k\rangle\rbrace$ basis and using the triangle inequality for complex numbers (we will see in Sec.~\ref{sec:DFTsAreOptimal} that this inequality is saturated by some cyclic permutation in every system). Thus, systems with sufficiently large ($\gg O(d^{1/2})$) trace of $\uh(t)$ would possess no cyclic permutation satisfying aperiodicity. Correspondingly, it is natural to call a quantum system aperiodic if it admits an aperiodic cyclic permutation.}
We then have the following definitions of ergodicity and aperiodicity, pertaining to a dynamical (i.e. time-domain) version of quantum ergodicity (which is distinct from the use of ``quantum ergodicity'' in the mathematical literature to refer to the delocalization of energy eigenstates in a given basis~\cite{Zelditch, Anantharaman}):

\begin{definition}[\textbf{Ergodicity and aperiodicity of a quantum system}]
We call a quantum system (dynamically) \textbf{ergodic} \editb{with precision $\zprec(d)$} in the energy subspace $\Eshell{d}$ within a time $T > 0$, if it admits at least one cyclic permutation \editb{that is ergodic with precision $\zprec(d)$, satisfying} $t_0 d < T$. Similarly, the system is \textbf{aperiodic} \editb{in $(t_1,t_2)$ with precision $\zprec(d)$} in $\Eshell{d}$ within the time $T$ if it admits at least one cyclic permutation aperiodic \editb{in $(t_1,t_2)$ with precision $\zprec(d)$, satisfying} $t_0 d < T$.
\label{def:QuantumErgodicity}
\end{definition}
\editb{While this definition is stated in general terms, we will assume the physically motivated precision and time interval specified by Eqs.~\eqref{eq:ggnotation}, \eqref{eq:lesssimnotation} and \eqref{eq:timeintervalnotation} in all physical applications considered later.}

As a simple example, every system is always ergodic and not aperiodic in any subspace $\Eshell{1}$ consisting of a single energy level. For typical quantum systems, we will implicitly assume  a choice of $T$ that is as large as possible while being much less than the quantum recurrence time scale \newedit{for a generic quantum system, e.g. evolution by a (Haar) random unitary in $\Eshell{d}$}. In general, identifying which energy subspaces $\Eshell{d}$ of the system satisfy these properties would provide an observable-independent characterization of the ergodicity of a quantum system. \edit{In anticipation of Sec.~\ref{sec:DFTsAreOptimal}, we emphasize that Definition \ref{def:QuantumErgodicity} is insensitive to unitary transformations of $\Eshell{d}$ (which would simply map cyclic permutations in $\Eshell{d}$ to each other), and consequently independent of any observables or measurement bases that one may consider; whether a system is ergodic or aperiodic in the above sense is then determined entirely by the unitary invariants of $\uh(t_0)$ within this subspace: the energy eigenvalues $\lbrace E_n\rbrace_{n=0}^{d-1}$}.

\subsection{Quantum bounds from the single-step error}
\label{sec:quantumerrorbounds}
\edit{Similar to the classical case, we can rigorously prove ergodicity or disprove aperiodicity for a cyclic permutation given just the single-step error $\vareps_C(1,t_0)$. In the quantum case, this is made possible by noting that the single-step deviation $\uerr = \uc^\dagger \uh(t_0)$ of time evolution from a cyclic permutation is unitary, corresponding to a (complex) rotation in the Hilbert space $\Eshell{d}$;  it can thus be described by effective angles of deviation $\theta_k(\ptau) = \arccos z_k(\ptau,t_0)$ of each $\uh(\ptau t_0)\lvert C_k\rangle$ from $\uc^\ptau\lvert C_k\rangle$. A simple application of the triangle inequality in $\Eshell{d}$ (see Appendix~\ref{app:q_erg_errors}) shows that the $p$-step angle of deviation cannot exceed the sum of the corresponding single-step angles (with the bound being saturated for a 2D rotation of a vector in successive steps of the $\theta_k(p)$):}
\begin{equation}
    \min\left\lbrace\theta_k(p), \frac{\pi}{2}\right\rbrace \leq \sum_{q=0}^{p-1} \theta_{k+q}(1).
    \label{eq:q_persistence_bound}
\end{equation}
\edit{Using $\vareps_C(\ptau,t_0) = \sin^2 \theta_k(p)$, it follows from here that $\vareps_C(1,t_0) < \pi^2/d^2$ implies ergodicity and $\vareps_C(1,t_0) < \pi^2/(4d^2)$ implies non-aperiodicity for $\mathcal{C}$ (for large $d$); we emphasize that these are one-way implications. These results are of interest to the extent that they are the direct extension of the classical bounds of Refs.~\cite{KatokStepin1, KatokStepin2} (see Sec.~\ref{sec:cl_cyclic}) to pure state cyclic permutations; however, they are similarly restricted in their applicability to special systems that admit sufficiently small single step errors as in the classical case. In the next section, we will show that an analysis in terms of the energy level statistics of $\Eshell{d}$ allows one to characterize ergodicity and aperiodicity in much more general terms in the quantum case.}

\section{Optimal cyclic permutations and energy level statistics}
\label{sec:levelstatistics}

\edit{In this section, we explicitly identify how specific measures of energy level statistics determine the ergodicity and aperiodicity of a quantum system. Sec.~\ref{sec:DFTsAreOptimal} constructs certain cyclic permutations directly related to the energy eigenstates (specifically, their discrete Fourier transforms) and establishes results concerning their optimality for determining ergodicity and aperiodicity. Building on these results, Sec.~\ref{sec:DFTexpressions}} shows that ergodicity is directly determined by the mode fluctuation distribution~\cite{Aurich1, Aurich2, Lozej2021} of the energy levels, while aperiodicity is directly determined by the spectral form factor~\cite{Haake}.

\subsection{Optimizing ergodicity and aperiodicity with Discrete Fourier Transforms of energy eigenstates}
\label{sec:DFTsAreOptimal}



\subsubsection{Discrete Fourier Transforms of energy eigenstates}

\edit{To establish the ergodicity or aperiodicity of a general system in a subspace $\Eshell{d}$, given the corresponding energy levels $\lbrace E_n\rbrace_{n=0}^{d-1}$, it is convenient to explicitly identify ``optimal'' cyclic permutations in $\Eshell{d}$ which are the ``most likely'' to be ergodic and/or aperiodic in a system. A special role in this regard is played by the set of cycling operators which are diagonal in the energy eigenbasis:}
\begin{equation}
    \uc[q] = \sum_{n=0}^{d-1}e^{-2\pi i n/d}\lvert E_{q(n)}\rangle\langle E_{q(n)}\rvert,
    \label{eq:DFTuc}
\end{equation}
where $q:\mathbb{Z}_d \leftrightarrow \mathbb{Z}_d$ ranges over all permutations of the indices $n \in \mathbb{Z}_d = \lbrace 0,\ldots,d-1\rbrace$. This corresponds to cyclic permutations $\mathcal{C}$ which can be written as a discrete Fourier transform (DFT) of the energy eigenstates,
\begin{equation}
    \lvert C_k(q,\varphi_n)\rangle = \frac{1}{\sqrt{d}}\sum_{n=0}^{d-1}e^{-2\pi i n k / d}e^{-i\varphi_n}\lvert E_{q(n)}\rangle,
    \label{eq:DFTbasis}
\end{equation}
for arbitrary phases $\varphi_n$ that don't influence the persistence amplitudes.

\edit{In Sec.~\ref{sec:ergoptimalcyclic}, we show that the minimum value $\vareps_{\min}(\ptau,t_0)$ of $\vareps_C(\ptau,t_0)$ for a given $p$, among all cyclic permutations $\mathcal{C}$ in $\Eshell{d}$, occurs when the corresponding $\uc$ has the form in Eq.~\eqref{eq:DFTuc} by 1. Theorem~\ref{thm:dftoptimal} when $\vareps_{\min}(\ptau,t_0) < 2/d$ (``small'' errors), and 2. a heuristic argument for ``generic'' energies $\lbrace E_n\rbrace_{n=0}^{d-1}$ when $\vareps_{\min}(\ptau,t_0) > 2/d$ (``large'' errors). This shows that an ergodic system is most likely to possess an ergodic cyclic permutation among the set satisfying Eqs.~\eqref{eq:DFTuc} and \eqref{eq:DFTbasis}.
Further, in Sec.~\ref{sec:apdoptimalcyclic}, we prove that the system is aperiodic in $\Eshell{d}$ \textit{if and only if} any and all cyclic permutations of the form in Eq.~\eqref{eq:DFTbasis} are aperiodic. Thus, the ``most likely'' aperiodic cyclic permutation is also given by the above set. The involvement of energy eigenstates in Eqs.~\eqref{eq:DFTuc} and \eqref{eq:DFTbasis} naturally connects ergodicity and aperiodicity to functions of the energy levels $\lbrace E_n\rbrace_{n=0}^{d-1}$. Later, in Sec.~\ref{sec:DFTexpressions}, we discuss how these properties are determined by concrete measures of energy level statistics.}

\edit{We note that Eq.~\eqref{eq:DFTbasis} allows an intuitive interpretation of why DFTs of energy eigenstates correspond to optimal cyclic permutations: states of this form (particularly when $q$ sorts the $E_{q(n)}$ in ascending order with $n$) are the closest one can get to defining approximate ``time eigenstates'' $\lvert C_k\rangle$ with quantized ``time eigenvalue'' $k$ (of a fictitious ``time operator'' conjugate to the energy in $\Eshell{d}$, keeping in mind the Fourier relation between e.g. canonically conjugate position and momentum variables in quantum mechanics~\cite{ShankarQM}). Time evolution by $\uh(t_0)$ should then ideally cause the time eigenvalue $k$ to be incremented by $1$ in such a state, essentially functioning as a cyclic permutation of the $\lvert C_k\rangle$ (up to small errors caused by discretization). In other words, $\uh(t_0) \approx \uc$ with arguably minimal error for such ``time'' eigenstates.}


\subsubsection{Optimizing ergodicity via persistence amplitudes}
\label{sec:ergoptimalcyclic}


The optimality of cyclic permutations satisfying Eq.~\eqref{eq:DFTuc} for small errors is formalized by the following theorem.

\begin{theorem}[\textbf{Optimal cyclic permutations}]
If the system (in some energy subspace $\eshell_d$) admits some cyclic permutation $\mathcal{C}'$ with $p$-step error $\vareps_{C'}(\ptau, t_0) \leq (2/d)$ for a given $p$ and $t_0$, then $\vareps_C(\ptau, t_0)$ attains its minimum value among all cyclic permutations for a cyclic permutation $\mathcal{C}$, whose cycling operator $\uc$ satisfies
\begin{equation}
    \lim_{\delta \to 0}\left[\uh(t_0)e^{i\delta\hat{Y}},\uc\right] = 0.
    \label{eq:dftoptimal}
\end{equation}
Here, $\hat{Y}$ is any fixed Hermitian operator (which effectively selects a unique eigenbasis of $\uh(t_0)$ if the latter is degenerate). In particular, the global minimum of the error is achieved by one such $\uc$ for every choice of $\hat{Y}$.
\label{thm:dftoptimal}
\end{theorem}
\begin{proof}[Outline of proof]
\edit{The proof of this statement is outlined below in the special case of $p=1$ (with $\uh \equiv \uh(t_0)$); the full proof may be found in Appendix~\ref{app:q_cyc_dft}.}

We note that unitary transformations $\hat{V}$ on $\Eshell{d}$, being the most general orthonormality preserving linear transformations, can generate all possible cyclic permutations from a given $\mathcal{C}$ via $\lvert C_k\rangle \to \hat{V}\lvert C_k\rangle$, which induces a transformation $\uc \to \hat{V} \uc \hat{V}^\dagger$. Our initial objective is to identify the global maximum of the $1$-step mean persistence,
\begin{equation}
    \meanz(1,t_0) \equiv \frac{1}{d}\sum_{k=0}^{d-1}z_k(1,t_0),
    \label{eq:outline_mean_persistence}
\end{equation}
over all cyclic permutations.
We have $\lvert \Tr(\uc^\dagger \uh)\rvert/d \leq \meanz(1,t_0)$ (as is evident from expressing the trace in the $\mathcal{C}$ basis); on the other hand, there always exists a unitary transformation $\lvert C_k\rangle \to \lvert C''_k\rangle = e^{i\varphi_k}\lvert C_k\rangle$
to a basis $\mathcal{C}''$, leaving the $z_k(1,t_0)$ unchanged, such that $\lvert \Tr(\ucb^\dagger \uh)\rvert/d = \meanz(1,t_0)$. Thus, a cyclic permutation that maximizes the trace overlap also necessarily maximizes $\meanz(1,t_0)$, with the same maximum value:
\begin{equation}
    \max_{\text{all}\ \hat{V}}\ \frac{1}{d}\left\lvert\Tr\left[\hat{V} \uc^\dagger\hat{V}^\dagger \uh \right]\right\rvert = \max_{\text{all}\ \mathcal{C}}\ \meanz(1,t_0).
\end{equation}
The maxima of the trace overlap can occur only when it is stationary with respect to small variations of $\uc$ --- effected by $\hat{V} = e^{i\hat{X}}$ for all small, Hermitian $\hat{X}$. Correspondingly, imposing stationarity via $\lvert \Tr(e^{i\hat{X}} \uc^\dagger e^{-i\hat{X}} \uh)\rvert = \lvert \Tr( \uc^\dagger \uh)\rvert + O(X^2)$ gives
\begin{equation}
    \uh\uc^\dagger-\uc^\dagger \uh = \hat{F}e^{i\alpha} \text{ for some } \hat{F} = \hat{F}^\dagger,
    \label{eq:outline_stationary_points}
\end{equation}
where $\alpha = \arg[\Tr(\uc^\dagger \uh)]$, as a necessary condition for a given cyclic permutation $\mathcal{C}$ with cycling operator $\uc$ to maximize both $\lvert \Tr(\uc^\dagger \uh)\rvert$ and $\meanz(1,t_0)$. We then have two distinct cases of interest.

\begin{enumerate}[wide, labelwidth = 0pt, labelindent = 0pt, label = \textbf{\arabic*}.]
    \item $\hat{F} = 0$: Solutions to Eq.~\eqref{eq:outline_stationary_points} with $\hat{F} = 0$ (equivalently, those satisfying Eq.~\eqref{eq:dftoptimal}) exist in every system, corresponding precisely to the set of cyclic permutations satisfying Eq.~\eqref{eq:DFTuc}. For such solutions, we have $z_k(1,t_0) = \meanz(1,t_0)$ for all $k$ as $\langle C_{k+\ell}\rvert \uh\lvert C_{j+\ell}\rangle = \langle C_{k}\rvert \uh\lvert C_{j}\rangle$. If the global maximum $\meanz_{\max}(1,t_0)$ of the mean persistence occurs at such a solution, it must also have the smallest error by Eq.~\eqref{eq:q_err_def}, as $\min_k z_{k,\max}(1,t_0) = \meanz_{\max}(1,t_0)$ in this case, while
    \begin{equation}
        \min_k z_k(1,t_0) \leq \meanz(1,t_0) \leq \meanz_{\max}(1,t_0)
    \end{equation}
    for any cyclic permutation.
    
    \item $\hat{F} \neq 0$: The status of solutions with $\hat{F} \neq 0$ is much less clear, with respect to whether they exist in a given system, and if so, whether such solutions can include the global maximum of $\meanz(1,t_0)$. To analyze this further, we write
    \begin{equation}
        \uh \uc^\dagger = e^{i\hat{A}}e^{i\alpha} \text{ and } \uc^\dagger \uh = e^{i\hat{B}}e^{i\alpha},
        \label{eq:UDeltaParametrization}
    \end{equation}
    for \textit{Hermitian} $\hat{A}, \hat{B}$. We then have the following properties when $\uc$ is a solution to Eq.~\eqref{eq:outline_stationary_points}:
    \begin{align}
        \lbrace\text{Eigenvalues of } e^{i\hat{A}}\rbrace &= \lbrace\text{Eigenvalues of } e^{i\hat{B}}\rbrace, \label{eq:AB1} \\
        \Tr(\sin \hat{A}) &= \Tr(\sin \hat{B}) = 0, \label{eq:AB2}\\
       \sin\hat{A} &= \sin \hat{B}, \label{eq:AB3}
    \end{align}
    with the first following from the definition of $e^{i\hat{A}}$ and $e^{i\hat{B}}$ as products of the same two operators ordered differently (which must have the same eigenvalues~\cite{MatrixProductEigenvalues}), the second from the definition of $\alpha$, and the third from $\hat{F} = \hat{F}^\dagger$. From these, we can show that any $\hat{F} \neq 0$ solution must have $\vareps_C(1,t_0) > (2/d)$; this is done in Appendix~\ref{app:q_cyc_dft}, but here we describe a more intuitive argument showing that a sufficiently small error rules out $\hat{F}\neq 0$. If $\hat{A}$, $\hat{B}$ and the error are ``infinitesimal'' (which is allowed by Eqs.~\eqref{eq:AB1}, \eqref{eq:AB2}), we have $\hat{F} \approx i(\hat{A}-\hat{B}) = 0$ to leading order by Eq.~\eqref{eq:AB3}. In other words, the difference $\hat{F}$ between two unitaries $e^{i\hat{A}}$, $e^{i\hat{B}}$ sufficiently close to $\mathds{1}$ must be (almost) \textit{anti}-Hermitian, while Eq.~\eqref{eq:outline_stationary_points} requires that $\hat{F}$ be Hermitian; the two requirements are only consistent for $\hat{F} = 0$.
\end{enumerate}

On extending these considerations to general $\ptau$ (see Appendix~\ref{app:q_cyc_dft}), we get Theorem~\ref{thm:dftoptimal}.

\end{proof}

\edit{Based on the above outline, we can also describe a heuristic argument for why we expect Eq.~\eqref{eq:DFTuc} to include the optimal cyclic permutation for ``generic'' $\uh(t_0)$. $\hat{F} = 0$ solutions to Eq.~\eqref{eq:outline_stationary_points} always exist without imposing any constraints on the \text{eigenvalues} of $\uerr = \uc^\dagger \uh = e^{i\hat{B}+i\alpha}$ whatsoever; but for Eq.~\eqref{eq:AB3} to not reduce to $\hat{A} = \hat{B}$ (and therefore, $\hat{F} = 0$) requires $\sin\hat{B}$ to have at least one degeneracy (on account of Eq.~\eqref{eq:AB1}). We expect that this additional fine-tuning of the eigenvalues required for $\hat{F} \neq 0$ makes it unlikely for the optimal cyclic permutation to occur with this condition for a generic $\uh(t_0)$. If this is the case, Eq.~\eqref{eq:DFTuc} generically includes the optimal cyclic permutation for small or large error\footnote{In numerical experiments with small $d$ (up to $d=5$) and arbitrarily chosen eigenvalues $E_n t_0$ of $\uh(t_0)$, we see that the global minimum of the error over all unitary transformations $\uc \to \hat{V} \uc \hat{V}^\dagger$ does appear to occur for a cyclic permutation satisfying Eq.~\eqref{eq:DFTuc} for both small and large optimal error, providing some support for this argument.}.}


\subsubsection{Optimizing aperiodicity}
\label{sec:apdoptimalcyclic}

\edit{Using $\langle C_k\rvert \uh(t)\lvert C_k\rangle = \langle C_j \rvert \uc^{j-k} \uh(t) \uc^{k-j}\lvert C_j\rangle = \langle C_j \rvert \uh(t) \lvert C_j\rangle$ for DFT cyclic permutations from Eq.~\eqref{eq:DFTbasis} (or equivalently, Eq.~\eqref{eq:DFTuc} or $[\uh(t), \uc^{k-j}] = 0$), we have
\begin{equation}
    \frac{1}{d}\sum_k \left\lvert\langle C_{k}\rvert \uh(t)\lvert C_k\rangle\right\rvert = \frac{1}{d}\left\lvert \Tr\left[\uh(t)\right]\right\rvert.
    \label{eq:dft_aperiodicity}
\end{equation}
This shows that the DFT cyclic permutations of Eq.~\eqref{eq:DFTbasis} \textit{all} saturate the lower bound of Eq.~\eqref{eq:apd_trace_bound}. Thus, the aperiodicity of any (and every) one of these DFT cyclic permutations is a necessary and sufficient condition for the system to be aperiodic in $\Eshell{d}$.}

\subsection{Ergodicity, aperiodicity and energy level statistics}
\label{sec:DFTexpressions}
\label{sec:modefluctuations}

In a DFT basis, the $\ptau$-step persistence amplitudes defined in Eq.~\eqref{eq:q_persistence_def}
are equal, $z_k(\ptau, t_0) = z(\ptau, t_0)$, and the persistence amplitudes can be expressed directly in terms of the energy eigenvalues $E_n$:
\begin{align}
    z(\ptau, t_0)[q] &= \left\lvert\frac{1}{d}\Tr\left[\uh(\ptau t_0)\uc^{-\ptau}[q]\right]\right\rvert \nonumber \\
    &= \left\lvert\frac{1}{d}\sum_{n=0}^{d-1}\exp\left[i\ptau \left(\frac{2\pi n}{d}-E_{q(n)}t_0\right)\right]\right\rvert.
    \label{eq:q_dft_persistence}
\end{align}
The corresponding $p$-step errors are given by $\vareps_C(p,t_0)[q] = 1-z^2(p,t_0)[q]$, as per Eq.~\eqref{eq:q_err_def}, among which the global minimum $\vareps_{\min}(\ptau, t_0)$ is attained for some choice of $q$ (if this minimum is less than $(2/d)$, as guaranteed by Theorem~\ref{thm:dftoptimal}).



We will regard the $z(p,t_0)[q]$ as measures of the energy level statistics within $\Eshell{d}$, in particular the deviation of the (permuted) energy levels from a regularly spaced spectrum. Namely, let
\begin{equation}
    \Delta_n(t_0,q) = \left(\frac{t_0 d}{2\pi}E_{q(n)}\right)-n
    \label{eq:deltamodefluctuation}
\end{equation}
represent the deviation of the $q(n)$-th level in a rescaled spectrum from the integer $n$. The persistence as a function of time as given by
\begin{align}
    z^2(\ptau,t_0)[q] &= \left\lvert\frac{1}{d}\sum_n e^{-i (2\pi\ptau/d) \Delta_n(t_0,q)} \right\rvert^2 \nonumber \\
    &\xrightarrow{d \to \infty} \left\lvert\int \diff\Delta\  f(\Delta; t_0,q)e^{-i(2\pi \ptau/d)\Delta}\right\rvert^2,
    \label{eq:persistence_modefluctuations}
\end{align}
where $f(\Delta; t_0,q)$ is the probability density function of the $\Delta_n(t_0,q)$ in the $d\to \infty$ limit (or sufficiently large $d$).

Intuitively, the persistence at any time $\ptau$ would be maximized when the $\Delta_n$ are minimized. A practically reasonable choice of $t_0$ and $q$ to estimate the global minimum of the $1$-step error, for uniform density of states $\Omega(\eshell_d) = (d-1)/(E_{\max}-E_{\min})$ (i.e. appearing uniform over large energy windows~\cite{Haake} within $\Eshell{d}$), is one in which the rescaled levels $t_0 d E_{q(n)}/(2\pi)$ are each close to the $n$-th integer. In other words, $t_0 \approx 2\pi \Omega(\eshell_d)/d$, with $q$ chosen to be the \newedit{sorting} permutation $\overline{q}$ that sorts the energy levels in ascending order of the phase $(E_n t_0 \bmod 2\pi)$: 
\begin{equation}
    \lbrace (E_{\overline{q}(n)} t_0 \bmod 2\pi) > (E_{\overline{q}(m)} t_0 \bmod 2\pi)\rbrace \implies \overline{q}(n)>\overline{q}(m),
\end{equation}
\newedit{(essentially, $E_{\overline{q}(0)}$ is the lowest energy level in $\Eshell{d}$, $E_{\overline{q}(1)}$ the next higher level and so on until the highest level $E_{\overline{q}(d-1)}$, for this choice of $t_0$)}, \edit{ensuring that $t_0 d E_{\overline{q}(n)}/(2\pi)$ remains reasonably close to $\overline{q}(n)$ for some choice of $t_0$}. For a given $t_0$, it is shown in Appendix~\ref{app:optimalsorting} that Eq.~\eqref{eq:persistence_modefluctuations} is indeed maximized at $\ptau=1$ when $q = \overline{q}$, among a certain class of ``small'' permutations when $\Delta_n \ll d$. In other words, the sorting permutation is a (discrete version of a) local minimum for the error.

In this case, the $\Delta_n$ are essentially what have been called mode fluctuations~\cite{Aurich1, Aurich2, Lozej2021} in the spectrum\footnote{The term ``mode fluctuations'' has been used with at least two different meanings in the literature~\cite{Aurich1,Aurich2, Lozej2021}. In Refs.~\cite{Aurich1, Aurich2} and related works cited there, it refers to the fluctuations of the \textit{spectral staircase} around a straight line. Our usage is in the sense of Ref.~\cite{Lozej2021}, referring to deviations of the levels themselves from a straight line. The two are different in general, but show close agreement in their statistical properties for Wigner-Dyson random matrix ensembles~\cite{DeltaStar1, DeltaStar2} (see also Sec.~\ref{sec:RMTergodicity}).}; the Gaussianity of their distribution has been conjectured to be a ``signature of chaos''~\cite{Aurich1, Aurich2}. A minor, but important, technical distinction between $\Delta_n$ and conventional mode fluctuations is that there is no unfolding~\cite{Haake, Mehta} --- \newedit{a convenient modification} of the energy levels to make $\Omega(\eshell_d)$ appear uniform \newedit{over large energy scales while preserving shorter range correlations} --- prior to calculating the $\Delta_n$. Such a procedure, while \newedit{useful for \textit{combining} short or medium range level statistics from different parts of the spectrum for improved statistical quality} in numerical studies, \newedit{manually alters the long-range structure of the spectrum and does not preserve the dynamics of the system in the time domain}. \newedit{This makes unfolding unsuitable for any analytical approach aiming to relate the \textit{unmodified} dynamics of a system to its energy levels, including the present study}. Given this qualifier, Eq.~\eqref{eq:persistence_modefluctuations} naturally states that the Fourier components of mode fluctuation distributions, obtained \textit{without unfolding}, directly determine the optimal persistence of cyclic permutations. \newedit{As we will see more quantitatively in the discussion around Eq.~\eqref{eq:SFFschematic}, this means that the persistence may be large enough to show cyclic ergodicity only when $\Eshell{d}$ is contained in a sufficiently narrow energy shell, within which larger scale variations in $\Omega(\Eshell{d})$ can readily be neglected.}

Another relevant (and extensively studied) measure is the spectral form factor (SFF)~\cite{Haake} (here, of the energy levels within $\Eshell{d}$), defined by
\begin{equation}
     K(t) \equiv \left\lvert \frac{1}{d}\Tr\left[\uh(t)\right]\right\rvert^2 = \frac{1}{d^2}\sum_{n,m} e^{i(E_n-E_m)t}.
     \label{eq:sffdef}
\end{equation}
\edit{The SFF is usually the central analytically tractable quantity in studies of ``quantum chaotic'' systems, as far as level statistics is concerned~\cite{BerrySpectralRigidity, Haake}. Excluding $O(1)$ transients at early times (the ``slope''~\cite{BlackHoleRandomMatrix}), a high degree of spectral rigidity in such systems is indicated by significantly suppressed late-time quantum fluctuations~\cite{PrangeSFF} in $K(t)$, over a length of time, to well below its ``natural'' yet small average value $\langle K(t)\rangle_t = d^{-1}$ seen over the longest time scales, e.g. to around $K(t) \sim [t/\Omega(\Eshell{d})]d^{-1}$ when $t \ll \Omega(\Eshell{d})$ for Wigner-Dyson statistics (these suppressed fluctuations form the ``ramp''~\cite{BlackHoleRandomMatrix}). For low spectral rigidity, there is weaker or no suppression, e.g. for Poisson spectra, virtually all late-time fluctuations in $K(t)$ oscillate strongly around $d^{-1}$.}

\edit{Calculations of $K(t)$ in several ``quantum chaotic'' systems (with various approximations) suggest that the low magnitude of the ramp is determined by generic randomness properties and low levels of recurrence/periodicity of certain physical processes, rather than any direct notion of ergodicity --- e.g. an appropriate ``uniform''\footnote{This distribution of periodic orbits, called the Hannay-Ozorio de Almeida sum rule~\cite{HOdA}, is often motivated in terms of its similarity to ergodicity~\cite{Haake}. It is worth emphasizing that in spite of the mathematical similarity, the two are logically and conceptually distinct, partly due to the fact that isolated periodic orbits are of measure zero in Eq.~\eqref{eq:cl_ergodic}; as noted in Ref.~\cite{HOdA}, KAM tori are ergodic systems that do not satisfy this sum rule.} and minimally correlated distribution of strictly isolated periodic orbits in systems with a chaotic classical limit~\cite{HOdA, BerrySpectralRigidity, Haake, argaman}, analogous properties of closed Feynman paths in random Floquet systems~\cite{KosProsen2018, ChanScrambling, GarrattChalker}, or small return probabilities in diffusive processes~\cite{argaman, ChalkerReturnProbability, WinerHydro}. Indeed, from Eqs.~\eqref{eq:dft_aperiodicity} and \eqref{eq:sffdef}, it is clear that the SFF is most directly associated with aperiodicity. Nevertheless, in Sec.~\ref{sec:TypicalCyclic}, we will show how the behavior of the SFF ramp influences \textit{cyclic ergodicity} within $\Eshell{d}$ --- connecting an observable-independent notion of quantum ergodicity to these somewhat better understood recurrence properties of specific physical processes in some systems, while remaining applicable to more general systems.}


\edit{In terms of the measures $z(p,t_0)[q]$ and $K(t)$, we have the following \textit{direct} requirements on the energy level statistics within $\Eshell{d}$, for ergodicity (Eq.~\eqref{eq:q_cyclic_ergodicity}) and aperiodicity (Eq.~\eqref{eq:q_cyclic_aperiodicity}) as per Definition~\ref{def:QuantumErgodicity}:
\begin{enumerate}
    \item \textit{Ergodicity in $\Eshell{d}$:}
    \begin{equation}
    z^2(p,t_0)[q] \gg O(d^{-1}),\ \forall\ \lvert p\rvert \leq \frac{d}{2}
    \label{eq:dft_cyc_ergodicity}
\end{equation}
is a sufficient condition for ergodicity if satisfied \newedit{at all $\lvert p\rvert \leq d/2$ for at least one choice of $q$ (usually $q=\overline{q}$)} by Eq.~\eqref{eq:q_dft_persistence}. It is also necessary \newedit{that Eq.~\eqref{eq:dft_cyc_ergodicity} be satisfied by at least one $q = q_p$ for each $\lvert \ptau\rvert \leq d/2$ in a cyclic ergodic system,} if the heuristic argument for the large error version of Theorem~\ref{thm:dftoptimal} holds.
\item \textit{Aperiodicity in $\Eshell{d}$:}
\begin{equation}
    K(t) \lesssim O(d^{-1}),\ \forall\ t_0 \ll \lvert t\rvert = O(t_0 d) \label{eq:dft_cyc_aperiodicity}
\end{equation}
is a necessary and sufficient condition for aperiodicity, by Eqs.~\eqref{eq:dft_aperiodicity} and \eqref{eq:sffdef}.
\end{enumerate}
}

A further question is if the simplest i.e. $p=1$ measures of level statistics can be used to study these dynamical properties for all times $p$, say for a given $q$.To enable this, we first have the formal bound of Eq.~\eqref{eq:q_persistence_bound}, which can be expressed in terms of $\vareps_C(1,t_0)$ as a lower bound on the decay of the persistence (via $z = \cos \theta$):
\begin{equation}
    z(\ptau,t_0) \geq  
    \begin{dcases}
    \cos\left(\ptau\sqrt{\vareps_C(1,t_0)}\right),\ &\lvert \ptau\rvert < \pi/\sqrt{4\vareps_C(1,t_0)}, \\
    0,\ &\lvert \ptau\rvert  \geq \pi/\sqrt{4\vareps_C(1,t_0)},
    \end{dcases}
\label{eq:q_cyc_persistence_bound}
\end{equation}
neglecting $O[\ptau\vareps_C^{3/2}(1,t_0)]$ contributions.





\begin{figure*}[!htb]
\centering
\subfloat[][Cyclic ergodicity and aperiodicity (staying closer to the boundary than $O(d^{-1/2})$ for more than one and less than two full rotations); data for a single Circular Unitary Ensemble (CUE~\cite{Haake, Mehta}) random matrix with $E_n$ spanning $[0,2\pi)$.]{\includegraphics[width=0.47\textwidth]{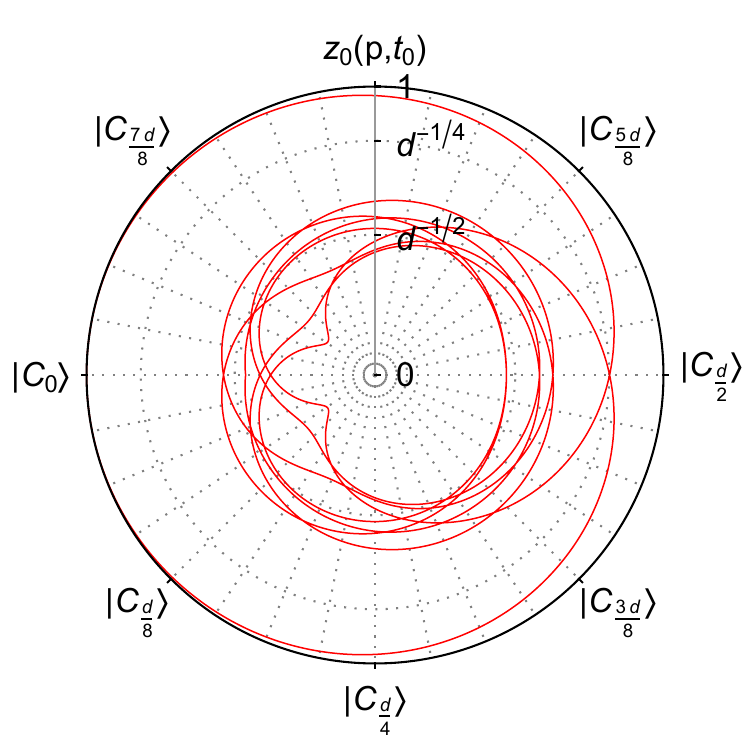}
    \label{fig:quantumergodicityData}}
    \qquad
\subfloat[][Non-ergodicity (staying closer to the boundary than $O(d^{-1/2})$ for less than one full rotation); data for a single realization of Poisson/uncorrelated energy levels spanning $[0,2\pi)$.]{\includegraphics[width=0.47\textwidth]{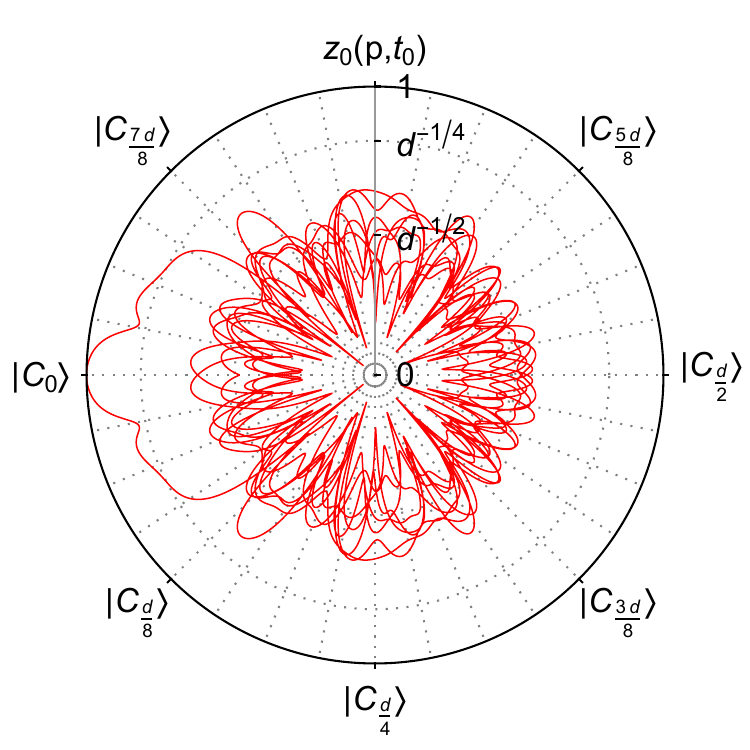}
    \label{fig:quantumnonergodicityData}}
    
    \caption{Exact numerical data for a polar representation of (a) cyclic ergodicity and aperiodicity, as well as (b) non-ergodicity, via the trajectory of $\lvert C_0\rangle$ in a Hilbert space $\eshell_d$ with $d=2048$, and $t_0 = 2\pi/\Delta E = 2\pi \Omega(\Eshell{d})/d = 1$.  Essentially, the cyclic permutation basis elements $\lvert C_k\rangle$ are points on the boundary (with $p=k$) of the polar representation, and the permutation is ergodic if the actual trajectory of any such point remains close to the boundary for a full rotation of the angular coordinate (including future and past evolution). The angular coordinate is $\theta = 2\pi p/d$ (depicted here for $p \in \mathbb{Z}$), and the radial coordinate represents $z_0(\ptau,t_0)$ via the map $r = g(z_0(\ptau,t_0))/g(1)$ with $g(x) = \lbrace 1+\tanh[\ln(x^2 d/2)/6]\rbrace$. The trajectories extend up to $\lvert p\rvert = 4d$. The chosen cyclic permutation in both cases is the sorted DFT cyclic permutation. This figure anticipates the ergodicity of Wigner-Dyson level statistics and non-ergodicity of Poisson level statistics (Sec.~\ref{sec:TypicalCyclic}). The central region $z_0(\ptau,t_0) \in O(d^{-1/2})$ of Haar-random persistence again corresponds to nearly all of the Haar volume of $\eshell_d$ by canonical typicality~\cite{tumulka_CT, CanonicalTypicalityPSW}, and is where the trajectory typically remains for long times with the exception of occasional recurrences near the boundary. See Fig.~\ref{fig:RMT_ergodicity} for a different depiction of similar data.}
    \label{fig:qcycpermutationExact}
\end{figure*}

In general, by the reciprocal relation of Fourier variables, a slow decay of the persistence corresponds to a narrow distribution $f(\Delta; t_0, q)$ e.g. as measured by the variance $\sigma^2_\Delta \equiv \langle f^2\rangle_\Delta - \langle f\rangle_{\Delta}^2$, which is related to $z(1,t_0)$ via Eq.~\eqref{eq:persistence_modefluctuations}. When $q$ is a sorting permutation, a low $\sigma^2_{\Delta}$ essentially implies a high rigidity of the spectrum . While the precise connection between ergodicity, aperiodicity and $\sigma^2_\Delta$ depends on the functional form of $f(\Delta; t_0, q)$, \edit{it is natural to pay special attention to the case of an ideal Gaussian distribution of mode fluctuations, as has been seen in typical ``quantum chaotic'' systems~\cite{Aurich1, Aurich2, Lozej2021}. In this case, we get $z(p,t_0) \approx e^{-\gamma p^2}$ (up to random fluctuations) and $\vareps_C(1,t_0) = 2\gamma = 4\pi^2\sigma^2_\Delta/d^2$ (for $\vareps_C \ll 1$) from Eq~\eqref{eq:persistence_modefluctuations}. More significant is the following proposition, which follows from substituting the Gaussian form of $z(\ptau,t_0)$ in Eq.~\eqref{eq:dft_cyc_ergodicity}, as well as in Eq.~\eqref{eq:dft_cyc_aperiodicity} after identifying $z^2(d,t_0) = K(t_0 d)$ (detailed in Sec.~\ref{sec:gaussianpropositionderivation}):}

\begin{proposition}[\textbf{Ergodicity, aperiodicity and Wigner-Dyson spectral rigidity}]
If $z(\ptau,t_0) = e^{-\gamma p^2}+O(d^{-1/2})$ with $\gamma > 0$ when $\lvert p\rvert \in O(d)$ for some DFT cyclic permutation $\mathcal{C}$, then \newedit{as $d\to\infty$},
\begin{equation}
    \text{Cyclic ergodicity and aperiodicity of } \mathcal{C} \iff \left[\sigma_\Delta^2 = \frac{\alpha^2}{4\pi^2} \ln d, \text{ with } \alpha \in (1,2], \newedit{\text{ and Eq.~\eqref{eq:dft_cyc_aperiodicity} holds}}\right].
    \label{eq:WDproposition}
\end{equation}
\label{prop:GaussWD}
\end{proposition}

This form of $\sigma^2_\Delta$ is precisely that of the Wigner-Dyson circular random matrix ensembles --- COE, CUE, CSE respectively corresponding to maximal ensembles of symmetric, unconstrained, or self-dual unitaries~\cite{Haake, Mehta} --- for $\alpha = 1, \sqrt{2}, 2$, which also satisfy Eq.~\eqref{eq:dft_cyc_aperiodicity}.
\edit{Moreover, in typical ``quantum chaotic'' systems, the energy levels show Wigner-Dyson statistics (in particular, possessing the spectral rigidity of the circular Wigner-Dyson ensembles and Gaussian mode fluctuations)~\cite{Haake, Mehta, Aurich1, Aurich2, Lozej2021}, if $\Eshell{d}$ is chosen to be a sufficiently narrow energy shell spanning $[E,E+\Delta E]$ with $\Delta E \lesssim 2\pi /t _{\text{ramp}}(E)$; here, $t_{\text{ramp}}(E)$ can be directly identified~\cite{ShenkerThouless} as the time beyond which the ramp appears in the SFF around the energy $E$ after eliminating the effect of early-time transients \newedit{as described below}\footnote{As an illustration, a Hamiltonian CUE-class Wigner-Dyson system with an energy spectrum of width $\mathcal{E}$ has $\lvert \Tr \uh(t)\rvert^2 = \mathcal{E} t/(2\pi)$ (up to fluctuations~\cite{PrangeSFF}) for sufficiently small $t \gtrsim t_{\text{ramp}}$~\cite{ShenkerThouless}. Each (disjoint) energy shell $\eshell(k)$ (with unitary $\uhd{(k)}(t)$) of width $\Delta E_t = 2\pi/t$ contributes phases $E_n(k) t$ that span $[0,2\pi)$ with $\lvert \Tr \uhd{(k)}(t)\rvert^2 = 1$ (\textit{irrespective} of the number of levels $d_k$ in the shell), effectively behaving like a CUE random unitary~\cite{Haake,Mehta} at the time $t$. The direct sum of all $M = \mathcal{E}/\Delta E = \mathcal{E} t/(2\pi)$ such CUE-like unitaries has the individual traces combine with random phases, giving $\lvert \Tr \uh(t)\rvert^2 = M$ (see also Refs.~\cite{ShenkerThouless, WinerHydro}). In other words, the quantitative form of the Wigner-Dyson ramp in Hamiltonian systems is due to all energy shells of some width $\Delta E \lesssim 2\pi/t_{\text{ramp}}$ behaving like circular ensemble random unitaries at the time $t = 2\pi/\Delta E$.}. This ramp time is system specific, e.g. ranging from $t_{\text{ramp}} = 1$ in some quantized chaotic maps~\cite{QuantumGraphs1D}, $O(1)$ to growing logarithmically with system size in some Floquet many-body systems~\cite{ShenkerThouless, KosProsen2018, BertiniProsen, ChanScrambling}, or the time scale of diffusion in some disordered systems~\cite{ShenkerThouless}. In general, the definite width $\Delta E$ of $\Eshell{d}$ introduces a large $\sinc^2(t\Delta E/2)$ transient (see also Ref.~\cite{Reimann2016}) owing to the Fourier sum in Eq.~\eqref{eq:sffdef}, due to which the SFF effectively takes the following form \newedit{when $\Delta E \lesssim 2\pi/ t_{\text{ramp}}$}:
\begin{equation}
    K(t) = \sinc^2\left(t \frac{\Delta E}{2}\right) + \text{ramp}(t).
    \label{eq:SFFschematic}
\end{equation}
Thus, by choosing $t_0 = 2\pi/\Delta E$ ($= 2\pi \Omega(\Eshell{d})/d$, in agreement with our earlier estimate after Eq.~\eqref{eq:persistence_modefluctuations}; we also note that this is the minimum time scale for nontrivial dynamics in an energy window of width $\Delta E$, suggested by the energy-time uncertainty principle) so that integer steps $\ptau t_0$ coincide with the zeros of the transient, we can eliminate the influence of this transient and obtain cyclic permutations that are directly determined by the intrinsic spectral rigidity of the system represented by the ramp (this will be justified further in Sec.~\ref{sec:TypicalCyclic}). In addition, this means that the period $t_0 d$ of such cyclic permutations is given precisely by the Heisenberg time $t_{\htime} \equiv 2\pi \Omega(\Eshell{d})$ of $\Eshell{d}$, at which the individual energy levels typically dephase completely \newedit{marking the end of the ramp}~\cite{Haake}.}

Overall, this proposition suggests that all energy shells of width $\Delta E \lesssim 2\pi/t_{\text{ramp}}$ in systems with Wigner-Dyson level statistics are ergodic and aperiodic with $t_0 = 2\pi/\Delta E$ (by Definition~\ref{def:QuantumErgodicity}), as the sorted DFT cyclic permutation is both ergodic and aperiodic, which is anticipated by the numerical data in Fig.~\ref{fig:qcycpermutationExact} for CUE and Poisson level statistics.  \edit{From a dynamical standpoint, that the ergodicity of such systems only holds within thin energy shells is not surprising --- this just reflects the fact that Hamiltonian systems with a classical limit are only ergodic in thin energy shells, and not over phase space volumes covering a wide range of energies}.

\section{Cyclic permutations for typical systems}
\label{sec:TypicalCyclic}

\edit{In this section, we study the behavior of DFT cyclic permutations for a ``typical'' system with sufficiently random fluctuations in the energy levels. Based on a general decomposition of $\uh(\ptau t_0)$ into a periodic part that follows the cyclic permutation and orthogonal random fluctuations in Sec.~\ref{sec:PeriodicRandom}, we motivate a Gaussian estimate for the time dependence of persistence amplitudes for sufficiently early $\ptau$ in typical systems in Sec.~\ref{sec:GaussianEstimate}. In Sec.~\ref{sec:errorboundSFF}, we show how the Gaussian estimate can be used to derive a lower bound for the error from the SFF for the system sampled at discrete times $\ptau t_0$ given by Eq.~\eqref{eq:SFFerror_relation}, directly connecting cyclic ergodicity to the size of quantum fluctuations in the SFF ramp (which is often analytically tractable). In Sec.~\ref{sec:GaussWD}, we discuss detailed analytical and numerical evidence for Proposition~\ref{prop:GaussWD}, showing that in the ideal case where the Gaussian estimate remains valid for longer times ($\ptau t_0 \lesssim t_{\htime}$) in an individual system, cyclic ergodicity and aperiodicity are equivalent to requiring the spectral rigidity of energy levels to be in the range spanned by the Wigner-Dyson circular ensembles: COE, CUE and CSE.}



\subsection{Periodic and random parts of time evolution}
\label{sec:PeriodicRandom}
For a cyclic permutation $\mathcal{C}$ with cycling operator $\uc$ that commutes with $\uh$ (i.e. a cyclic permutation of a DFT basis), the $\ptau$-step persistence probability $z^2(\ptau, t_0)$ is given by the analogue of the SFF for the error unitary $\uerr = \uc^{-1}\uh(t_0)$:
\begin{equation}
    z^2(\ptau, t_0) = \left\lvert\frac{1}{d} \Tr\left[\uerr^\ptau\right]\right\rvert^2.
\end{equation}
To study the development of the persistence over time, it is convenient to write a general expression for $\uerr^\ptau$ in terms of the $\ptau$-step errors $\vareps_C(\ptau, t_0)$. On account of $[\uerr, \uc] = 0$, we have $\langle C_{k+\ell}\rvert \uh\lvert C_{j+\ell}\rangle = \langle C_{k}\rvert \uh\lvert C_{j}\rangle$, due to which $\uerr^\ptau$ can be expressed simply in terms of powers of $\uc$:
\begin{equation}
    \uerr^\ptau = \left[\left(\sqrt{1-\vareps_C(\ptau, t_0)}\right) \idop + \left(\sqrt{\vareps_C(\ptau, t_0)}\right)\sum_{m=1}^{d-1} \nu_m(\ptau) \uc^m\right] e^{i\phierr(\ptau)},
    \label{eq:q_periodic_random}
\end{equation}
for some phases $\phierr(\ptau)$ and complex error coefficients $\nu_m(\ptau)$. Unitarity $\uerr^\dagger \uerr = \uerr \uerr^\dagger = \idop$ translates to nonlinear constraints on the $\nu_m(\ptau)$:
\begin{equation}
    \sum_{m=1}^{d-1}\lvert \nu_m(\ptau)\rvert^2 = 1,
    \label{eq:nu_constraint1}
    \end{equation}
    \begin{equation}
    \nu_m(\ptau) + \nu^\ast_{-m}(\ptau) = -g_\ptau\sum_{k=1}^{d-1}\nu_k^\ast(\ptau)\nu_{k+m}(\ptau), \text{ for } m \neq 0,
    \label{eq:nu_constraint2}
\end{equation}
where $\nu_0(\ptau) \equiv 0$, and $g_\ptau \equiv \sqrt{\vareps_C(\ptau, t_0)/[1-\vareps_C(\ptau,t_0)]}$.

As a matter of nomenclature, we call the first term proportional to $\idop$ in Eq.~\eqref{eq:q_periodic_random} the ``periodic part'', and the remaining terms involving $\uc^m$ (orthogonal to the periodic part) the ``random part'', of time evolution. This is because the former becomes a term proportional to $\uc^\ptau$ in $\uh(\ptau t_0)$ which is periodic in $\ptau$, while we expect the $\nu_m(\ptau)$ to generally (but not necessarily) look ``random''. In fact, (a subset of) the $\nu_m(\ptau)$ are directly related to the SFF of $\uh(\ptau t_0)$ within the subspace $\eshell_d$, via:
\begin{equation}
    K(\ptau t_0) = \left\lvert \frac{1}{d}\Tr\left[\uh(\ptau t_0)\right]\right\rvert^2 = \vareps_C(\ptau, t_0) \lvert \nu_{-\ptau}(\ptau)\rvert^2,
    \label{eq:nu_sff}
\end{equation}
and the expectation of randomness in the $\nu_m(\ptau)$ reflects the randomness in the SFF ramp~\cite{PrangeSFF, SSS, KosProsen2018} (more precisely, particularly in the phases of $\Tr[\uh(t)]$). Additionally, $K(\ptau t_0)$ serves as a (rather weak) lower bound for the $\ptau$ step errors. In particular, $\vareps_C(\ptau, t_0) = O(1)$ if $K(\ptau t_0) = O(1)$, establishing the impossibility of finding cyclic permutations that are reasonably close to $\uh(t_0)$, when $t_0 \ll t_{\text{ramp}}$ i.e. in the early-time ``slope'' regime of the SFF. To refine this bound, we will need a generic expression for the $\ptau$-dependence of the right hand side, derived in the following subsection.

\subsection{Gaussian estimate for persistence amplitudes}
\label{sec:GaussianEstimate}
Using Eq.~\eqref{eq:q_periodic_random}, one can readily express the persistence at arbitrary time $\ptau$ in terms of the $1$-step parameters $\vareps_C(1,t_0)$ and $\nu_m(1)$. The resulting expression involves a complicated multinomial expansion in the $\nu_m(1)$ (with $\binom{\ptau}{s}$ representing binomial coefficients), 
\begin{equation}
    \uerr^{\ptau} e^{-i\ptau \phierr(1)} = \left[\left(1-\vareps_C(1,t_0)\right)^{\ptau/2} \sum_{s=0}^{\ptau} \binom{\ptau}{s} g_1^s \sum_{m_1,\ldots, m_s} \nu_{m_1}(1)\ldots\nu_{m_s}(1) \uc^{m_1+\ldots+m_s}\right],
    \label{eq:dpt0}
\end{equation}
which is hard to extract general predictions out of. To simplify the expression, we invoke a heuristic argument that relies on the expected randomness of the $\nu_m$.

Specifically, we assume that each $\nu_m(1)$ is well described by an ensemble of complex numbers with a fixed magnitude and random phases, subject to the constraints Eq.~\eqref{eq:nu_constraint1} and Eq.~\eqref{eq:nu_constraint2}. Further, if one neglects $O(\sqrt{\vareps_C(\ptau, t_0)})$ corrections to the $\nu_m$, Eq.~\eqref{eq:nu_constraint2} essentially becomes
\begin{equation}
    \nu_m(\ptau) \approx -\nu_{-m}^\ast(\ptau).
    \label{eq:nu_symmetry}
\end{equation}
Thus, pairings of $\nu_{m}(1)$ and $\nu_{-m}(1)$ in Eq.~\eqref{eq:dpt0} have a definite phase and generate contributions that potentially interfere constructively, while the remaining random terms add out of phase. This suggests following a strategy similar to methods based on the pairing of closed Feynman paths in studies of generic semiclassical~\cite{BerrySpectralRigidity, Berry227, HaakePO, HaakePO2, Haake} and quantum~\cite{KosProsen2018, GarrattChalker} ``chaotic'' systems: we evaluate the contribution from terms dominated by pairings of $\nu_{m}(1)$ and $\nu_{-m}(1)$ with at most one free $\nu_{m_k}(1)$, assuming (with no proof beyond the above argument) that the remaining terms are negligible. As is common with these methods, other contributions would eventually dominate at large enough times, when $\vareps_C(\ptau, t_0)$ is sufficiently large and $\nu_m(\ptau)$ is sufficiently random, invalidating Eq.~\eqref{eq:nu_symmetry} for such $\ptau$.

The assumed dominance of paired error coefficients can be used to derive a general form of $\uerr^\ptau$ for small $\ptau$, and from there an estimate for $z(\ptau, t_0)$ using a recurrence relation; this is detailed in Appendix~\ref{app:errorcoefficientpairing}, with numerical evidence for error coefficient pairing. For $\vareps_C(1,t_0) \ll 1$ and $\ptau \ll 1/\sqrt{\vareps_C(1,t_0)}$, the general form is
\begin{equation}
    \uerr^{\ptau} e^{-i\ptau \phierr(1)} \approx \frac{z(\ptau, t_0)}{\sqrt{1-\vareps_C(1,t_0)}}\left[\sqrt{1-\vareps_C(1,t_0)}\idop + \ptau \sqrt{\vareps_C(1,t_0)}\sum_{r=1}^{d-1}\nu_r(1)\uc^r\right].
    \label{eq:uerrT_approx}
\end{equation}
In other words, time evolution for small $\ptau$ simply manifests as a relative growth of the random part in comparison to the periodic part, up to an overall phase. This gives a simple Gaussian expression for the persistence amplitude (in the same regime of small error and time):
\begin{equation}
    z(\ptau, t_0) \approx \exp\left[-\frac{1}{2}\frac{\vareps_C(1,t_0)}{1-\vareps_C(1,t_0)}\ptau^2-\frac{1}{2}\vareps_C(1,t_0)\lvert \ptau\rvert \right].
    \label{eq:zGaussianEstimate}
\end{equation}
The second (linear) term in the exponent is negligible until $\lvert \ptau\rvert \sim 1/\vareps_C(1,t_0)$, and we will simply drop it in further calculations. The Gaussian follows the sinusoidal lower bound in Eq.~\eqref{eq:q_cyc_persistence_bound} rather closely, suggesting that typical cyclic permutations are surprisingly close to saturating the lower bound. In other words, $\uerr^\ptau$ remains close to a 2D rotation in Hilbert space, until a time $\ptau \sim 1/\sqrt{\vareps_C(1,t_0)}$ when the cyclic permutation develops a large ($\sim 1$) error.


\subsection{Lower bound on the error from the SFF}
\label{sec:errorboundSFF}
Now we are in a position to quantitatively analyze the connection between the SFF ramp and the persistence of cyclic permutations. The $1$-step error coefficients $\nu_m(1)$ can be related to the SFF $K(\ptau t_0)$ in the $\ptau \ll 1/\sqrt{\vareps_C(1,t_0)}$ regime, using Eqs.~\eqref{eq:q_periodic_random}, \eqref{eq:nu_sff} and \eqref{eq:uerrT_approx}:
\begin{equation}
    \lvert \nu_{-\ptau}(1)\rvert^2 \approx \frac{1-\vareps_C(1,t_0)}{z^2(\ptau, t_0)\vareps_C(1,t_0)}\frac{K(\ptau t_0)}{\ptau^2}.
\end{equation}
Summing over $\ptau = -\overline{\ptau}$ to $\overline{\ptau}$ excluding $0$, the left hand side can be at most $1$ on account of the normalization constraint, Eq.~\eqref{eq:nu_constraint1}. Expanding $z^2(\ptau,t_0) = 1-O(\vareps_C(1,t_0)\ptau^2)$ and using $K(t) = K(-t)$, we get
\begin{equation}
    \sum_{\ptau = 1}^{\overline{\ptau}} K(\ptau t_0)\left\lbrace\frac{1}{\ptau^2}+O[\vareps_C(1,t_0)]\right\rbrace \lessapprox \frac{\vareps_C(1,t_0)}{2(1-\vareps_C(1,t_0))}.
    \label{eq:SFFerror_relation}
\end{equation}
Every term on the left hand side is positive. Considering only the first term and choosing the largest possible $\overline{\ptau}$ for which the second term is negligible then gives a reasonably restrictive lower bound on $\vareps_C(1,t_0)$. Correspondingly, we take $\overline{\ptau} = 1/(M\sqrt{\vareps_C(1,t_0)})$ where $M$ is some large number satisfying $M = O(1) \geq 1$.

\edit{Eq.~\eqref{eq:SFFerror_relation} is the main relation of interest connecting the recurrence properties represented by the SFF to cyclic ergodicity via $\vareps_C(1,t_0)$ (assuming typical $\nu_m(\ptau)$). It demonstrates that the suppression of recurrences indicated by the small magnitude of the SFF ramp is essential for the dynamics to be able to closely follow a cyclic permutation, essentially due to the conservation of probability (Eq.~\eqref{eq:nu_constraint1}). As it involves only integer steps $\ptau \in \mathbb{Z}$ of time $\ptau t_0$, we see directly that choosing $t_0 = 2\pi/\Delta E$ for an energy shell prevents the $\sinc^2(t\Delta E/2)$ transient in Eq.~\eqref{eq:SFFschematic} from influencing ergodicity}.

We can also derive explicit bounds for specific cases. As such sums of the SFF over time are generally self-averaging (i.e. fluctuations of the ramp average out to give a more steady sum)~\cite{KosProsen2018}, we replace $K(\ptau t_0)$ with a smooth power law expression for the ramp: $K(t) = \lambda t^\gamma$ for $t_0 \leq t \ll t_0 d$, $\gamma \geq 0$, and with $\lambda \ll 1$, which accounts for the behavior of a wide variety of systems\footnote{In the following sense: $\gamma = 0$ and $\lambda = d^{-1}$ corresponds to generic integrable systems with Poisson statistics~\cite{Haake, BerryTabor}; $\gamma = 1$ corresponds to generic ``chaotic'' systems when $\lambda \in O(d^{-2})$~\cite{Mehta,Haake}, and those with macroscopic conserved quantities for larger magnitudes of $\lambda$~\cite{ChalkerSum, WinerHydro}; integer $\gamma > 1$ with $\lambda \in O(d^{-2})$ corresponds to tensor products of $\gamma$ independent chaotic systems, as well as the $\gamma$-particle sectors of single-particle chaotic systems with $\lambda \in O(\gamma! d^{-2})$ (for large $d$), in which the many-particle SFF shows an exponential ramp~\cite{ExpRamp1, ExpRamp2, ExpRamp3}.}; \newedit{all such systems have $K(t_0 \leq t = O(t_0 d)) \lesssim O(d^{-1})$ and are consequently aperiodic}. Evaluating the sum in Eq.~\eqref{eq:SFFerror_relation} for this power law (Appendix~\ref{app:sfferror}) gives the following constraints on the error:
\begin{equation}
    \vareps_C(1,t_0) \gtrapprox \begin{dcases}
    2\lambda t_0^\gamma\zeta(2-\gamma),\ &\text{for}\ 0 \leq \gamma < 1, \\
    \lambda t_0^\gamma\ln \frac{1}{\lambda},\ &\text{for}\ \gamma = 1, \\
    \left[2\lambda t_0^\gamma\frac{\gamma-1}{M^{\gamma-1}}\right]^{\frac{2}{\gamma+1}},\ &\text{for}\ \gamma > 1,
    \end{dcases}
\end{equation}
where $\zeta(z)$ is the Riemann zeta function. Now we consider the most important (i.e. typical) cases of practical interest. Poisson statistics~\cite{Haake} corresponds to $\lambda = d^{-1}$ and $\gamma = 0$, for which we obtain
\begin{equation}
    \vareps_C(1,t_0)\rvert_{\text{Poisson}} \gtrapprox \frac{\pi^2}{3d}.
    \label{eq:Poisson_minerror}
\end{equation}
Together with the conditions for Eq.~\eqref{eq:dftoptimal}, this implies that every (DFT and non-DFT) cyclic permutation for a system with Poisson level statistics has $\vareps_C(1,t_0) > (2/d)$. \edit{As long as $z(\ptau, t_0)$ is not drastically different from a Gaussian in $\ptau$, as expected from the typicality considerations of Sec.~\ref{sec:GaussianEstimate}, it follows that the persistence decays to the Haar random value by $\ptau \lesssim O(d^{1/2}) \ll d/2$, and no DFT cyclic permutation is even remotely close to being ergodic for a typical system with Poisson statistics}. On the other hand, the circular Wigner-Dyson ensembles~\cite{Haake, Mehta} have $\gamma = 1$ and $\lambda = 2/(\beta d^2)$ with $\beta = 1,2,4$ for COE, CUE, CSE respectively. With $t_0 = 1$ ($=2\pi\Omega/d$), the error satisfies
\begin{equation}
    \vareps_C(1,1)\rvert_{\text{Wigner-Dyson}} \gtrapprox \frac{4}{\beta d^2}\ln d, \text{ with } \beta \in \lbrace 1,2,4\rbrace.
    \label{eq:WD_minerror}
\end{equation}
These relations encode the following property: any system admits cyclic permutations with large error, but only sufficiently rigid spectra can admit cyclic permutations with small error, quantifying the discussion in Sec.~\ref{sec:modefluctuations}. For instance, if a system is known to have a cyclic permutation with error smaller than $(2/d)$, we can rule out Poisson statistics for that system.

\subsection{Cyclic permutations and Wigner-Dyson level statistics}
\label{sec:GaussWD}
\subsubsection{Spectral rigidity for ergodic, aperiodic systems with almost exactly Gaussian persistence amplitudes}
\label{sec:gaussianpropositionderivation}
From the viewpoint of the Gaussian estimate, an idealized situation is when the persistence amplitude $z(\ptau, t_0)$ remains exactly Gaussian as it decays all the way through to the random state (order of magnitude) value $z(\ptau, t_0) \in O(d^{-1/2})$. Writing $g_1^2 = \vareps_C(1,t_0)/[1-\vareps_C(1,t_0)]$, we can solve for $g_1$ corresponding to ergodic or non-aperiodic evolution by imposing:
\begin{equation}
    \exp\left[-\frac{1}{2\alpha^2}g_1^2 d^2\right] \geq c d^{-1/2},
    \label{eq:exactGaussian}
\end{equation}
where $\alpha = 2$ for ergodicity and $\alpha = 1$ for non-aperiodicity (from Eqs.~\eqref{eq:q_cyclic_ergodicity}, \eqref{eq:q_cyclic_aperiodicity}), while $c$ is some $O(1)$ positive constant.
From Eq.~\eqref{eq:persistence_modefluctuations}, we also obtain a Gaussian distribution for mode fluctuations given some $g_1$ (assuming that the DFT cyclic permutation under discussion corresponds to a level permutation function $q$),
\begin{equation}
    f(\Delta; t_0, q)  = \frac{1}{\sqrt{2\pi \sigma_\Delta^2}}\exp\left[-\frac{1}{2\sigma_{\Delta}^2}\Delta^2\right],
\end{equation}
with variance $\sigma^2_\Delta = g_1^2 d^2 / (4\pi^2)$. Requiring ergodicity and aperiodicity therefore gives:
\begin{equation}
    \sigma^2_{\Delta} = \frac{\alpha^2}{4\pi^2} \ln d + O(1), \text{ with } \alpha \in (1,2].
    \label{eq:ergodic_aperiodic_rigidity}
\end{equation}
This amounts to a derivation of Eq.~\eqref{eq:WDproposition}.

The logarithmic growth of the variance of mode fluctuations with the dimension $d$ of the energy subspace is a direct consequence of the Gaussianity of the persistence. In less idealized situations, it is possible to have a non-Gaussian tail in Eq.~\eqref{eq:zGaussianEstimate}, for $\ptau \gtrsim 1/\sqrt{\vareps_C(1,t_0)}$, even if the Gaussian estimate holds for smaller times. It is worth noting that non-Gaussian tails at long times would show up as non-Gaussianities near $\Delta \approx 0$ in the mode fluctuation distribution; such deviations from Gaussianity are largely determined by the complicated correlations between the errors $\nu_m$, partly encoded in the fluctuations of the SFF $K(\ptau t_0)$. The main takeaway here is instead the extremely specific numerical range $\alpha \in (1,2]$, of the coefficient multiplying the logarithm, demanded by ergodicity and aperiodicity. For non-Gaussian tails, one would have a similarly specific range of some other parameter.

\subsubsection{Ergodicity and aperiodicity for Wigner-Dyson spectral rigidity}
\label{sec:RMTergodicity}


\begin{figure*}[!htb]
\subfloat[][COE level spacings]{\includegraphics[width=0.28\textwidth]{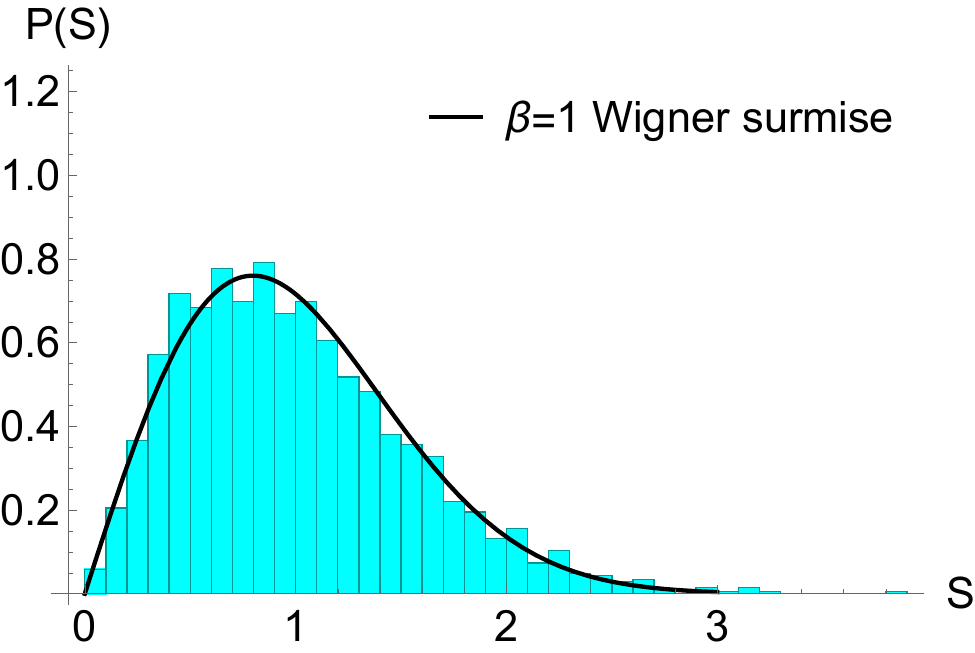}
    \label{fig:coeps}}
\hfill
\subfloat[][COE persistence (linear scale); $\ptau t_0 = \text{time}$.]{\includegraphics[width=0.28\textwidth]{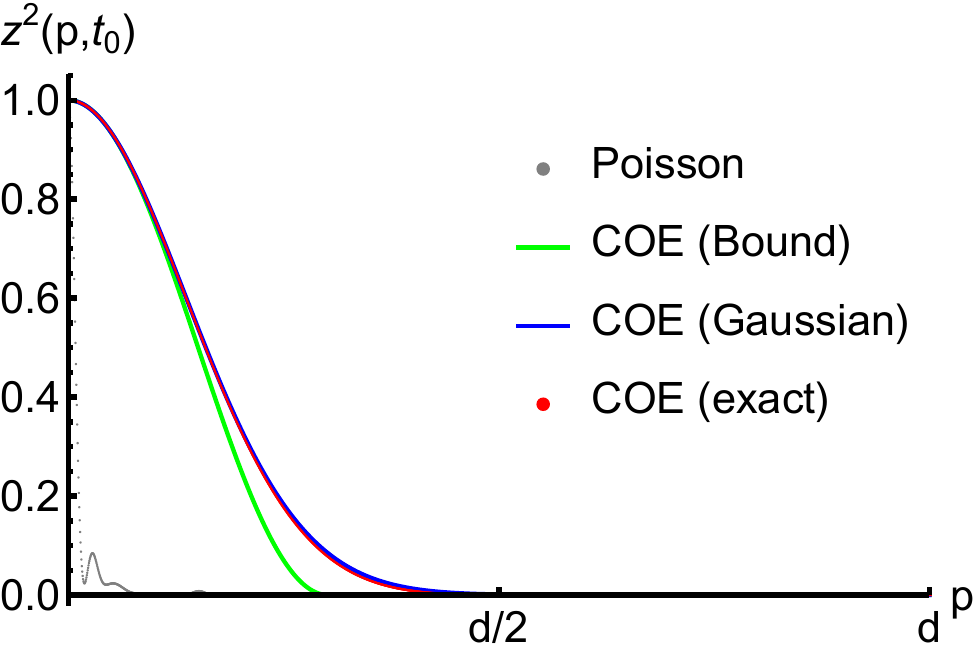}
    \label{fig:coepl}}
\hfill
\subfloat[][COE persistence (log-linear)]{\includegraphics[width=0.28\textwidth]{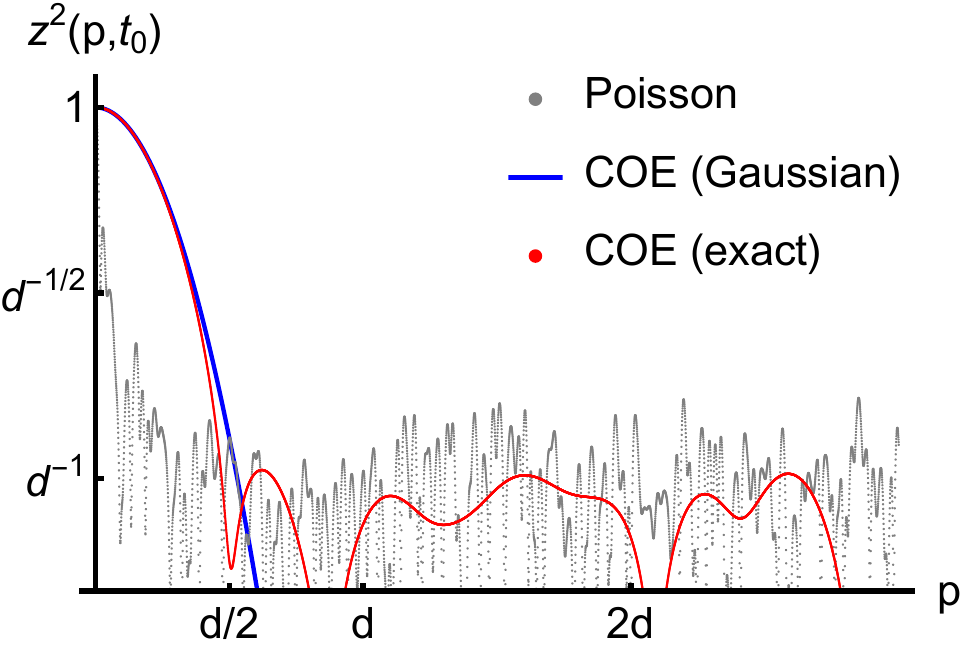}
    \label{fig:coepll}}

\subfloat[][CUE level spacings]{\includegraphics[width=0.28\textwidth]{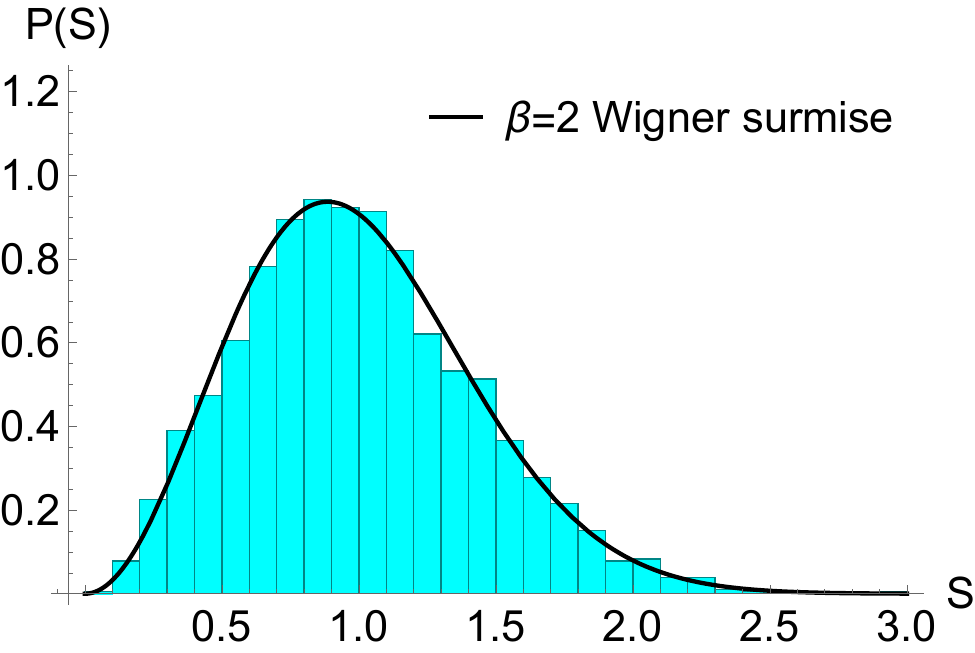}
    \label{fig:cueps}}
\hfill
\subfloat[][CUE persistence (linear)]{\includegraphics[width=0.28\textwidth]{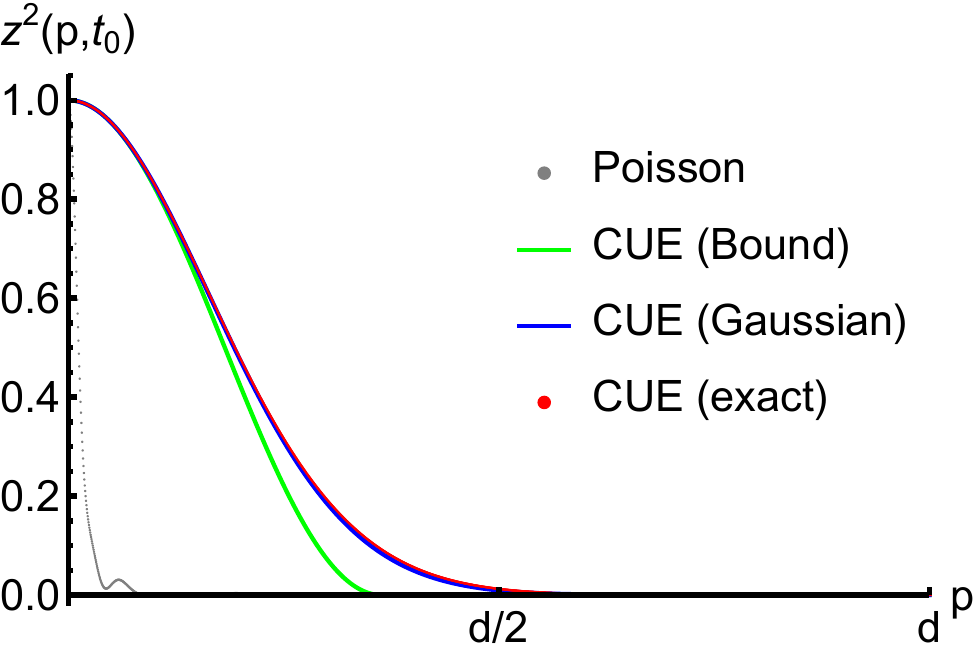}
    \label{fig:cuepl}}
\hfill
\subfloat[][CUE persistence (log-linear)]{\includegraphics[width=0.28\textwidth]{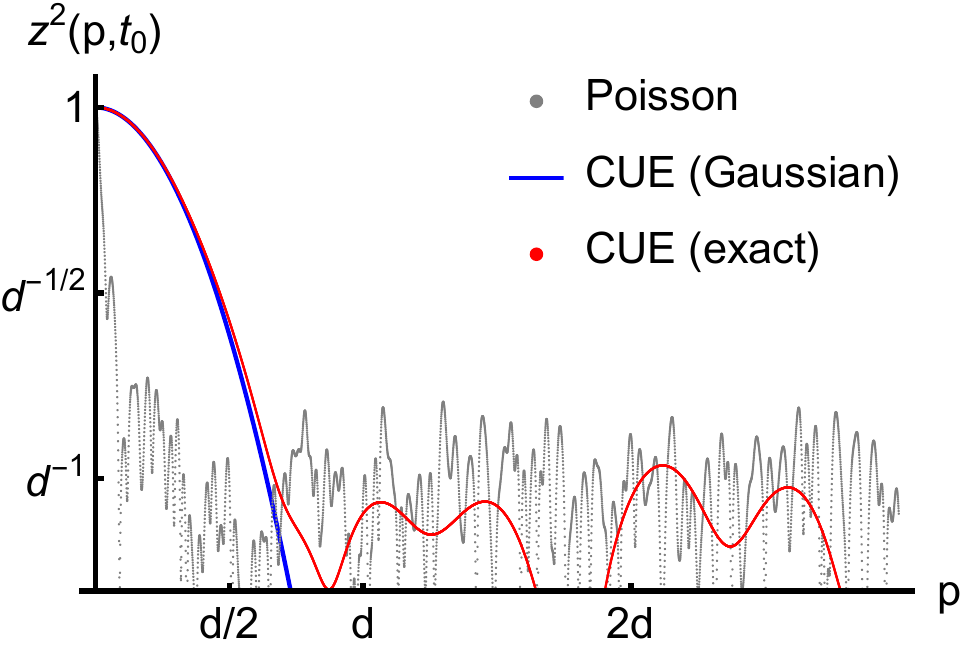}
    \label{fig:cuepll}}

\subfloat[][CSE level spacings]{\includegraphics[width=0.28\textwidth]{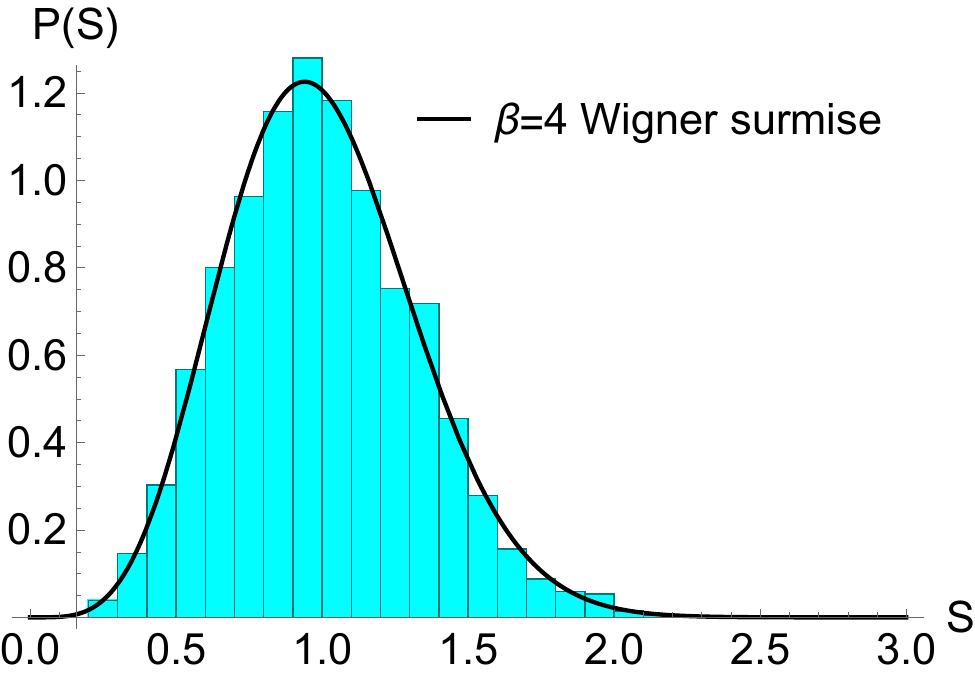}
    \label{fig:cseps}}
\hfill
\subfloat[][CSE persistence (linear)]{\includegraphics[width=0.28\textwidth]{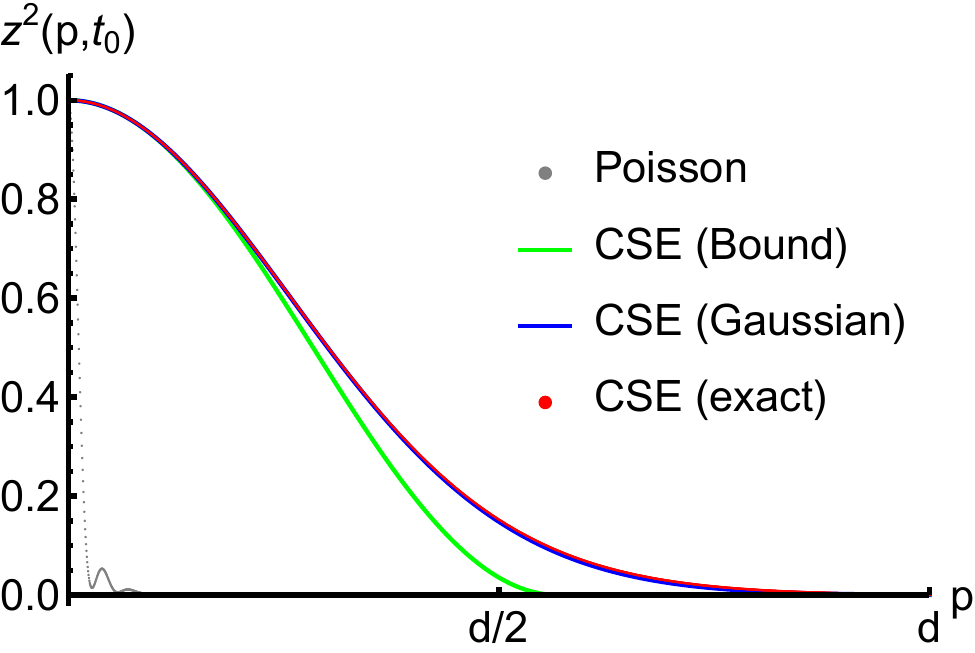}
    \label{fig:csepl}}
\hfill
\subfloat[][CSE persistence (log-linear)]{\includegraphics[width=0.28\textwidth]{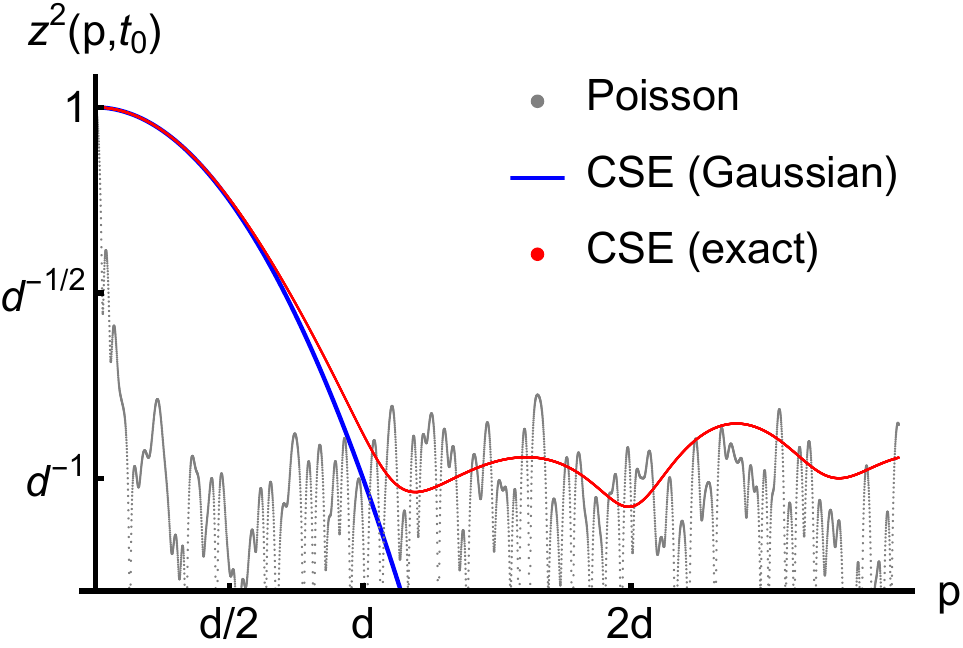}
    \label{fig:csepll}}
    
    \caption{Numerical support for ergodicity and aperiodicity of realizations of Wigner-Dyson random matrix ensembles, for $d=2048$, $t_0 = 1$ and $q$ being the sorting permutation. Level spacing data, showing the distribution of nearest neighbor level spacings $S \in \lbrace E_{\overline{q}(k+1)}-E_{\overline{q}(k)}\rbrace$, depicts the closeness of the realization to an ideal Wigner-Dyson distribution as given by the appropriate Wigner surmise~\cite{Haake,Mehta}. Persistence is plotted (in red) in terms of persistence probabilities $z^2(\ptau,t_0)$. The lower bound (``Bound'', green) of Eq.~\eqref{eq:q_cyc_persistence_bound} is satisfied, and good agreement is seen with the Gaussian estimate (``Gaussian'', blue) of Eq.~\eqref{eq:zGaussianEstimate} including the tail at long times; both are calculated based on the numerical value of $\vareps_C(1,t_0)$ for the realization. The Poisson persistence probability fluctuations (for a sorted, uncorrelated distribution of points in the same range of energies/eigenphases; in gray) are included to provide a visual reference for the range of persistence probabilities that should be considered $O(d^{-1})$ for random states, while simultaneously substantiating the non-ergodicity of Poisson statistics. The time scales when the random matrix persistence amplitudes reach $O(d^{-1})$ are consistent with $\ptau = d/2$ for COE, $\ptau = d/\sqrt{2}$ for CUE and $\ptau = d$ for CSE as predicted by Eqs.~\eqref{eq:exactGaussian}, \eqref{eq:ergodic_aperiodic_rigidity} and \eqref{eq:WignerDyson_rigidity}.}
    \label{fig:RMT_ergodicity}
\end{figure*}

Now we consider Wigner-Dyson random matrix ensembles as well as individual systems with Wigner-Dyson level statistics within energy subspaces $\Eshell{d}$ with $\Delta E \lesssim 2\pi/t_{\text{ramp}}$.
There is numerical evidence that the mode fluctuation distribution is exactly Gaussian~\cite{Aurich1,Aurich2, Lozej2021} (as well as an analytical proof of Gaussianity for a closely related measure, number fluctuations~\cite{GRMGaussian}) especially near $\Delta \approx 0$, suggestive of an almost Gaussian persistence even at late times. In Refs.~\cite{DeltaStar1, DeltaStar2}, the leading behavior of the variance $\sigma^2_\Delta$ (there called $\Delta^\ast$) for Wigner-Dyson ensembles has been shown to be equal to that of the (spectrum or ensemble averaged) spectral rigidity parameter $\Delta_3(d)$~\cite{DysonMehtaSR, Mehta, BerrySpectralRigidity} --- measuring the variance of the ``spectral staircase'' around a best fit straight line --- when $t_0 = 2\pi/\Delta E$ is determined by the slope of the straight line and $q = \overline{q}$ is the sorting permutation. Moreover, $\Delta_3(d)$ can be calculated exactly~\cite{Mehta, Berry227, BerrySpectralRigidity} by using the appropriate Wigner-Dyson ensemble averaged SFF $\overline{K_{\beta}(t)}$ for the energy subspace.

In fact, the leading contribution for large $d$ comes only from the ramp at $t \ll t_{\htime}$, given by $\overline{K_{\beta}(t)} \approx t/(\beta \pi \Omega d)$ with $\beta = 1,2,4$ respectively for COE, CUE and CSE (much like in the derivation of Eq.~\eqref{eq:WD_minerror}). The result is a logarithmic dependence of $\sigma^2_{\Delta}$ on $d$ to leading order (for $t_0 = 1 $ and $q$ being the sorting permutation),
\begin{equation}
    \left.\sigma^2_{\Delta}\right\rvert_{\text{Wigner-Dyson}} = \frac{1}{\beta \pi^2} \ln d+O(1), \text{ with } \beta \in \lbrace 1,2,4\rbrace.
    \label{eq:WignerDyson_rigidity}
\end{equation}
This precisely corresponds (via $\sigma^2_{\Delta} = g_1^2 d^2/(4\pi^2)$ ) to an error that saturates the lower bound in Eq.~\eqref{eq:WD_minerror}, providing an important sanity check. Comparing this with Eq.~\eqref{eq:ergodic_aperiodic_rigidity} (or Eq.~\eqref{eq:WDproposition}), we see that the Wigner-Dyson ensembles span exactly the range of allowed coefficients for ergodic, aperiodic systems with a Gaussian persistence. CUE is well within this range, whereas COE is at the upper bound and barely ergodic while CSE is at the lower bound and barely aperiodic (here, it is worth noting that for CSE, we have considered only one non-degenerate half of the doubly-degenerate spectrum as is conventional~\cite{Mehta, Haake}).

It remains to be verified that $z(\ptau,t_0)$ for Wigner-Dyson random matrix ensembles is indeed well approximated by a Gaussian all the way until $\ptau \in [d/2,d]$ as suggested by the ideal Gaussian distribution of mode fluctuations, so that the identification between Eq.~\eqref{eq:WignerDyson_rigidity} and Eq.~\eqref{eq:ergodic_aperiodic_rigidity} can be made with some confidence. We provide numerical support for this statement in Fig.~\ref{fig:RMT_ergodicity} for $d=2048$.

\section{Cyclic ergodicity and spectral rigidity in 2D KAM tori}
\label{sec:KAMtori}

\edit{In this section, we study the quantum ergodic properties of linear flows on 2D tori, a class of non-mixing dynamical systems for which the framework of quantum cyclic permutations allows an analytical proof of spectral rigidity when the system is ergodic --- thereby rigorously demonstrating the link between cyclic ergodicity and spectral rigidity at the level of an individual system, without relying on ensembles or heuristic arguments. It is convenient to classify these tori into the ergodic KAM tori (with an irrational frequency ratio) and non-ergodic rational tori (with a rational frequency ratio). For some physical context, we note that both occur as invariant subsets of classically integrable systems, with almost all KAM tori remaining stable under small perturbations~\cite{Ott, Goldstein}. It is particularly interesting to note that KAM tori, \newedit{though ergodic, possess neither periodic orbits nor random matrix (-like) level statistics as required in semiclassical explanations of spectral rigidity~\cite{BerrySpectralRigidity, Haake, argaman}; the results of this section (together with Sec.~\ref{sec:TypicalCyclic} concerning random matrix level statistics) therefore suggest} a wider applicability of cyclic permutations as a way to \newedit{understand spectral rigidity in terms of ergodic properties}.}

\edit{In Sec.~\ref{sec:torusquantization}, we discuss a simple quantization procedure for these tori, motivated by the standard quantization of integrable systems. In Sec.~\ref{sec:torus_cyc}, building on the example of irrational rotations discussed in Sec.~\ref{sec:cl_cyc_examples}, we prove that all 2D KAM tori are both classical and quantum cyclic ergodic (if suitably quantized), while all rational tori are non-ergodic. Finally in Sec.~\ref{sec:torus_spectrum}, we show using the corresponding cyclic permutations and Theorem~\ref{thm:dftoptimal} that all 2D KAM tori possess much higher spectral rigidity than Poisson and even Wigner-Dyson statistics (as indicated by a smaller $\vareps_C(1,t_0)$ or $\sigma_\Delta^2$), and verify this result numerically. We note that this remains consistent with the Berry-Tabor conjecture of Poisson statistics in typical integrable systems~\cite{BerryTabor}, as the latter typically contain several uncorrelated KAM tori overlapping with each other in an energy shell, leading to enhanced spectral fluctuations (similar to Refs.~\cite{WinerHydro, WinerSpinGlass}) and low spectral rigidity of the integrable system.}


\subsection{Quantization of linear flows on a torus}
\label{sec:torusquantization}

The Hamiltonian of a linear flow with frequencies $\boldsymbol{\omega} = (\omega_x,\omega_y)$ on a 2D torus is given by
\begin{equation}
    H = \mathbf{J}\cdot\boldsymbol{\omega} = J_x \omega_x + J_y \omega_y,
    \label{eq:torusHamiltonian}
\end{equation}
with angle variables $\boldsymbol{\theta} = (\theta_x,\theta_y) \in [0,2\pi)^2$ conjugate to the action variables $\mathbf{J} = (J_x,J_y)$. The equation of motion of the linear flow is $\mathcal{T}^t \boldsymbol{\theta} = \boldsymbol{\theta} + \boldsymbol{\omega}t$.

The ergodicity of this system on the 2D phase space $\mathcal{P}_{\mathbf{J}} = \lbrace{(\theta_x,\theta_y)\rbrace}$ with fixed $\mathbf{J}$ is characterized~\cite{Sinai1976, SinaiCornfield} by the ratio $\alpha=\omega_y/\omega_x$. When $\alpha$ is irrational, the dynamics is ergodic on this phase space, and the system may be called a KAM torus\footnote{There is an additional formal requirement that $\lvert \alpha-m/n\rvert \geq c(\lvert n\rvert+\lvert m\rvert)^{-\gamma}$ for some constant $c$, with $\gamma \geq 2$, and all $m,n \in \mathbb{Z}-\lbrace{0}\rbrace$ in the strict mathematical definition of a KAM torus, required for stability under perturbations~\cite{KAMtorusdef}. All but a negligible fraction of irrationals satisfy this property (especially as $\gamma$ can be arbitrarily large), and we will ignore this restriction when using the terminology.}\cite{Ott,KAMtorusdef}; but when $\alpha$ is rational, $\mathcal{P}_{\mathbf{J}}$ decomposes into an infinite number of invariant ergodic and periodic subsets, which share the same period. In both cases, there is no mixing. We emphasize that the system of Eq.~\eqref{eq:torusHamiltonian} is different from a free particle moving on a torus. The latter is never ergodic in its \textit{phase space}, but possibly (depending on initial conditions) merely visits all points among its position coordinates with conserved momentum. The Hamiltonian of the free particle is quadratic in $\mathbf{J}$ rather than linear, and its level statistics has been found to be Poissonian~\cite{ParticleTorus1, ParticleTorus2} (see also Ref.~\cite{Aurich2} for its mode fluctuations) in accordance with the Berry-Tabor conjecture~\cite{BerryTabor} for fully integrable systems. 

The quantization of Eq.~\eqref{eq:torusHamiltonian} can be motivated by considering how such tori occur as invariant subsets of integrable systems~\cite{Ott, BerryTabor, Goldstein}. The Hamiltonian of a 2D integrable system with constants of motion $\mathbf{I} = (I_x,I_y)$ is given by $H = H(\mathbf{I})$, with quantization restricting each $\hat{I}_k \in \mathbf{N}_0$ to the non-negative integers due to the periodicity of the conjugate angle variables. Classically, expanding $\mathbf{I} = \mathbf{I}_0 + \mathbf{J}$ in the neighborhood of $\mathbf{I}_0$ (say, with $J_x,J_y \geq 0$) gives Eq.~\eqref{eq:torusHamiltonian} to leading order (up to an $\mathbf{I}_0$-dependent constant) with $\boldsymbol{\omega}(\mathbf{I}_0) = [\partial H/\partial \mathbf{I}](\mathbf{I}_0)$.  Thus, a natural way to quantize Eq.~\eqref{eq:torusHamiltonian} in terms of operators $\hat{\mathbf{J}}$, $\hat{\boldsymbol{\theta}}$ is to restrict the eigenvalues $J_x, J_y$ of $\hat{J}_x, \hat{J}_y$ each to a finite (but large) set of consecutive integers; up to additive constants in $H$, this is equivalent to working in the Hilbert space $\Eshell{d}$ spanned by
\begin{equation}
    \left\lbrace \lvert\mathbf{J}\rangle : J_x \in \mathbb{Z}_{d_x}, J_y \in \mathbb{Z}_{d_y}\right\rbrace,
\end{equation}
with $d = d_x d_y$. Of particular interest is the following manifold of $\boldsymbol{\theta}$-states corresponding to points on the torus:
\begin{equation}
    \lvert \boldsymbol{\theta}\rangle = \frac{1}{\sqrt{d_x d_y}} \sum_{J_x \in \mathbb{Z}_{d_x}}\sum_{J_y \in \mathbb{Z}_{d_y}} e^{-i \mathbf{J}\cdot \boldsymbol{\theta}}\lvert \mathbf{J}\rangle\ \in \Eshell{d}.
    \label{eq:quantum_torus_manifold}
\end{equation}
We can construct infinitely many complete orthonormal bases for $\Eshell{d}$ in this manifold, each comprised of $d$ such states, such as $\lbrace (\theta_x,\theta_y) = (2\pi n_x/d_x, 2\pi n_y/d_y) : n_x \in \mathbb{Z}_{d_x}, n_y \in \mathbb{Z}_{d_y}\rbrace$. It is straightforward to see that time evolution reproduces the classical equation of motion in this manifold:
\begin{equation}
    \uh(t)\lvert \boldsymbol{\theta}\rangle \equiv e^{-i\hat{H}t}\lvert \boldsymbol{\theta}\rangle = \lvert \boldsymbol{\theta}+\boldsymbol{\omega}t\rangle,
    \label{eq:quantum_torus_flow}
\end{equation}
which provides a natural way to map classical cyclic permutations onto $\Eshell{d}$.


\subsection{Classical and quantum cyclic ergodicity}
\label{sec:torus_cyc}

In this and the next subsection, we will see how cyclic permutations can be used to rigorously prove spectral rigidity for KAM tori. First, we show that all 2D KAM tori admit a classical cyclic permutation (sequence) that is cyclic ergodic and not aperiodic. This is made possible by identifying the discretized flow on the torus as a sequence of irrational rotations, allowing us to use the cyclic permutations constructed for the latter in Sec.~\ref{sec:cl_cyc_examples}. Using Eqs.~\eqref{eq:quantum_torus_manifold} and \eqref{eq:quantum_torus_flow} to adapt this construction to a quantum cyclic permutation of an orthonormal basis in $\Eshell{d}$, we will show the following:
\begin{theorem}[\textbf{Quantum cyclic ergodicity and non-aperiodicity of 2D KAM tori}]
For every linear flow on a 2D torus with irrational frequency ratio $\alpha = \omega_y/\omega_x$, there exists an infinite sequence of energy subspaces $\Eshell{d}$ with $d\to \infty$ (such that $d_x, d_y\to \infty$) in which the system is ergodic and not aperiodic, within any time $T > \min(2\pi d_y/ \omega_x, 2\pi d_x/\omega_y)$.
\label{thm:KAMcycerg}
\end{theorem}
Thus, the quantum dynamical properties of 2D KAM tori reflect their classically cyclic ergodic and non-aperiodic nature, which is further consistent with their ergodicity and lack of mixing.

\subsubsection{Classical cyclic permutation}

\begin{figure*}[!hbt]
\centering
\subfloat[][The classical cyclic permutation of Eq.~\eqref{eq:torus_cl_cyc_permutation} with $n=49$, depicting the $C_k$ (blue-edged squares; thicker edges for $k = 0$ to $14$). The evolution $\mathcal{T}^{\ptau t_0}C_0$ of the set $C_0$ (gray,filled) and the trajectory $\mathcal{T}^{t}\boldsymbol{\theta}$ of the point $\boldsymbol{\theta} = (0,0)$ (red) are depicted respectively for $\ptau = 0$ to $14$ and $t = 0$ to $14t_0$. The corresponding $\ptau$-step errors are determined by the portion of each highlighted square that does not intersect a gray region.]{\includegraphics[width=0.45\textwidth]{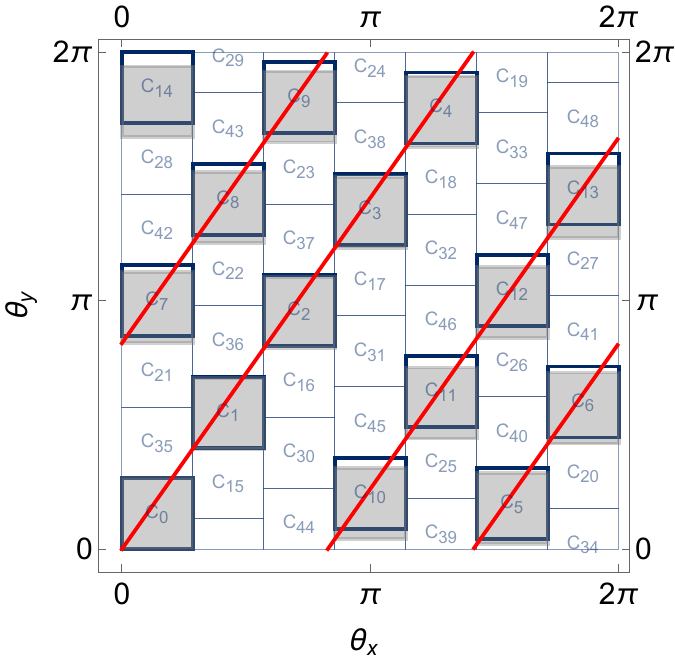} \label{fig:torus_cyclic_cl}}
\qquad
\subfloat[][The quantum cyclic permutation of Eq.~\eqref{eq:torus_q_cyclic_permutation} with $d=49$, overlaid on a depiction of the corresponding classical cyclic permutation as per Fig.~\ref{fig:torus_cyclic_cl}. Filled disks represent pure states in $\lvert \boldsymbol{\theta}\rangle$-space. The depicted pure states are the $\lvert C_k\rangle$ (blue; highlighted/opaque for $k = 0$ to $14$, transparent otherwise) and $\uh(\ptau t_0)\lvert C_0\rangle$ for $\ptau  =0$ to $14$ (red, lying on classical trajectory).]{\includegraphics[width=0.45\textwidth]{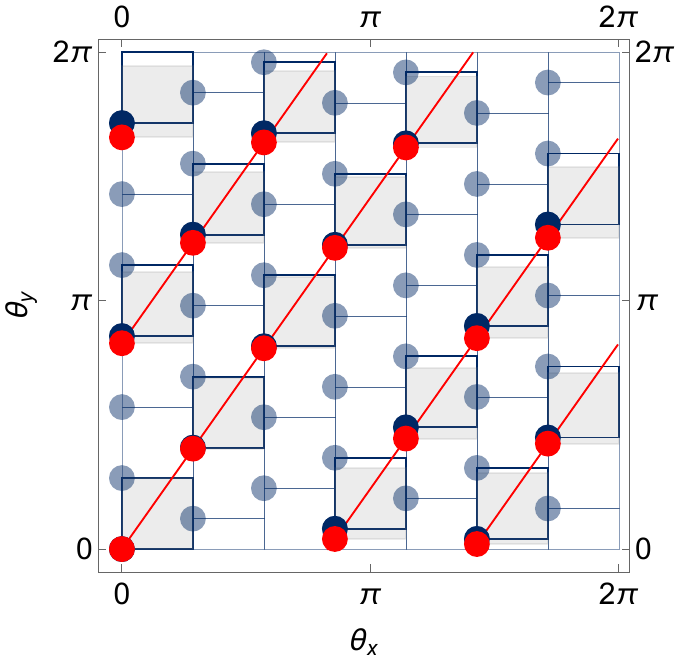} \label{fig:torus_cyclic_q}}

    \caption{The (a) classical cyclic permutation of Eq.~\eqref{eq:torus_cl_cyc_permutation} with $n_x = n_y = 7$, $m_y = 10$ and $t_0 = 2\pi/(\omega_x n_x)$, and (b) the corresponding quantum cyclic permutation of Eq.~\eqref{eq:torus_q_cyclic_permutation} with $d_x = d_y = 7$,  $m_y = 10$ and $t_0 = 2\pi/(\omega_x d_x)$, for an irrational flow on the 2D torus with $\alpha = \sqrt{2}$. For these parameters, we have $\delta_{n_y} \approx 0.704/n_y < 1/n_y$ (and likewise with $d_y$), making both the classical and quantum cyclic permutation ergodic and non-aperiodic. Trajectories starting from $\boldsymbol{\theta} = (0,0)$ are depicted up to $\ptau = 2d_x = 14$, which corresponds to $r=2$ steps of the irrational rotation $\mathcal{T}^{r T_x}\theta_y$ at $\theta_x = 0$.}
    \label{fig:torus_cyclic}
\end{figure*}

For the classical construction, we first observe that for $T_x \equiv 2\pi/\omega_x$ (the period along the $\theta_x$ direction), the map $\mathcal{T}^{r T_x}(\theta_x,\theta_y) = (\theta_x,\theta_y + 2\pi \alpha r)$ with $r\in \mathbb{Z}$ is a discrete rotation of any $\theta_x = \text{const.}$ circle in the $\theta_y$ direction, with $\alpha = \omega_y/\omega_x$. Writing $\alpha = (m_y+\delta_{n_y})/n_y$ as in Sec.~\ref{sec:cl_cyc_examples} with coprime integers $m_y$ and $n_y$, we know that every rotation admits an $n_y$-element cyclic permutation $\mathtt{C}_y$ with error $\lvert \delta_{n_y}\rvert$ by Eq.~\eqref{eq:cl_irr_rotation_cyclic}. To adapt this to the torus, we divide the $\theta_x$ direction into $n_x$ elements and consider time steps in units of $t_0 = T_x/n_x$, such that each step captures a $1/n_x$ fraction of the error generated in the full rotation by $T_x = n_x t_0$ (see Fig.~\ref{fig:torus_cyclic_cl}). Formally, our cyclic permutation $\mathtt{C}$ for the 2D torus consists of:
\begin{align}
    C_k = &\left[2\pi\frac{k}{n_x}, 2\pi \frac{k+1}{n_x}\right]_{\theta_x} \times \left[2\pi \frac{\tfrac{k}{n_x}m_y}{n_y}, 2\pi \frac{\tfrac{k}{n_x}m_y+1}{n_y}\right]_{\theta_y}, \nonumber \\
    &\text{ with error } \eps_C\left(t_0 = \frac{2\pi}{\omega_x n_x}\right) = \frac{\lvert \delta_{n_y}\rvert}{n_x}.
    \label{eq:torus_cl_cyc_permutation}
\end{align}
It is easily verified that the intersection of the $C_{qn_x}$ (for $q \in \mathbb{Z}$) with $\theta_x = 0$ gives precisely the cyclic permutation of Eq.~\eqref{eq:cl_irr_rotation_cyclic} for the rotation $\mathcal{T}^{rT_x}$. From the discussion of irrational rotations in Sec.~\ref{sec:cl_cyc_examples}, we can find a sequence of coprime $m_y,n_y\to \infty$ such that $\lvert \delta_{n_y}\rvert < 1/n_y$ for all irrational $\alpha$. Thus, we have $\eps_C(t_0) < 1/n$ with $n\to\infty$ for all $n_x$, establishing the existence of a \textit{classical} cyclic permutation showing cyclic ergodicity and non-aperiodicity (by the bounds discussed in Sec.~\ref{sec:cl_cyclic}) for all 2D KAM tori.

There are two straightforward extensions of this construction that are worth mentioning: 1. choosing $t_0 = 2\pi/(\omega_y n_y)$ and considering irrational rotations in the $\theta_y$ direction gives essentially similar results, and 2. for rational $\alpha$, the non-ergodicity of the flow implies that no cyclic permutation with the $C_k$ reducing to (neighborhoods of) points in the $n\to \infty$ limit is ergodic or aperiodic, as can be shown by suitably adapting the corresponding discussion for rational rotations in Sec.~\ref{sec:cl_cyc_examples}.

\subsubsection{Quantum cyclic permutation}

We note that among the $\boldsymbol{\theta}$-states of Eq.~\eqref{eq:quantum_torus_manifold}, any two states $\lvert \boldsymbol{\theta}\rangle$, $\lvert \boldsymbol{\theta}'\rangle$ are orthogonal if $(\theta_x - \theta'_x) = 2\pi \ell/d_x$ or $(\theta_y-\theta'_y) = 2\pi \ell/d_y$ for some $\ell \in \mathbb{Z}$. A pure state cyclic permutation corresponding to Eq.~\eqref{eq:torus_cl_cyc_permutation} can then be constructed using these states, by formally setting $d_x = n_x$, $d_y = n_y$ and considering the set of points given by the $d$ ``lower'' corners of the classical $C_k$ in Eq.~\eqref{eq:torus_cl_cyc_permutation} (see also Fig.~\ref{fig:torus_cyclic_q}), namely:
\begin{equation}
    \lvert C_k\rangle = \left\lvert \boldsymbol{\theta} = \left(2\pi \frac{k}{d_x}, 2\pi \frac{\tfrac{k}{d_x} m_y}{d_y}\right)\right\rangle,
    \text{with } \uc \lvert \boldsymbol{\theta}\rangle = \left\lvert \boldsymbol{\theta}+\left(\frac{2\pi}{d_x},\frac{2\pi m_y}{d_x d_y}\right)\right\rangle, \label{eq:torus_q_cyclic_permutation}
\end{equation}
and $t_0 = 2\pi/(\omega_x d_x)$. From Eqs.~\eqref{eq:quantum_torus_flow} and \eqref{eq:torus_q_cyclic_permutation}, it is clear that $[\uh(t),\uc] = 0$ as both $\uh(t)$ and $\uc$ are additive maps on the torus; consequently, the above cyclic permutation corresponds to a DFT of the energy eigenstates.

Using Eqs.~\eqref{eq:quantum_torus_manifold} and \eqref{eq:quantum_torus_flow}, it is straightforward to evaluate the $\ptau$-step persistence probabilities $z^2(\ptau, t_0) = \lvert\langle C_{k+1}\rvert \uh(\ptau t_0)\lvert C_k\rangle\rvert^2$ in terms of the frequency ratio $\alpha = (m_y + \delta_{d_y})/d_y$. We obtain the following exact expression for $\ptau \neq 0$:
\begin{equation}
    z^2\left(\ptau, t_0 = \frac{2\pi}{\omega_x d_x}\right) = \frac{\sin^2\left[\ptau \pi \frac{\delta_{d_y}}{d_x}\right]}{d_y^2\sin^2\left[\ptau \pi \frac{\delta_{d_y}}{d_x d_y}\right]}.
    \label{eq:2DKAMpersistence}
\end{equation}
To discuss ergodicity and aperiodicity in the quantum case, we need a slightly more refined implication of Eq.~\eqref{eq:irrationalDiophantine} concerning Diophantine approximations than in the classical case, namely that every irrational $\alpha$ admits an infinite sequence of $m_y,d_y$ with $\lvert \delta_{d_y}\rvert < c/d_y$ for some constant $c < 1$ (this constant ensures that $\lvert \delta_{d_y} \rvert$ does not even approach $1/d_y$ asymptotically, as $d_y \to \infty$). 
\newedit{Using $(\sin x)^2 \leq x^2$ in the denominator of Eq.~\eqref{eq:2DKAMpersistence}, we get that $z^2(\ptau t_0) \geq \sinc^2[\ptau \pi \delta_{d_y}/d_x]$; as the smallest zeros of the right hand side occur at $\lvert \ptau\rvert = d_x/\delta_{d_y}$, $z^2(\ptau, t_0)$ is} guaranteed to remain $\Theta(1)$, i.e., no less than $O(1)$ for any $d_y$ belonging to the above sequence, until some $\lvert \ptau \rvert > d_x d_y/c > d$ (as $c<1$).
It follows that $z^2(\ptau, t_0) \gg O(d^{-1})$ for $\lvert \ptau \rvert \leq d$, making the constructed cyclic permutation ergodic and non-aperiodic (the latter via $z^2(\pm d, t_0) = K(\pm t_0 d)$). As this is a DFT cyclic permutation, it follows from the considerations of Sec.~\ref{sec:apdoptimalcyclic} that no cyclic permutation in $\Eshell{d}$ is aperiodic.
On using these results in Definition~\ref{def:QuantumErgodicity} with $t_0 d = 2\pi d_y/\omega_x$, while noting that analogous results would hold if we swap the $\theta_x$ and $\theta_y$ directions, we directly get Theorem~\ref{thm:KAMcycerg}.

When $\alpha = m/\ell$ is rational with coprime $m$ and $\ell$, the motion in $\lvert \boldsymbol{\theta}\rangle$-space is periodic with period $T_{\alpha} = 2\pi \ell/\omega_x$. In particular, we have
\begin{equation}
    K(r T_{\alpha}) = 1,\ \forall\ r \in \mathbb{Z},
\end{equation}
immediately ruling out aperiodicity in any $\Eshell{d}$ over any time $T > T_{\alpha}$. Further, from Eq.~\eqref{eq:nu_sff}, we get $z_k(\ptau,t_0) = 0$ for every cyclic permutation in some $\Eshell{d}$ with $\ptau t_0 = r T_{\alpha}$, which rules out ergodicity for any cyclic permutation with $t_0 = r T_\alpha /\ptau$ with $(\ptau \in \mathbb{Z}) \leq d/2$. Ruling out ergodicity for more general $t_0$ would require a more detailed study of spectral properties.

\subsection{Spectral rigidity}
\label{sec:torus_spectrum}

\begin{figure*}[!hbt]

\subfloat[][Level spacings.]{\includegraphics[width=0.31\textwidth]{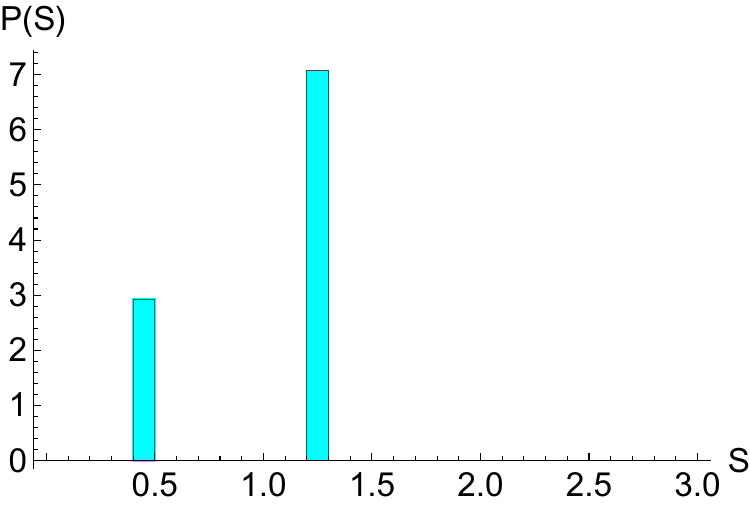}
    \label{fig:torus1}}
\hfill
\subfloat[][Non-uniform density of states in $H$.]{\includegraphics[width=0.31\textwidth]{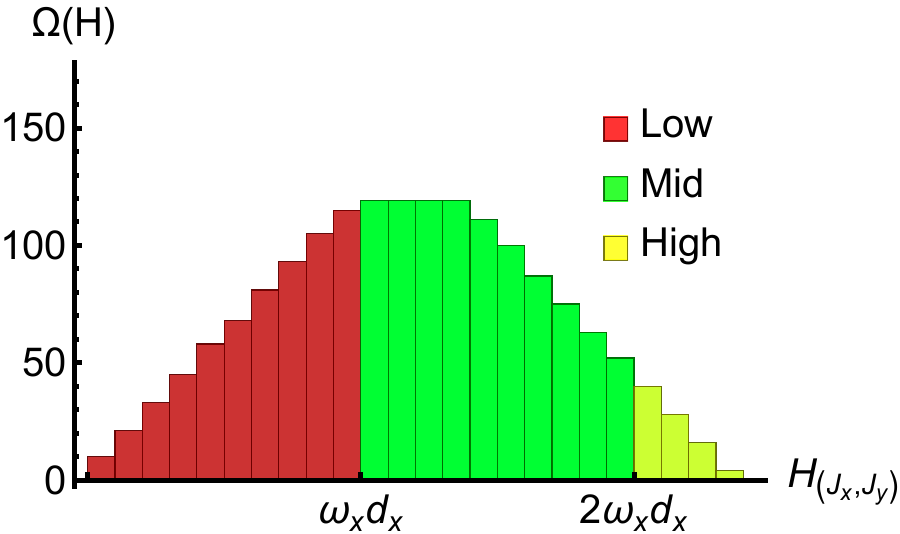}
    \label{fig:torus2}}
\hfill
\subfloat[][Uniform density of states in $E$.]{\includegraphics[width=0.31\textwidth]{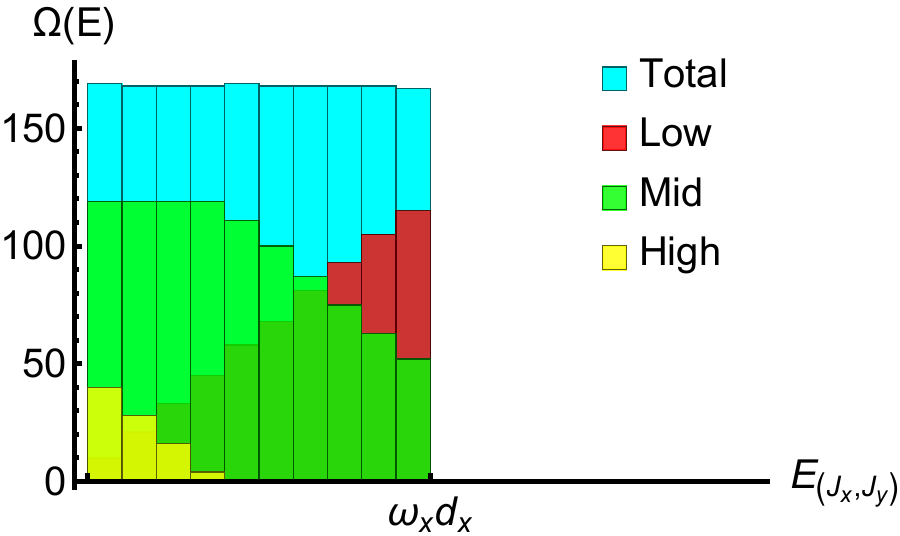}
    \label{fig:torus3}}

\subfloat[][Analytically derived error ($q=q_{m_y}$) vs numerical sorted DFT error ($q=\overline{q}$); $1 \leq d_x= d_y \leq 100$ (with $d$ ranging up to $10^4$), excluding points where $m_y$ and $d_y$ are not coprime.]{\includegraphics[width=0.31\textwidth]{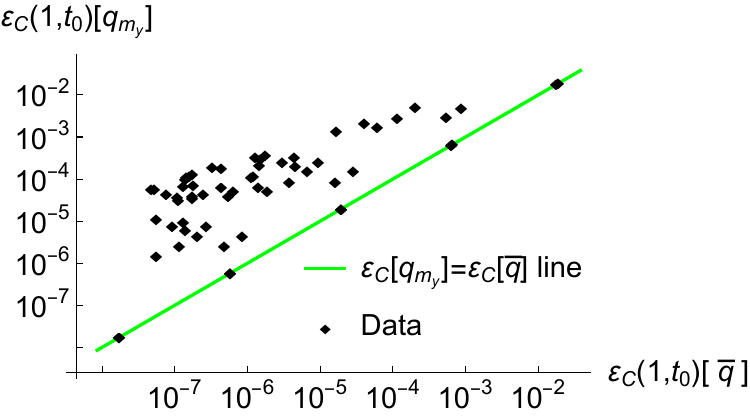}
    \label{fig:torus4}}
\hfill
\subfloat[][Cyclic ergodicity in terms of the analytically derived persistence and the numerical sorted DFT persistence (linear), showing extremely close agreement.]{\includegraphics[width=0.31\textwidth]{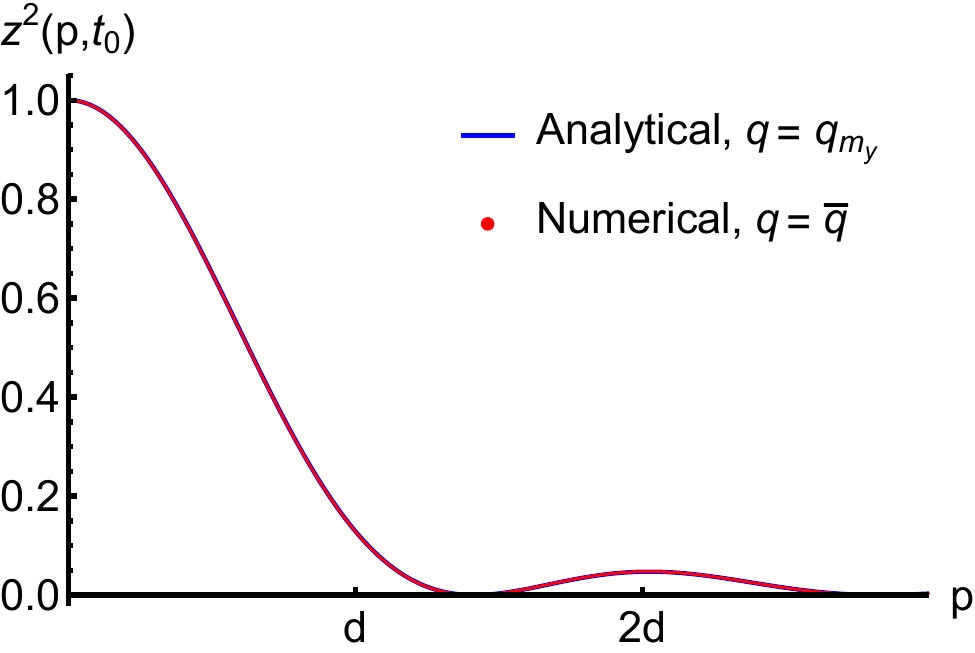}
    \label{fig:torus5}}
\hfill
\subfloat[][Numerical SFF (linear), showing non-aperiodicity.]{\includegraphics[width=0.31\textwidth]{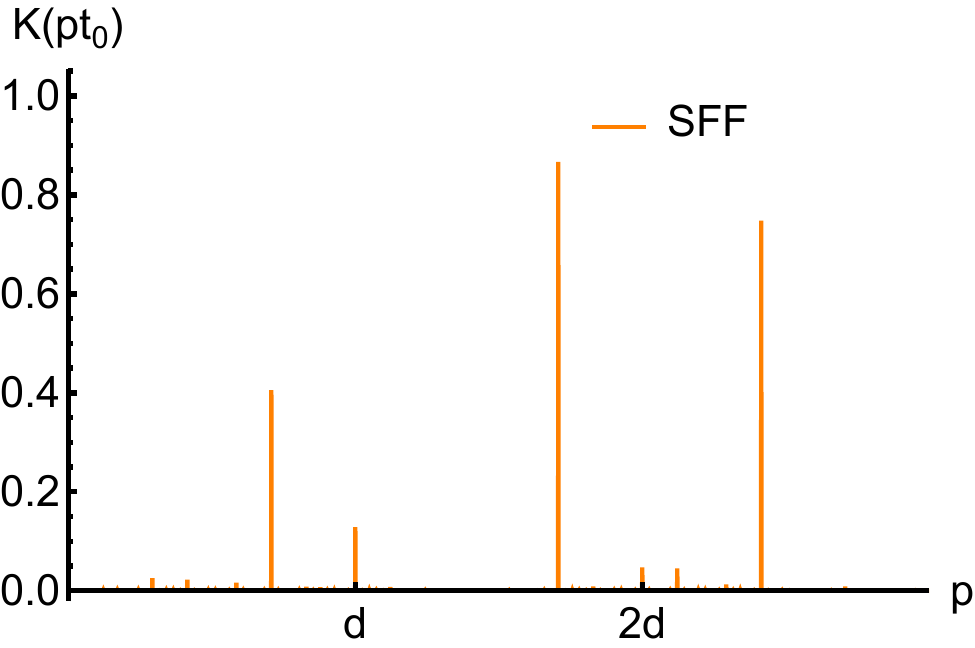}
    \label{fig:torus6}}
    
   \caption{Spectral properties of 2D KAM tori with $\alpha = \sqrt{2}$, compared with analytical predictions corresponding to the cyclic permutation with $t_0 = 2\pi/(\omega_x d_x)$ constructed in Sec.~\ref{sec:torus_cyc}; $d_x = d_y = 41$ and $d=1681$, except in (d) which considers a range of $d_y$. (a) Level spacings, required to be at most 3 delta functions. (b) and (c): comparison of unwrapped and wrapped density of states, showing how the choice of $t_0 = 2\pi/(\omega_x d_x)$ leads to a uniform ``wrapped'' density of states, when ``low'', ``mid'' and ``high'' energy portions of the spectrum demarcated by multiples of $\omega_x d_x$ are considered separately. (d) Comparison of the analytical prediction of Eq.~\eqref{eq:KAMtorus_spectralrigidity}, where $m_y$ is numerically chosen to be the nearest integer to $d_y \alpha$, with the sorted DFT error. (e) The persistence amplitude of Eq.~\eqref{eq:2DKAMpersistence} for $m_y = 58$ (with $\delta_{d_y} \approx 0.707/d_y$) compared with the sorted DFT persistence obtained numerically; this also shows that $d_x = d_y = 41$ belongs to the infinite sequence of ergodic, non-aperiodic $\Eshell{d}$ in Theorem~\ref{thm:KAMcycerg}. (f) Numerically computed SFF, showing a peak at $\ptau = d$ (as expected analytically) and elsewhere, confirming non-aperiodicity.}
    \label{fig:irr_torus}
\end{figure*}

Having established the classical and quantum cyclic ergodicity of 2D KAM tori, as well as non-aperiodicity in both the irrational and rational cases, we now turn to an analytical and numerical study of the spectral rigidity of the quantized systems, with the main results depicted in Fig.~\ref{fig:irr_torus}. The distribution of the energy levels depends sensitively on $\alpha$ (via $H = (J_x+J_y\alpha)\omega_x$ for $J_x \in \mathbb{Z}_{d_x}$, $J_y \in \mathbb{Z}_{d_y}$), and this dependence has been studied numerically and analytically in the context of the 2D harmonic oscillator in e.g. Refs.~\cite{BerryTabor, PandeyLevels1, PandeyLevels2}. Their main analytical result is that the nearest-neighbor level spacings $S \in \lbrace E_{\overline{q}(k+1)}-E_{\overline{q}(k)} \rbrace$ within sufficiently small energy windows takes the form of at most 3 delta functions, $P(S) = \sum_{j=1}^{3} c_j \delta(S-S_j)$ with $c_j, S_j \geq 0$; this behavior is shown in our context in Fig.~\ref{fig:torus1}. This distribution is quite unlike the smooth level spacing distributions of Wigner-Dyson (compare with Fig.~\ref{fig:RMT_ergodicity}) or Poisson level statistics~\cite{Haake}, and these results leave open the question of how the spectrum may be quantitatively characterized in direct relation to some dynamical property of the system.

The cyclic permutation of Eq.~\eqref{eq:torus_q_cyclic_permutation} provides precisely such a characterization, via the persistence probabilities in Eq.~\eqref{eq:2DKAMpersistence}. As discussed in Sec.~\ref{sec:modefluctuations}, these quantities are themselves measures of energy level statistics when the cyclic permutation is a DFT of energy eigenstates, due to Eq.~\eqref{eq:q_dft_persistence}. The associated map $q$ that permutes the energy levels before comparing them to regularly spaced levels is determined by the choice of $m_y$; for convenience, we denote this map $q_{m_y}$ for the cyclic permutation of Eq.~\eqref{eq:torus_q_cyclic_permutation}. Then, setting $d_x = d_y$ for convenience and using Eq.~\eqref{eq:2DKAMpersistence} with $\delta_{d_y} < 1/d_y$ to obtain the single-step DFT error $\vareps_C(1,t_0)[q_{m_y}] = 1-z^2(1,t_0)[q_{m_y}]$, we get
\begin{equation}
    \vareps_C\left(1,t_0 = \frac{2\pi}{\omega_x d_x}\right)\left[q_{m_y}\right] < \frac{\pi^2}{3d^2} + O(d^{-3}),
    \label{eq:KAMtorus_spectralrigidity}
\end{equation}
where $d$ belongs to the infinite sequence occurring in Theorem~\ref{thm:KAMcycerg} (with $d_x = d_y$), for every irrational $\alpha$. By the discussion in Sec.~\ref{sec:errorboundSFF}, we expect $\vareps_C(1,t_0)[q] \gtrsim O(d^{-1})$ for Poisson statistics, and $\vareps_C(1,t_0)[q] \gtrsim O(d^{-2}\ln d)$ for Wigner-Dyson statistics, for all $q$; Eq.~\eqref{eq:KAMtorus_spectralrigidity} thus constitutes an analytical proof that all 2D KAM tori have \textit{significantly} ``more rigid'' spectra than both Wigner-Dyson and Poisson statistics in the appropriate energy subspaces. In Appendix~\ref{app:toruscyclic}, we argue that rational tori are non-ergodic over a range of $t_0$ and show spectral rigidity of the same order of magnitude as Poisson statistics.

Another measure of spectral rigidity is the sorted DFT single-step error $\vareps_C(1,t_0)[\overline{q}]$, which is most directly related to the conventionally studied mode fluctuations according to Sec.~\ref{sec:modefluctuations}. Given our choice of $t_0$, the permutation $q=\overline{q}$ corresponds to a direct sorting of the ``wrapped'' energy levels:
\begin{align}
    E_{(J_x,J_y)}t_0 &= H(J_x,J_y)\frac{2\pi}{\omega_x d_x} \bmod 2\pi \nonumber \\
    &= 2\pi\left[\left(\frac{J_x}{d_x} + \frac{J_y \alpha}{d_x}\right)\bmod 1\right],
\end{align}
which are essentially the eigenphases of the time evolution operator $\uh(t_0 = 2\pi/[\omega_x d_x])$. Indeed, while the density of states is highly nonuniform with linearly growing and decaying parts when expressed in terms of $H(J_x,J_y)$, it naturally appears uniform in terms of the wrapped energies $E_{(J_x,J_y)}$ at $t_0$ (or the eigenphases of $\uh(t_0)$); see Figs.~\ref{fig:torus2}, \ref{fig:torus3}. It is this ``wrapping'' that allows the spectrum to have energy levels close to a regularly spaced spectrum (and correspondingly high spectral rigidity) as per the discussion in Sec.~\ref{sec:modefluctuations}.

We should also expect the analytically calculated DFT error of Eq.~\eqref{eq:KAMtorus_spectralrigidity} to not be less than the sorted DFT error (see also the corresponding discussion in Sec.~\ref{sec:modefluctuations}):
\begin{equation}
    \vareps_C(1,t_0)[q_{m_y}] \geq \vareps_C(1,t_0)[\overline{q}].
\end{equation}
We numerically verify this to be the case for $\Eshell{d}$ (with $d_x = d_y$) both belonging to and outside the infinite sequence to which Theorem~\ref{thm:KAMcycerg} applies; however, we specifically appear to have $\vareps_C(1,t_0)[q_{m_y}] \approx \vareps_C(1,t_0)[\overline{q}]$ for several $\Eshell{d}$ including all those belonging to the infinite sequence (i.e. with $\lvert \delta_{d_y}\rvert < 1/d_y$). This is shown in Fig.~\ref{fig:torus4}. Further, we find that the persistence amplitudes $z^2(\ptau,t_0)[\overline{q}]$ are also in close numerical agreement with Eq.~\eqref{eq:2DKAMpersistence} for $\Eshell{d}$ in this sequence, as seen in Fig.~\ref{fig:torus5}. Finally, the SFF is shown in Fig.~\ref{fig:torus6}, with a large $O(1)$ peak expected at $\ptau = \pm d$ from Eq.~\eqref{eq:2DKAMpersistence}.

In conclusion, we have analytically proven and numerically verified that all quantized 2D KAM tori admit an infinite sequence of energy subspaces that are cyclic ergodic and non-aperiodic, with an exact expression for the spectral rigidity that is of a higher order of magnitude than even Wigner-Dyson level statistics (i.e. has lower error). On the other hand, quantized rational tori were shown to be non-aperiodic and argued (in Appendix~\ref{app:toruscyclic}) to have spectral rigidity comparable to Poisson statistics, making them unlikely to be ergodic in any subspace. These quantum properties directly reflect the discretized classical ergodic properties of these systems. Some additional numerical evidence (outside the scope of this paper) further indicates that it is possible, and perhaps even typical, to have cyclic ergodic and non-aperiodic subspaces for KAM tori even when $\Eshell{d}$ is not in the infinite sequence for which we are able to prove these properties analytically. These subspaces often show an almost ideal Gaussian persistence and an apparent $O(d^{-2} \ln d)$ error (e.g., for $\alpha = \sqrt{2}$), but we do not yet have a quantitative understanding of this behavior.



\section{Discussion: Dynamical ergodicity and thermalization}
\label{sec:discussion}

In classical dynamical systems, it is usually unnecessary to differentiate between two equivalent notions of ergodicity. The first is ``dynamical ergodicity'', by which we mean the dynamical behavior where almost all initial states in phase space visit the neighborhood of every other state. The second is ``thermal ergodicity'', referring to the initial-state-independence of time-averaged expectation values of any observable, as quantified by Eq.~\eqref{eq:cl_ergodic}. The direct equivalence~\cite{Sinai1976, HalmosErgodic} between the dynamical and thermal property is at the foundation of classical statistical mechanics~\cite{PlatoErgodic}.

The analogous picture for quantum dynamical systems is the main subject of this section, which we will approach in a semi-qualitative manner. As noted in Sec.~\ref{sec:intro}, ETH has been proposed as a quantum analogue (or at least, a sufficient condition) for the thermal ergodicity of physical observables, and is largely a statement about energy eigenstates in some ``physical'' basis (with minimal conditions on the energy eigenvalues themselves, namely non-degeneracy and rational incommensurability)~\cite{DAlessio2016, deutsch2018eth, subETH}. The detailed study of quantum cyclic permutations in the previous sections strongly suggests that quantum cyclic ergodicity, which directly decodes spectral rigidity into a dynamical property in Hilbert space, is the appropriate quantum counterpart to dynamical ergodicity. The open problem of the connection between dynamical and thermal properties in quantum statistical mechanics is therefore essentially equivalent (through the framework of cyclic permutations) to that of the connection between the observable-independent energy eigenvalues and observable- (or basis-)dependent energy eigenstates of a quantum system.


The close correspondence between quantum and classical cyclic ergodicity suggests that one may explore these connections in systems with a classical limit, and apply it to a more general class of quantum systems that show spectral rigidity and ETH. In particular, we may treat a quantum cyclic permutation $\mathcal{C}$ in some $\Eshell{d}$ as representing a fictitious phase space (or energy shell) $\mathcal{P}[\mathcal{C}]$ in systems with or without a classical limit. By Weyl's law~\cite{Haake}, an ensemble of $\mu(A) d$ pure states in $\mathcal{C}$ represents a region $\mathcal{P}[\mathcal{C}]$ of measure $\mu(A)$ (see Sec.~\ref{sec:quantumcyclic}, and Refs.~\cite{StechelMeasure, StechelHeller}). Such an ensemble corresponds to a mixed state of the form
\begin{equation}
    \hat{\rho}(A) = \frac{1}{\mu(A)d} \sum_{k \in C(A)} \lvert C_k\rangle \langle C_k\rvert,
    \label{eq:phasespacemixedstates}
\end{equation}
where $C(A) \subseteq \mathbb{Z}_d$ is a set of indices of size $\lvert C(A)\rvert = \mu(A) d$ (with $\lvert \cdot \rvert$ denoting the cardinality of a finite set). Interestingly, while such a connection between mixed states and phase space regions could in principle be used to quantitatively connect the quantum error $\vareps_C(1,t_0)$ to that of the classical cyclic permutations, it is not clear how to do this in practice. This is because the dominant contribution to the classical error is in the lack of overlap of these mixed states, which is determined more strongly by their support in Hilbert space~\cite{Halmos2sub, Intro2sub} than the much smaller contribution from the error in the overlap of the constituent pure states, except in the trivial $\mu(C_k) \to 1/d$ pure state limit.

It is also worth noting that not all $\mathcal{P}[\mathcal{C}]$ correspond to the actual phase space $\mathcal{P}$ in systems with a classical limit, but likely only those that satisfy the \textit{generating property} with respect to classical phase space regions (see Sec.~\ref{sec:cl_cyclic}). At a strictly formal level, any $n$-element partition $B_k(n)$ of $\mathcal{P}$ can be used to define non-degenerate observables $b[B_k(n)]$ that take distinct values on each $B_k(n)$ as $n\to\infty$ (e.g. $b$ is essentially $\boldsymbol{\theta}$ for the tori of Sec.~\ref{sec:KAMtori}); such observables admit an orthonormal eigenbasis $\mathcal{B} = \lbrace\lvert B_k\rangle\rbrace$ by the standard procedure of quantization, such as the postulates in Ref.~\cite{ShankarQM}. The quantum analogue of the generating property then appears to be
\begin{equation}
    \left(\mathcal{C} = \mathcal{B}\right) \implies \left(\mathcal{P}[\mathcal{C}] = \mathcal{P}\right),
    \label{eq:quantumgeneratingproperty}
\end{equation}
for some such $\mathcal{B}$,
and is therefore an observable-dependent eigenstate property, unlike cyclic ergodicity.

\subsection{Quantum analogues of classical ergodicity and mixing}
\label{sec:q_erg_mix}
If such a generating $\mathcal{C}$ has error $\vareps_C(1,t_0)$, we can approximate $\uh(\ptau t_0) \approx  \uc^\ptau$ whenever $\lvert \ptau\rvert \ll 1/\sqrt{\vareps_C(1,t_0)}$ (from the bound of Eqs.~\eqref{eq:q_persistence_bound}, \eqref{eq:q_cyc_persistence_bound}). As we will see below, this approximation implies that late-time quantum dynamics in the fictitious phase space basis $\mathcal{P}[\mathcal{C}]$ can be described in nearly classical terms up to extremely long times for sufficiently rigid spectra (e.g. $t_{\htime}/\sqrt{\ln(t_{\htime}/t_{\text{ramp}})}$ with $t_0 = t_{\text{ramp}}$, $t_0 d = t_{\htime}$ for Wigner-Dyson spectra), standing in contrast to rapid ``quantum'' thermalization processes such as scrambling that cause an initial state to rapidly spread in some ``local'' basis over much shorter times (closer to $t_{\text{ramp}}$) without any significant influence of spectral rigidity beyond $t>t_{\text{ramp}}$~\cite{ProsenErgodic, ShenkerThouless, ChanScrambling, GoogleScrambling, ThoulessRelaxation, WinerHydro, Reimann2016}.

Using the above approximation with Eq.~\eqref{eq:phasespacemixedstates} for $A,B \subseteq \mathcal{P}$, we have
\begin{equation}
    \mu(\mathcal{T}^{\ptau t_0}A \cap B) \approx \frac{1}{d}\left\lvert C(\mathcal{T}^{\ptau t_0}A) \cap C(B)\right\rvert.
    \label{eq:q_ergodic_hierarchy}
\end{equation}
In this way, thermal properties of fictitious phase space regions $A$,$B$ can be represented over this time scale as simple correlation functions of subsets $C(A),C(B)$ of $\mathbb{Z}_d$ under relative shifts, which may obey e.g. Eq.~\eqref{eq:cl_ergodic} or Eq.~\eqref{eq:cl_mixing}. We further expect that not all such regions $A,B$ are ``physically relevant'' in the classical limit: some may correspond to e.g. scattered points or regions that do not contain the neighborhood of any point in the appropriate physical variables (such as the highlighted set $\lbrace C_k \rbrace_{k=0}^{14}$ in Fig.~\ref{fig:torus_cyclic}). The definition of such regions requires considerably higher resolution (of measure $\sim 1/d$) in the physical variables than classically accessible, and arguably become unphysical in the classical limit. Thus, if all \textit{physical} regions $A,B,\ldots$ satisfy Eq.~\eqref{eq:cl_ergodic} we have the emergence of effectively classical ergodic behavior, and if they further satisfy Eq.~\eqref{eq:cl_mixing} we have mixing behavior, in the fictitious phase space $\mathcal{P}[\mathcal{C}]$.

We note that such statements, about the physicality of fictitious phase space regions with respect to observables, is again an observable-dependent eigenstate property and not determined by spectral rigidity. This suggests that the classical ergodic hierarchy is significantly encoded in the energy eigenstates (with spectral rigidity determining the time scale of applicability), while it is their discretized counterparts (cyclic ergodicity and aperiodicity) that are fully encoded in the energy eigenvalues. Such a conclusion is consistent with the mathematical observation~\cite{KatokSinaiStepin} that classical cyclic permutation errors cannot fully identify classical mixing behaviors, which instead requires an ``incongruity'' of the $C_k$ when expressed in terms of the physical variables in the $n\to\infty$ continuum limit (\newedit{essentially, the $C_k$ must be arranged erratically in phase space such that} non-constant eigenfunctions \newedit{$f_{m \neq 0}(x \in \mathcal{P})$} of the classical unitary $\mathcal{U}_\mathcal{T} \approx \mathcal{U}_\mathcal{T_C}$, \newedit{which satisfy $f_m(x \in C_k) \approx \exp(-2\pi i k m/n)$ for $0 < \lvert m\rvert \ll n$, do not exist as smooth, well-behaved functions of the phase space coordinates $x$ in the $n\to\infty$ limit}; see also Sec.~\ref{sec:cl_ErgodicHierarchy}). An identical conclusion is suggested by the numerical observation~\cite{Lozej2021, CasatiWang} that similar spectral properties may be seen in mixing and merely ergodic systems, and one may require explicitly identified physical variables to distinguish them \newedit{(see also a related discussion of physical bases and the role of eigenstates in Ref.~\cite{Borgonovi2016})}. Nevertheless, in Sec.~\ref{sec:PoincareETH}, we argue that a connection between spectral rigidity and eigenstate properties can be made if we introduce an appropriate dynamical criterion to identify physical observables.



\subsection{Poincar\'{e} recurrences and eigenstate thermalization}
\label{sec:PoincareETH}


It is natural to ask if we can characterize a physical phase space $\mathcal{P}[\mathcal{C}]$ directly based on some physical property, without relying on observables such as $\mathcal{B}$ (which may only be possible to identify in quantum systems with an explicit phase space description~\cite{PolkovnikovPhaseSpace}, and even then, may be complicated by the non-classical behavior of e.g. Wigner quasiprobability distributions~\cite{WignerNonnegativity, ShepelyanskyEhrenfest, ChirikovEhrenfest}, especially for small $\mu(C_k)$). In this subsection, we consider a quantum analogue of Poincar\'{e} recurrences as a candidate for this role, noting that it is satisfied by the physical phase space in all classical systems --- ergodic and non-ergodic\footnote{We recall that quantum recurrences~\cite{QuantumRecurrences} of phases in the energy eigenbasis occur over times possibly exponentially large in $d$~\cite{BrownSusskind2}, and are not directly relevant at earlier times.}.

We introduce the following \textit{ad hoc} definition based on the classical statement of the Poincar\'{e} recurrence theorem~\cite{Sinai1976, SinaiCornfield, HalmosErgodic} (see also Sec.~\ref{sec:ErgodicReview}): any subspace $\mathcal{H}(A) \subseteq \eshell_d$ with projector $\proj(A)$ (and density matrix $\hat{\rho}(A) = \proj(A)/\Tr[\proj(A)]$) is Poincar\'{e} recurrent if there exists an orthonormal basis $\mathcal{A}$ for $\mathcal{H}(A)$, such that any pure state $\lvert \psi\rangle \in \mathcal{A}$ returns to have a larger-than-random overlap with the subspace at some time $t$, with $t_0 \ll \lvert t\rvert \lesssim O(t_0 d)$:
\begin{equation}
    \Tr\left[\proj(A) \uh(t) \lvert \psi\rangle \langle \psi\rvert \uh^\dagger(t)\right] \gg O(d^{-1}).
\end{equation}
We note the similarity of the restriction on the range of $t$ to that in the definition of cyclic aperiodicity (Eq.~\eqref{eq:q_cyclic_aperiodicity}).




For simplicity, we assume that it is sufficient to consider DFT cyclic permutations as representing any (sub-)regions in phase space. We also assume that the persistence amplitude is the only greater-than-random component of any state with respect to a DFT basis of interest (in other words, none of the terms $\vareps_C^{1/2}(\ptau,t_0) \nu_m(\ptau)$ exceed $O(d^{-1/2})$ in magnitude), as is typically the case for e.g. Wigner-Dyson or Poisson statistics (partly due to Eq.~\eqref{eq:nu_sff}). Let $t_R(\mathcal{C})$ then represent the randomization time of a DFT cyclic permutation $\mathcal{C}$ --- the smallest time for which we have $z(t_R/t_0, t_0) \in O(d^{-1/2})$, which generically decreases with increasing $\vareps_C(1,t_0)$.

The time $t_R$ determines the minimum dimension of Poincar\'{e} recurrent subspaces $\mathcal{H}(A)$ that have the diagonal form in Eq.~\eqref{eq:phasespacemixedstates}. Specifically, only subspaces with
\begin{equation}
    \dim \mathcal{H}(A) = \mu(A) d \geq \frac{t_0 d}{t_R(\mathcal{C})},
\end{equation}
are guaranteed to be Poincar\'{e} recurrent. Non-aperiodic cyclic permutations have $t_R(\mathcal{C}) > t_0 d$ and every subspace is recurrent; for ergodic ones, $t_R(\mathcal{C}) > t_0 d/2$, and any subspace with $\mathcal{H}(A) \geq 2$ (in other words, every subspace that is not a pure state) is recurrent.

If we impose Poincar\'{e} recurrence for all regions of the form of Eq.~\eqref{eq:phasespacemixedstates} containing more than one pure state ($\mu(A) > 1/d$)\footnote{If we imagine pure states as corresponding to regions with negligible ($= 1/d$) phase space measure, they are effectively measure zero sets which are not required to satisfy Poincar\'{e} recurrence even classically, unless they are on e.g. periodic orbits. More realistically, we should expect recurrence times to increase with decreasing measure.}, as a criterion to identify a physical phase space $\mathcal{P}[\mathcal{C}] = \mathcal{P}$, it follows that the only DFT cyclic permutations that can satisfy this requirement are ergodic ones. In other DFT bases, Poincar\'{e} recurrence fails for certain regions $A$ with $2/d \leq \mu(A) < t_0/t_R(\mathcal{C})$. Thus, if we want to construct a fictitious phase space for some energy subspace $\eshell_d$ that satisfies Poincar\'{e} recurrence for arbitrarily small regions (of greater than the smallest volume $1/d$), there are two possibilities given our assumptions:
\begin{enumerate}
    \item $\eshell_d$ is itself ergodic, and a corresponding ergodic (DFT) cyclic permutation $\mathcal{C}$ allows the definition of mixed states corresponding to Poincar\'{e} recurrent phase space regions according to Eq.~\eqref{eq:phasespacemixedstates}.
    \item $\eshell_d$ must be decomposed into $M$ ergodic subspaces $\eshell_{d_1}(1),\ldots,\eshell_{d_M}(M)$, each spanned by $d_k$ energy levels (respectively) that add up to $d$. Poincar\'{e} recurrent phase space regions can then be defined according to Eq.~\eqref{eq:phasespacemixedstates} on the combination of their respective ergodic (DFT) cyclic permutations $\mathcal{C}(1),\ldots,\mathcal{C}(M)$. The previous case corresponds to $M=1$.
\end{enumerate}
This is the key component of our argument: imposing an \textit{ad hoc} Poincar\'{e} recurrence requirement to identify a physical phase space basis $\mathcal{P}[\mathcal{C}] = \mathcal{P}$ forces this basis to be a collection of ergodic cyclic permutations. Thus, we obtain a quantum analogue of the classically more trivial statement that the phase space $\mathcal{P}$ is either itself ergodic or decomposes into ergodic subsets~\cite{Sinai1976,SinaiCornfield,HalmosErgodic}.

In either case, the projectors $\proj(A)$ can be written as
\begin{equation}
    \proj(A) = \sum_{m=1}^{M}\sum_{k\in C_m(A)}\lvert C_k(m)\rangle\langle C_k(m)\rvert
\end{equation}
where each $C_m(A)$ is a set of $n_m(A) = \lvert C_m(A)\rvert$ indices of elements of $\mathcal{C}(m)$, and $\sum_m n_m(A) = n(A) \approx \mu(A) d$.

The connection to ETH emerges if one asks for the matrix elements of these projectors in the energy eigenbasis. Let $\lbrace \lvert E_k(m)\rangle\rbrace_{k=0}^{d_m-1}$ be the energy eigenstates contained in $\eshell_{d_m}(m)$, whose form is explicitly known as DFTs of the $\lvert C_k(m)\rangle$. Assuming that (in the generic case) each $C_m(A)$ is randomly distributed on $\mathcal{C}(m)$ (so the phases of the DFT can be taken to be random for $1 \ll n_m \ll d_m$), we have
\begin{equation}
    \langle E_{k}(m)\rvert \proj(A) \lvert E_{j}(m)\rangle = \frac{n_m(A)}{d_m}\delta_{kj} + O\left(\frac{\sqrt{n_m(A)}}{d_m}\right) R_{kj}(m),
    \label{eq:RecurrentProjectorMatrices}
\end{equation}
for some random $d_m \times d_m$ Hermitian matrix $R_{kj}(m)$ with $O(1)$ matrix elements (with weak correlations ensuring $\proj^2 = \proj$). On the other hand, the statement of ETH for an $n$-dimensional projector $\proj$ may be motivated by random matrix arguments (e.g. similar to Refs.~\cite{DAlessio2016, subETH, pSFF}; see also Ref.~\cite{AnzaETH} for a discussion of the role of highly degenerate projectors), giving
\begin{equation}
    \langle E_k\rvert \proj\lvert E_j\rangle = \frac{n}{d}\delta_{kj} + O\left(\frac{\sqrt{n}}{d}\right)R_{kj},
    \label{eq:projectorETH}
\end{equation}
in the full energy subspace $\eshell_d$. We see that Eq.~\eqref{eq:RecurrentProjectorMatrices} and Eq.~\eqref{eq:projectorETH} are guaranteed to agree for $M=1$. For $M>1$, the first (diagonal) terms of Eq.~\eqref{eq:RecurrentProjectorMatrices} and \eqref{eq:projectorETH} can only agree through a statistically unlikely coincidence $n_m/d_m = n/d$; even in the rare instance that this holds, the second (fluctuation) term of the former typically has matrix elements of size $O[(\sqrt{M n})/d]$ (or zero when $k,j$ correspond to different $\eshell_{d_m}(m)$), and ETH can be satisfied only for $M\in O(1)$.

We have essentially argued (under some simplifications, e.g. focusing on DFT bases with typical random parts), that a quantum correspondence between dynamical ergodicity (equivalently, spectral rigidity) and thermal ergodicity (essentially, observable-dependent ETH or eigenstate properties) can be established within the framework of cyclic permutations, given a single additional principle to identify physical observables. In particular, given Poincar\'{e} recurrence, ETH is satisfied by generic projectors onto ``physical'' phase space regions (which then show quantum analogues of thermal ergodicity and mixing, albeit over long times $t \gg t_{\text{ramp}}$) only if the system is cyclic ergodic in $\Eshell{d}$.   It would be interesting to see if such an argument can be made more rigorous, and suitably generalized to systems without a classical limit where a similar correspondence between eigenvalue and eigenstate statistics is observed~\cite{DAlessio2016} but not yet understood. To do so, it would be necessary to determine if there's any link between a Poincar\'{e} recurrence requirement for the fictitious phase space observables considered here (relevant for $t\gtrsim t_{\text{ramp}}$ dynamics), and the \textit{local} or few-body observables (relevant for $t\lesssim t_{\text{ramp}}$) that are the subject of conventional ETH~\cite{deutsch1991eth,srednicki1994eth, srednicki1999eth, deutsch2018eth,DAlessio2016, subETH, Nandkishore}.



\section{Conclusions}

We have identified fully quantum dynamical properties in the Hilbert space that resemble discretized classical ergodic properties, namely cyclic ergodicity and aperiodicity, and shown how these are determined by precise measures of energy level statistics, thereby introducing genuine dynamical elements into the study of quantum ergodicity and ``chaos''. These properties were illustrated with four physically relevant types of level statistics: Wigner-Dyson, Poisson, and that of 2D KAM and rational tori. A key takeaway is that we can isolate the precise aspects of random matrix behavior --- an exact Gaussian distribution of mode fluctuations with variance $\sigma_\Delta^2$ in the range of Wigner-Dyson spectral rigidity --- that are sufficient conditions for cyclic ergodicity, \newedit{identifying which of the numerous measures of level statistics~\cite{Mehta} are physically important (from the present viewpoint)}. However, \newedit{random matrix behavior is not necessary for cyclic ergodicity}, as typified by 2D KAM tori. Our study of the latter included an analytical proof of spectral rigidity in an infinite sequence of subspaces, demonstrating that cyclic permutations may offer an intuitive path towards the open problem of rigorously connecting spectral and ergodic properties in individual systems~\cite{Zelditch, Anantharaman}. 

Our dynamical construction also clarifies that spectral rigidity should be most directly associated with (quantum) cyclic ergodicity \newedit{and aperiodicity}, rather than \newedit{stronger notions of} chaos as is the norm in the literature~\cite{Haake}. \newedit{This provides a general theoretical framework for justifying} numerical observations~\cite{GiraudWDwithoutPO, Lozej2021, CasatiWang} in systems with a non-chaotic, ergodic classical limit. At the same time, the precise relationship between these \newedit{eigenvalue-based ``cyclic'' dynamical properties} and more familiar (semi-)classical properties \newedit{(generally involving both eigenvalues and eigenstates in the classical limit)} merits further investigation, perhaps based on a classical limit of Eqs.~\eqref{eq:nu_sff} and \eqref{eq:SFFerror_relation} that relate quantum cyclic ergodicity to mean recurrence properties (such as, possibly, the semiclassical effect of periodic orbits~\cite{HOdA, BerrySpectralRigidity, Haake}) in a DFT cyclic permutation. As argued in Sec.~\ref{sec:discussion}, the classical limit may also provide inspiration for identifying fully quantum connections between dynamical (or eigenvalue) and thermal (or eigenstate) properties in arbitrary quantum systems.


The approach developed here may generalize to ``quantizing'' other dynamical properties of physical interest. Connections between quantum dynamical properties and a suitably defined Kolmogorov-Sinai (KS) entropy (a measure of chaos closely related to Lyapunov exponents~\cite{Ott}, see e.g. Ref.~\cite{AlickiFannes} for candidate definitions) may be accessible through non-cyclic permutations, which are related to the classical KS entropy in Refs.~\cite{KatokStepin2, SinaiCornfield}. Generalizing further to non-unitary dynamical structures could help associate precise dynamical properties with the recently observed analogues of spectral rigidity in autonomous dissipative quantum systems, whose evolution is generated by non-unitary Liouvillian superoperators~\cite{DissipativeChaos1, DissipativeChaos2, DissipativeChaos3}. Finally, the precise understanding of the spectral rigidity of 2D KAM tori in terms of their dynamical properties obtained in Sec.~\ref{sec:KAMtori} may serve as a starting point to adapt elements of KAM theory~\cite{Ott, KAMtorusdef} --- the classical study of the development of ergodicity under perturbations to integrable systems --- to fully quantum systems. The rapid development of ergodicity in many-particle systems even under small perturbations, while not well understood in quantum mechanics, is believed to be essential for the applicability of statistical mechanics~\cite{deutsch2018eth,KAM2}.

\subsubsection*{Acknowledgments}
This work was supported by the U.S. Department of Energy, Office of Science, Office of Basic Energy Sciences under Award No. DE-SC0001911. We thank Yunxiang Liao, Laura Shou and Michael Winer for useful discussions.

\begin{appendices}
\addtocontents{toc}{\protect\setcounter{tocdepth}{1}} 

\section{Classical cyclic permutations (review)}
\label{app:cl_erg_errors}

This proof essentially follows Ref.~\cite{KatokStepin2}. First, we discuss the bound for cyclic ergodicity. Assume that each of the (say) $M_C$ ergodic subsets $\lbrace\mathcal{P}_j\rbrace_{j=1}^{M_C}$ of $\mathcal{P}$ completely contains at least one element $C_{p(j)} \subseteq \mathcal{P}_j$ of the decomposition. As $\mathcal{T}^tC_{p(j)}\in \mathcal{P}_j$ for all $t$ by definition, we must have  $\mu[(\mathcal{T}^{[p(j+1)-p(j)]t_0}C_{p(j)})\cap C_{p(j+1)}] = 0$. An important exception to this behavior is when $M_C=1$, where there is no reason to impose a vanishing intersection. Thus,
\begin{align}
    \frac{1}{2}\sum_{j=1}^{M_C}\ \mu\left[(\mathcal{T}^{[p(j+1)-p(j)]t_0}C_{p(j)})\symdiff C_{p(j+1)}\right] = \frac{1}{n}M_C,
    \label{eq:cl_erg_bound}
\end{align}
for $M_C \geq 2$. Now, we need to know how the error in an $\ell$-step time evolution $(\mathcal{T}^{\ell t_0} C_k) \symdiff C_{k+\ell}$ is related to the error $(\mathcal{T}^{t_0}C_m)\symdiff C_{m+1}$ made in approximating each step. 
For this, we note that
\begin{align}
    &(\mathcal{T}^{(m+1)t_0} A) - C_{m+1}  \subseteq \left[(\mathcal{T}^{t_0}C_m)-C_{m+1}\right]\cup\mathcal{T}^{t_0}\left[(\mathcal{T}^{mt_0}A)-C_m\right],\ \forall\ A \subseteq \mathcal{P} \label{eq:cl_err_recurrence} \\
    &\implies \mu[(\mathcal{T}^{\ell t_0} C_k) \symdiff C_{k+\ell}] \leq \sum_{m=1}^\ell\ \mu[(\mathcal{T}C_{k+m-1})\symdiff C_{k+m}],
    \label{eq:cl_err_cascade}
\end{align}
where the second line follows from recursively applying the first line to $\mathcal{T}^{(m+1)t_0}C_k\symdiff C_{k+q}$ with $A = C_k$. Using this in Eq.~\eqref{eq:cl_erg_bound}, one obtains $\eps_C \geq (M_C/n)$ for $M_C\geq 2$, giving $\eps_C < 2/n$ for $M_C = 1$.

Cyclic aperiodicity is more straightforward. We impose $\mu[(\mathcal{T}^{nt_0}C_k) \cap C_k] = 0$ (up to possible corrections that vanish as $n\to\infty$), which implies $\eps_C \geq 1/n$ from Eq.~\eqref{eq:cl_err_cascade}.

\section{Quantum cyclic permutations}

\subsection{Fastest decay of persistence}
\label{app:q_erg_errors}

Given a cyclic permutation basis $\mathcal{C} = \lbrace \lvert C_j\rangle\rbrace_{j=0}^{d-1}$, consider some initial state $\lvert C_k\rangle$. After $\ptau$ steps of time evolution, it evolves into
\begin{equation}
    \uh(\ptau t_0) \lvert C_k\rangle = z_k(\ptau, t_0)e^{i\phi_k(\ptau, t_0)}\lvert C_{k+\ptau}\rangle +\sqrt{1- z_k^2(\ptau, t_0)}\lvert \nu_{k+\ptau}^{(k)}\rangle,
\end{equation}
where $\lvert\nu_{k+\ptau}^{(k)}\rangle$ is some normalized vector orthogonal to $\lvert C_{k+\ptau}\rangle$, and $\phi_k(\ptau, t_0)$ is an unimportant phase. This leads to a recurrence relation for the persistence amplitudes,
\begin{align}
    z_{k}(\ptau+1; t_0) e^{i\phi_k(\ptau+1;t_0)} 
    &= z_k(\ptau)e^{i\phi_k(\ptau, t_0)} \langle C_{k+\ptau+1}\rvert \uh(t_0)\lvert C_{k+\ptau}\rangle \nonumber \\
    &\hphantom{=}+\sqrt{1-z_k^2(\ptau, t_0)}\langle C_{k+\ptau+1}\rvert \uh(t_0)\lvert \nu_{k+\ptau}^{(k)}\rangle.
\end{align}
Using the triangle inequality for the magnitudes of these vectors gives
\begin{align}
    &\left\lvert z_k(\ptau, t_0) z_{k+\ptau}(1,t_0) - \sqrt{1-z_k^2(\ptau, t_0)}\sqrt{1-z_{k+\ptau}^2(1,t_0)}\right\rvert \nonumber \\
    &\leq z_k(\ptau+1;t_0) \nonumber \\
    &\leq \left\lbrace z_k(\ptau, t_0)z_{k+\ptau}(1,t_0) + \sqrt{1-z_k^2(\ptau, t_0)}\sqrt{1-z_{k+\ptau}^2(1,t_0)}\right\rbrace,
    \label{eq:q_err_ineq_1}
\end{align}
on noting that $\uh(t_0)\lvert \nu_{k+\ptau}^{(k)}\rangle$ is orthogonal to $\uh(t_0)\lvert C_{k+\ptau}\rangle$, and consequently the inner product of the former with $\lvert C_{k+\ptau+1}\rangle$ cannot exceed $\sqrt{1-z_{k+\ptau}^2(1,t_0)}$ in magnitude.

The above inequalities can be simplified by defining $\theta_k(\ptau) = \arccos z_k(\ptau, t_0) \in [0,\pi/2]$. In terms of these variables, Eq.~\eqref{eq:q_err_ineq_1} becomes
\begin{equation}
    \min\left\lbrace \theta_k(\ptau)+\theta_{k+\ptau}(1), \frac{\pi}{2}\right\rbrace \geq \theta_k(\ptau+1) \geq \lvert \theta_k(\ptau)-\theta_{k+\ptau}(1) \rvert. 
\end{equation}
Summing $\theta_k(\ptau+1)-\theta_k(\ptau)$ from $\ptau = \ptau_1$ to $\ptau = \ptau_2-1$ gives
\begin{equation}
\sgn(\ptau_2)\min\left\lbrace\theta_k(\ptau_2),\frac{\pi}{2}\right\rbrace-\sgn(\ptau_1)\min\left\lbrace\theta_k(\ptau_1), \frac{\pi}{2}\right\rbrace \leq (\sgn(\ptau_2)-\sgn(\ptau_1))\sum_{\ptau=\ptau_1}^{\ptau_2-1}\theta_{k+\ptau}(1),
\end{equation}
which becomes Eq.~\eqref{eq:q_persistence_bound} when $p_1 = 0$. We see that the bound is saturated when $\uh(t_0)$ at the $\ptau$-th step acts like a 2D rotation by the angle $\theta_{k+\ptau}(1)$ in the same direction as the previous steps.

\subsection{Optimal errors for cyclic permutations}
\label{app:q_cyc_dft}

When $\uc$ is a cycling operator, $\uc^\ptau$ is generally a permutation operator on $\mathcal{C} = \lbrace \lvert C_k\rangle\rbrace_{k=0}^{d-1}$ that can be decomposed into a direct sum of cycling operators, each acting on a separate $[d/\mathcal{N}(d,\ptau)]$-sized subset of $\mathcal{C}$:
\begin{equation}
    \uc^\ptau = \bigoplus_{j=1}^{\mathcal{N}(d,\ptau)} \ucj{j}(\ptau).
\end{equation}
The number of cycling operators $\mathcal{N}(d,\ptau)$ is given by the greatest common divisor of $\ptau$ and $d$; in particular, $\mathcal{N}(d,\ptau) = 1$ when $\ptau$ and $d$ are coprime, including $\ptau = 1$. This is most easily seen in the eigenvalue structure of $\uc^\ptau$, which consists of $\mathcal{N}(d,\ptau)$ identical (degenerate) sets of distinct $[d/\mathcal{N}(d,\ptau)]$-th roots of unity. It is also convenient to consider \textit{twisted} versions of $\uc^{\ptau}$, in which each cycle acquires an additional phase $\alpha_j(\twi)$:
\begin{equation}
    \twi\lbrace\uc^{\ptau}\rbrace \equiv \bigoplus_{j=1}^{\mathcal{N}(d,\ptau)} e^{i\alpha_j(\twi)}\ucj{j}(\ptau).
\end{equation}
It is worth noting that the twisting functional $\twi$ affects only the eigenvalues of $\uc^\ptau$, lifting the degeneracy for most values of the $\alpha_j(w)$, while preserving at least one complete orthonormal set of its eigenvectors. Also, the $\ptau$-step persistence amplitudes $z_k(\ptau, t_0)$ are invariant under the action of $\twi$.

\subsubsection{Optimizing the error via the trace inner product}

Due to its non-negativity, the minimum persistence at a given $\ptau$ is bounded by the mean persistence at that time:
\begin{equation}
    \min_{j \in \mathbb{Z}_d} z_j(\ptau, t_0) \leq \frac{1}{d}\sum_{k=0}^{d-1}z_k(\ptau, t_0).
    \label{eq:mean_trace_bound}
\end{equation}
We also have the inequality,
\begin{equation}
    \left\lvert \frac{1}{d}\Tr\left[\twi\lbrace\uc^\ptau\rbrace^\dagger \uh(\ptau t_0)\right]\right\rvert \leq \frac{1}{d}\sum_{k=0}^{d-1} \left\lvert \langle C_k\rvert (\uc^{\ptau})^\dagger\uh(\ptau t_0)\lvert C_k\rangle\right\rvert,
    \label{eq:trace_vs_mean}
\end{equation}
for any $\twi$, where the right hand side is just the mean persistence at $\ptau$, expanded out.

Let us assume that for every $\mathcal{C}$ and given a $\ptau$, there exists a unitary $\hat{V}_C$ and a twisting functional $\twi$ such that
\begin{equation}
    \frac{1}{d}\Tr\left[\hat{V}_C \twi\lbrace\uc^\ptau\rbrace^{\dagger}\hat{V}^\dagger_C \uh(\ptau t_0) \right] = \frac{1}{d}\sum_{k=0}^{d-1} \left\lvert \langle C_k\rvert \twi\lbrace\uc^ \ptau\rbrace^\dagger \uh(\ptau t_0)\lvert C_k\rangle\right\rvert.
    \label{eq:mean_trace_assumption}
\end{equation}
If this holds, then on account of Eq.~\eqref{eq:trace_vs_mean} and the invariance of the $z_k(\ptau, t_0)$ under the action of $\twi$,
\begin{equation}
    \max_{\hat{V} \in \mathcal{U}(d)} \left\lvert \frac{1}{d}\Tr\left[\hat{V} \twi\lbrace\uc^\ptau\rbrace^{\dagger}\hat{V}^\dagger \uh(\ptau t_0) \right]\right\rvert = \max_{\text{all}\ \mathcal{C}} \frac{1}{d}\sum_{k=0}^{d-1}z_k(\ptau, t_0),
    \label{eq:b2maximizationconstraint}
\end{equation}
where $\twi$ is chosen so that Eq.~\eqref{eq:mean_trace_assumption} is satisfied for some $\mathcal{C}$ that maximizes the right hand side of Eq.~\eqref{eq:b2maximizationconstraint}. This follows as $\hat{V}_C$ becomes a special case of $\hat{V}$, and the only freedom to vary the orthonormal basis $\mathcal{C}$ is through its reorientations in Hilbert space --- precisely given by all possible unitary transformations $\hat{V} \in \mathcal{U}(d)$ acting on the energy subspace $\eshell_d$.


Now, we need to establish that Eq.~\eqref{eq:mean_trace_assumption} is indeed valid, and identify $\twi$. It is convenient to consider the two cases of nondegenerate and degenerate $\uc^\ptau$ separately.

\begin{enumerate}
    \item \textbf{Case 1: $\lvert \ptau\rvert$ and $d$ are coprime.}
In this case, $\uc^\ptau$ is itself a cycling operator. We separate the persistence inner product into an amplitude and phase,
\begin{equation}
    \langle C_k\rvert (\uc^{\ptau})^\dagger\uh(\ptau t_0)\lvert C_k\rangle = z_k(\ptau, t_0)e^{i\phi_k(\ptau, t_0)}.
    \label{eq:amplitudephase}
\end{equation}
Let $\overline{\phi}(\ptau, t_0) = \sum_{k=0}^{d-1}\phi_k(\ptau, t_0)$. Define a new cyclic permutation $\mathcal{C}'$ with basis vectors
\begin{equation}
    \lvert C'_k\rangle = e^{i\sum_{j=-1}^{(j+1)p = k} \left\lbrace\phi_{jp}(\ptau, t_0) - [\overline{\phi}(\ptau, t_0)/d]\right\rbrace}\lvert C_k\rangle,
\end{equation}
where $\sum_{j=-1}^{(j+1)p=k} \phi_{jp} = \phi_{-p} + \phi_{0} + \phi_{p}+\ldots+\phi_{k-p}$ is a sum over the index with steps of size $p$, and subtracting $\overline{\phi}(\ptau, t_0)/d$ from each term ensures the single-valuedness of the phases in the new basis. This induces a unitary transformation $\uc \to \ucp = \hat{V}_C \uc \hat{V}_C^\dagger$ (where $\ucp$ is required to satisfy Eq.~\eqref{eq:amplitudephase} with the $\lvert C_k\rangle$ replaced by $\lvert C'_k\rangle$), such that
\begin{equation}
    \langle C_k\rvert \hat{V}_C(\uc^{\ptau})^\dagger\hat{V}_C^\dagger \uh(\ptau t_0)\lvert C_k\rangle = z_k(\ptau, t_0)e^{i\overline{\phi}(\ptau, t_0)/d}.
    \label{eq:coprime_phase_adjustment}
\end{equation}
We see that Eq.~\eqref{eq:mean_trace_assumption} is then satisfied for a twisting functional $w$ with $\alpha_1(w) = -\overline{\phi}(\ptau, t_0)/d$ (however, this phase is inconsequential in this case, being absorbed by the absolute value in Eq.~\eqref{eq:b2maximizationconstraint}).

\item \textbf{Case 2: $\lvert \ptau\rvert$ and $d$ have a nontrivial common factor.} For this case, we can ensure that the analogue of Eq.~\eqref{eq:mean_trace_assumption} for each $[d/\mathcal{N}(d,\ptau)]$-element cycle is satisfied following the procedure leading up to Eq.~\eqref{eq:coprime_phase_adjustment}, with the total phase $\overline{\phi}(\ptau, t_0)$ replaced by that corresponding to the respective cycle, $\overline{\phi}_j(\ptau, t_0)$. Then, it follows that Eq.~\eqref{eq:mean_trace_assumption} is also satisfied overall for $\uc^\ptau$ with a twisting functional $w$ given by $\alpha_j(w) = -\overline{\phi}_j(\ptau, t_0)/d$.
\end{enumerate}

Thus, from Eq.~\eqref{eq:b2maximizationconstraint}, we can maximize the mean persistence by maximizing the magnitude of the trace
\begin{equation}
    f_p(\uc) = \left\lvert \Tr\left[\twi\lbrace \uc^\ptau\rbrace^\dagger \uh(\ptau t_0)\right]\right\rvert
    \label{eq:costfunc}
\end{equation}
with respect to reorientations $\uc \to \hat{V} \uc \hat{V}^\dagger$. In Sec.~\ref{sec:inflection}, this maximum is shown to occur for some $\uc$ satisfying
\begin{equation}
    \left[\uh(\ptau t_0), \twi\lbrace\uc^\ptau\rbrace^{\dagger} \right] = 0,
    \label{eq:dftcommutator1}
\end{equation}
as long as $f_p(\uc) \geq \sqrt{d(d-2)}$ at some such point.

If $\uh(\ptau t_0)$ and $\twi\lbrace \uc^\ptau\rbrace$ both have nondegenerate eigenvalues, each has a unique set of $d$ eigenvectors corresponding to the respective eigenvectors of $\uh(t_0)$ and $\uc$. Eq.~\eqref{eq:dftcommutator1} then implies that both sets of eigenvectors are identical, and $\uc$ must commute with $\uh(t_0)$ to achieve a local extremum of the mean persistence.

When there are degeneracies (in any of $\uh(t_0)$, $\uh(\ptau t_0)$ or $\twi\lbrace \uc^{\ptau}\rbrace$), we can nevertheless reach a similar conclusion by infinitesimally breaking the degeneracies. We can define $\uhd{(\delta_u)} = \uh(\ptau t_0)e^{i\delta_u \hat{Y}}$ where $\delta_u \to 0$ and $\hat{Y}$ is any finite Hermitian operator (i.e. with finite matrix elements in any orthonormal basis), such that $\uhd{(\delta_u)}$ has nondegenerate eigenvalues when $\delta_u \neq 0$. Similarly, we define $\twi_{(\delta_w)}$ by $\alpha_j(\twi_{(\delta_w)}) = \alpha_j(\twi)+\delta_w \gamma_j$ with $\delta_w \to 0$, with the $\gamma_j$ chosen so as to ensure the nondegeneracy of the eigenvalues of $\twi\lbrace \uc^\ptau\rbrace$ (essentially, infinitesimally twisting any degenerate $e^{i \alpha_{j}(w)}\ucj{j}(\ptau)$, $e^{i \alpha_{k}(w)}\ucj{k}(\ptau)$, $\ldots$ relative to each other). Re-expressing Eq.~\eqref{eq:b2maximizationconstraint} in terms of these variables, gives
\begin{equation}
    \max_{\hat{V} \in \mathcal{U}(d)} \left\lvert \frac{1}{d}\Tr\left[\hat{V} \twi_{(\delta_w)}\lbrace\uc^\ptau\rbrace^{\dagger}\hat{V}^\dagger \uhd{(\delta_u)}(\ptau t_0) \right]\right\rvert = \max_{\text{all}\ \mathcal{C}} \frac{1}{d}\sum_{k=0}^{d-1}z_k(\ptau, t_0)+O(\delta_u,\delta_w),
    \label{eq:b2degenerateconstraint}
\end{equation}
where $O(\delta_u,\delta_w)$ consists of terms of the form $(\delta_u)^a (\delta_w)^b y_{ab}$ with $a,b \geq 1$. As with Eq.~\eqref{eq:dftcommutator1}, the solution to the maximization on the left hand side must be among its local extrema, given by
\begin{equation}
    \left[\uhd{(\delta_u)}(\ptau t_0), \twi_{(\delta_w)}\lbrace\uc^\ptau\rbrace^{\dagger} \right] = 0.
    \label{eq:dftcommutator2}
\end{equation}
Now, each nondegenerate operator $\uhd{(\delta_u)}(\ptau t_0)$ and $\twi_{(\delta_w)}\lbrace\uc^\ptau\rbrace^{\dagger}$ has a unique set of $d$ eigenvectors, which the above equation asserts are identical. We can choose $\hat{Y}$ and $\gamma_j$ to break the degeneracy of $\uh(\ptau t_0)$ and $w\lbrace \uc^\ptau \rbrace$ in any desired way, i.e. to pick any complete orthonormal subset of each set of eigenvectors. By Eq.~\eqref{eq:b2degenerateconstraint}, any such choice is equally good for maximizing the mean persistence in the $\delta_u,\delta_w \to 0$ limit. In particular, we can pick $\twi_{(\delta_w)}\lbrace\uc^\ptau\rbrace^{\dagger}$ so that its eigenvectors are identical to those of $\hat{U}_C$; similarly, we can choose $\hat{Y}$ so that the eigenvectors of $\uhd{(\delta u)}(\ptau t_0)$ are identical to any complete orthonormal set of eigenvectors of $\uh(t_0)$. In other words, any choice of degeneracy breaking in the neighborhood of degenerate operators only infinitesimally affects the local extrema of the left hand side of Eq.~\eqref{eq:b2degenerateconstraint}.

Thus, the right hand side of Eq.~\eqref{eq:mean_trace_bound} attains its global maximum when the eigenvectors of $\uc$ are fixed to be any complete orthonormal set of eigenvectors of $\uh(t_0)$, with the only freedom remaining in the assignment of the distinct eigenvalues of $\uc$ to these eigenvectors. This can be concisely expressed as follows: the global maximum of the mean persistence occurs among the solutions to
\begin{equation}
    \lim_{\delta \to 0}\left[\uh(t_0)e^{i\delta\hat{Y}},\uc\right] = 0,
\end{equation}
for any Hermitian $\hat{Y}$. For any $\uc$ satisfying this property, all the $z_j(\ptau, t_0)$ are equal at any given $\ptau$. It follows that $\min_j z_j(\ptau, t_0)$ is also maximized, and the $\ptau$-step error minimized, by the same $\uc$ that maximizes the mean persistence. From the requirement $f_p(\uc) \geq \sqrt{d(d-2)}$, we get the condition $\vareps_C(\ptau,t_0) \leq (2/d)$ on such a minimum of the error.

\subsubsection{Local extrema, and the global maximum for large persistence amplitudes}
\label{sec:inflection}

For simplicity, let $\uone = \twi\lbrace \uc^\ptau\rbrace$ and $\utwo = \uh(\ptau t_0)$. We seek stationary points of the real valued function (from Eq.~\eqref{eq:costfunc})
\begin{equation}
    \left\lvert \Tr\left(\uone^\dagger \utwo\right)\right\rvert
    \label{eq:costfunc2}
\end{equation}
with respect to small reorientations of $\uone$ by $\hat{V}$, to first order. This would yield all the local maxima and minima (as well as saddle and inflection points) of the function except the global minima when the function attains the value $0$, where it is not differentiable. We write $\hat{V} = e^{i\hat{X}}$ with $\hat{X}$ near $0$, and require the phase of the $O(\hat{X})$ term in $\Tr[\hat{V}\uone^\dagger \hat{V}^\dagger \utwo]$ to be orthogonal to the phase of the $O(1)$ term (so that the first variation corresponds only to a change in phase and not in magnitude; alternatively, one could directly extremize the square of Eq.~\eqref{eq:costfunc2}). This gives
\begin{equation}
    \Tr\left(\hat{X}\left[ \uone^\dagger, \utwo\right]\right) = c(\hat{X}) \Tr\left(\uone^\dagger \utwo\right) \text{ for all Hermitian } \hat{X}
\end{equation}
with $c(\hat{X})$ required to be a real-valued function, for the stationary points. As can be verified by imposing this for each independent degree of freedom in the matrix elements of $\hat{X}$, this requires
\begin{equation}
    \left[ e^{-i\alpha_{12}}\uone^\dagger, \utwo\right] = \hat{F},
    \label{eq:tracecommutator_Hermitian}
\end{equation}
where $\hat{F}$ is some traceless Hermitian operator, and $\alpha_{12}$ is the phase of $\Tr(\uone^\dagger \utwo)$.

Up to this point, the unitarity of $\uone$ and $\utwo$ played no role. Now, we use the fact that their products are unitary, and write
\begin{equation}
    e^{-i\alpha_{12}} \uone^\dagger \utwo = e^{i\aot}, \text{ and } \utwo e^{-i\alpha_{12}}\uone^\dagger = e^{i \ato},
\end{equation}
for Hermitian $\aot$ and $\ato$. Formally defining sines and cosines of Hermitian operators through their Taylor series (which are also Hermitian), Eq.~\eqref{eq:tracecommutator_Hermitian} then gives
\begin{equation}
    \cos \aot - \cos \ato + i\left[\sin \aot - \sin \ato \right] = \hat{F}.
\end{equation}
The Hermiticity of $\hat{F}$ demands that the anti-Hermitian part of the left hand side vanishes, giving
\begin{equation}
    \sin \aot = \sin \ato.
\end{equation}
Let $\lbrace a(k)\rbrace_{k=0}^{d-1}$ be the eigenvalues of $\aot$ and $\ato$ (which must have identical eigenvalues up to irrelevant shifts of $2\pi$, as products of two unitaries have the same eigenvalues irrespective of the order~\cite{MatrixProductEigenvalues}). As long as it is known that $a(k) \in [-\pi/2,\pi/2]$, the sine is invertible  and $\ato = \aot$ (In fact, one gets $\aot = \ato$ and therefore $\hat{F} = 0$ for ``generic'' values of $a(k)$ such that $\sin \ato$ is not degenerate, i.e. the set $\lbrace a(k), \pi+a(k)\rbrace$ is non-degenerate). Consequently,
\begin{equation}
    \left\lbrace a(k) \in [-\pi/2,\pi/2],\ \forall\ k \vphantom{e^{i\alpha_{12}}}\right\rbrace \implies \left(\left[e^{-i\alpha_{12}} \uone^\dagger, \utwo\right] = 0\right)
    \label{eq:akimplication}
\end{equation}
at a stationary point. The vanishing commutator on the right side of the implication is precisely the condition of Eq.~\eqref{eq:dftcommutator1}.

The question of interest is now if there's a simple way to guarantee the restriction on $a(k)$ in Eq.~\eqref{eq:akimplication}. To see that there is, we note that $[\Tr(e^{i\aot})] \in \mathbb{R}$ by the definition of $\alpha_{12}$, which implies
\begin{align}
    \sum_{k} \cos a(k) &= \Tr\left(e^{i\aot}\right), \\
    \sum_{k} \sin a(k) &= 0. \label{eq:sinereality}
\end{align}
Let us maximize the multivariable function $b[a(k)] = \sum_k \cos a(k)$ with \textit{fixed} $a(0)$ (and free $a(k \neq 0)$) subject to the constraint in Eq.~\eqref{eq:sinereality} (and implicitly, non-negativity) using e.g. the method of Lagrange multipliers. The stationary points of $b[a(k)]$ occur at
\begin{equation}
    a(k \neq 0) = c + \pi \zeta_k, \text{ with } \zeta_k \in \lbrace{0,1}\rbrace,
\end{equation}
for some constant $c$. The global maximum of $b[a(k)]$ corresponds to $c \in [-\pi/2,\pi/2]$ and $\zeta_k = 0\ \forall k$. Imposing Eq.~\eqref{eq:sinereality} to fix $c$ in terms of $a(0)$, we get
\begin{equation}
    \sum_k \cos a(k) \leq b_{\max}[a(0)] \equiv \cos a(0) + (d-1) \sqrt{1-\frac{\sin^2 a(0)}{(d-1)^2}}.
\end{equation}
This is a monotonically decreasing function of $\lvert a(0)\rvert$ in its full domain $[0,\pi]$ for $d\geq 2$. In particular, if $\lvert a(0)\rvert > \pi/2$, then it is guaranteed that $b[a(k)] < b_{\max}[\pi/2] = \sqrt{d(d-2)}$. Re-expressing $b[a(k)]$ in terms of the trace of the relevant unitaries, we then have
\begin{equation}
    \left\lbrace\left\lvert \Tr\left(\uone^\dagger \utwo\right)\right\rvert \geq \sqrt{d(d-2)}\right\rbrace \implies \left\lbrace a(k) \in [-\pi/2,\pi/2],\ \forall\ k \vphantom{e^{i\alpha_{12}}}\right\rbrace.
\end{equation}
Combined with the implication in Eq.~\eqref{eq:akimplication}, it follows that maxima for which the trace is no smaller than $\sqrt{d(d-2)}$ occur for cycling operators that commute with time evolution, i.e. when $[\uone^\dagger, \utwo] = 0$. Such commuting operators remain local extrema of the trace in other cases, but it is unclear in the present analysis if the global maximum is among them. For comparison with the following subsection, we note that $a(k) = -2\pi p\Delta_k/d$, where $\Delta_k$ are the mode fluctuations used elsewhere (see Eq.~\eqref{eq:persistence_modefluctuations}) in the main text.

\subsection{Decrease of persistence for small permutations of sorted energy levels}
\label{app:optimalsorting}

When $\Delta_n \ll d$, assuming that the energies $E_n$ have been shifted by some additive constant so that $\sum_k \Delta_k = 0$, we have (representing $d$ times the persistence amplitude as per Eq.~\eqref{eq:persistence_modefluctuations})
\begin{equation}
    \sum_{k=0}^{d-1} e^{-2\pi i \Delta_k/d} = 1-\frac{2\pi^2}{d^2}\sum_{k=0}^{d-1}\Delta_k^2+O(\Delta^3_k d^{-3}).
    \label{eq:persistenceTaylor}
\end{equation}
For simplicity, we assume that the levels are already sorted i.e. $E_n < E_m$ when $n<m$. This further implies
\begin{equation}
    \Delta_n-\Delta_m > -\lvert n-m\rvert.
    \label{eq:sortingcondition}
\end{equation}
Any permutation $q(n)$ can be broken up~\cite{Mehta} into a set of cyclic permutations $q_{r}(n)$, each involving a subset of $N_r$ levels $\mathcal{E}[q_r] = \lbrace E_{r_k}\rbrace_{k=0}^{N_r-1}$. For the rest of the argument, we will require (where the subtraction of $r$ is on $\mathbb{Z}$ (linear), and not on $\mathbb{Z}_d$ (circular or modulo $d$))
\begin{equation}
    \lvert r_{k}-r_{j}\rvert < d/2,\ \forall\ k,j \in \mathbb{Z}_{N_r},
    \label{eq:smallpermutation}
\end{equation}
for each $q_r$; permutations $q$ satisfying this are what we refer to as ``small'' permutations. This will ensure that Eq.~\eqref{eq:persistenceTaylor} remains valid under these permutations without discrete shifts of some of the $\Delta_k$ by multiples of $(2\pi)$.

The new mode fluctuations after permutation are given by
\begin{equation}
    \Delta'_{r_{k}} =  \Delta_{r_{k+1}} + [r_{k+1}-r_{k}],
\end{equation}
for each cycle $q_r$. It follows that the mean is preserved, i.e.
\begin{equation}
    \sum_{k=0}^{N_r-1} \Delta'_{r_{k}} = \sum_{k=0}^{N_r-1} \Delta_{r_{k}}.
\end{equation}
Our goal is to show that the variance of the $\Delta'_k$ is larger than that of the $\Delta_k$, which would translate to a decreased persistence by Eq.~\eqref{eq:persistenceTaylor}. We have
\begin{equation}
    \sum_{k=0}^{N_r-1} (\Delta'_{r_{k}})^2 - \sum_{k=0}^{N_r-1}(\Delta_{r_{k}})^2 = 2\sum_{k=0}^{N_r-1} \Delta_{r_{k+1}}[r_{k+1}-r_{k}] + \sum_{k=0}^{N_r-1}[r_{k+1}-r_{k}]^2.
    \label{eq:variancedifference}
\end{equation}
In general, the $r_{k+1}$ are not in any simple (e.g. ascending or descending) order. We can split each difference $[r_{k+1}-r_{k}]$ in the first term on the right hand side into a sum of differences of the $r_{\ell}$ lying between (and inclusive of) them:
\begin{equation}
    \Delta_{r_{k+1}}[r_{k+1}-r_{k}] = \sum_{r_j \in [r_{k},r_{k+1}]} \zeta_k\Delta_{r_{k+1}} (r_{j+1}-r_j),
\end{equation}
where $\zeta_k= \sgn[r_{k+1}-r_{k}]$, and the $r_j$ are chosen to be sorted according to $j$. On including terms with different values of $k$, each interval $(r_{j+1}-r_j)$ occurs in an equal number of terms with positive $\zeta_k = +1$ ($r_{k+1}\geq r_{j+1}$) and negative $\zeta_k = -1$ ($r_{k+1} \leq r_{j})$. We can arbitrarily pair each positive term $r_{+}$ with a negative term $r_{-}$, and use Eq.~\eqref{eq:sortingcondition} for the difference $\Delta_{r_+}-\Delta_{r_-}$ noting that $r_{+} > r_{-}$. This amounts to replacing the equality with $\geq$, and each $\Delta_{r_{k+1}}$ with $-r_{k+1}$, in Eq.~\eqref{eq:variancedifference}. We therefore obtain
\begin{align}
    \sum_{k=0}^{N_r-1} (\Delta'_{r_{k}})^2 - \sum_{k=0}^{N_r-1}(\Delta_{r_{k}})^2 &\geq \sum_{k=0}^{N_r-1}[r_{k+1}-r_{k}]^2 + \left\lbrace-2\sum_{k=0}^{N_r-1} r_{k+1}[r_{k+1}-r_{k}]\right\rbrace, \nonumber \\
    \implies
    \sum_{k=0}^{N_r-1} (\Delta'_{r_{k}})^2 &\geq \sum_{k=0}^{N_r-1}(\Delta_{r_{k}})^2.
\end{align}
The second equation follows from simplifying the first. Adding all such equations from each $q_r$ together, we get
\begin{equation}
    \sum_{k=0}^{d-1} e^{-2\pi i \Delta'_k/d} \leq \sum_{k=0}^{d-1} e^{-2\pi i \Delta_k/d}+O(\Delta^3 d^{-3}).
\end{equation}
This shows that sorting the energy levels corresponds to the maximum persistence at $p=1$ for a given $t_0$ and small $\Delta_k$, at least among other possibilities that can be obtained as small permutations of the sorted levels. This is more like a discrete version of a local extremum. It would be interesting to check if ``larger'' permutations not subject to Eq.~\eqref{eq:smallpermutation} would lead to significantly better maxima; this is unlikely to be the case without some non-intuitive conspiracy between distant energy levels.

\section{Time dependence of persistence amplitudes}
\label{app:errorcoefficientpairing}

\subsection{Error coefficient pairing in discrete sum over paths}
\label{app:typicalpersistence}

We rewrite Eq.~\eqref{eq:q_periodic_random} for $\ptau = 1$ as
\begin{equation}
    \uerr e^{-i\phierr(1)} = (1-\vareps_1)^{1/2}\left[\idop + g_1 \sum_{m=1}^{d-1}\nu_m(1)\uc^m\right],
\end{equation}
where $\vareps_\ptau \equiv \vareps_C(\ptau, t_0)$ and $g_1 = \sqrt{\vareps_1/(1-\vareps_1)}$. We note that $g_1$ is also the coefficient that occurs on the right hand side of Eq.~\eqref{eq:nu_constraint2}. The $\ptau$-th power of the error unitary is
\begin{align}
    \uerr^{\ptau} e^{-i\ptau \phierr(1)} &= (1-\vareps_1)^{\ptau/2} \sum_{s=0}^{\ptau} \left[\binom{\ptau}{s} g_1^s \sum_{m_1,\ldots, m_s} \nu_{m_1}(1)\ldots\nu_{m_s}(1) \uc^{m_1+\ldots+m_s}\right],
    \nonumber \\
    &= (1-\vareps_1)^{\ptau/2} \sum_{r=0}^{d-1}\left(\sum_{s=0}^{\ptau} \dptG{s}{r}\right)\uc^r \label{eq:dpt1b}
\end{align}
where $\binom{\ptau}{s} = \ptau!/(s!(\ptau-s)!)$ is the binomial coefficient, and we recall that the sums are modulo $d$.
We have also defined
\begin{equation}
    \dptG{s}{r} = \binom{\ptau}{s} g_1^s \sum_{m_1,\ldots, m_s} \nu_{m_1}(1)\ldots\nu_{m_s}(1) \overline{\Theta}(m_1+\ldots+m_s = r),
    \label{eq:dpt2}
\end{equation}
with $\overline{\Theta}(x) = 1$ if $x$ is true and $0$ otherwise.
Each term with fixed $r$ in \eqref{eq:dpt1b} represents a sum over paths for the transition amplitude from any $\lvert C_{k}\rangle$ to $\lvert C_{k+r}\rangle$.

Now, we apply the assumption of error coefficient pairing, by considering only terms where \newedit{as many $m_j$ as possible are each paired with a corresponding $m_k = -m_j$}. For even $s$ in Eq.~\eqref{eq:dpt2}, restricting to such pairings necessarily implies that $m_1+\ldots+m_s =0$. For odd $s$, it is not possible to pair all error coefficients and a free error coefficient remains, whose index must necessarily be $r$ if the remaining coefficients are paired. Schematically (in the sense that we avoid explicitly enumerating the possible pairings), for non-negative integer $u$,
\begin{align}
    \dptG{2u}{r} \approx \delta_{r0} &\left\lbrace\binom{\ptau}{2u} g_1^{2u} \sum_{\text{pairings}} [\nu_{m_1}(1)\nu_{-m_1}(1)]\ldots[\nu_{m_u}(1)\nu_{-m_u}(1)]\right\rbrace, \label{eq:pairingsum1}\\
    \dptG{2u+1}{r} \approx g_1\nu_r(\ptau-2u) &\left\lbrace\binom{\ptau}{2u} g_1^{2u} \sum_{\text{pairings}} [\nu_{m_1}(1)\nu_{-m_1}(1)]\ldots[\nu_{m_u}(1)\nu_{-m_u}(1)]\right\rbrace. \label{eq:pairingsum2}
\end{align}
In the second line, we have accounted for $s=2u+1$ different ways of choosing the unpaired coefficient, and used $s \binom{\ptau}{s} = (\ptau+1-s)\binom{\ptau}{s-1}$.

For a given $u$, the sum over pairings and coefficients within the braces in Eqs.~\eqref{eq:pairingsum1} and \eqref{eq:pairingsum2} are identical, irrespective of the value of $r$. Treating $g_1$ as a formally independent parameter that we can take partial derivatives with respect to, we can further replace $(\ptau-2u)$ with $(\ptau-g_1 \vec{\partial}/\partial g_1)$ acting on its right in Eq.~\eqref{eq:pairingsum2}, which moves all the $u$ dependence to inside the braces. For even $\ptau$, this means that each sum over $s$ in Eq.~\eqref{eq:dpt1b} --- which is naturally restricted to even $s$ for $r=0$ and odd $s$ for $r\neq 0$ after pairing --- produces coefficients for all $r$ that are identical except for the operators outside the braces in Eqs.~\eqref{eq:pairingsum1} and \eqref{eq:pairingsum2}. If the time dependence is sufficiently slow, the result for odd $\ptau$ can be extrapolated (to a good approximation) in any convenient way between those for $\ptau\pm 1$. Thus, we have the approximate form
\begin{equation}
    \uerr^{\ptau} e^{-i\ptau \phierr(1)} \approx (1-\vareps_1)^{\ptau/2}\left[\idop \vphantom{\left(\ptau-g_1\frac{\vec{\partial}}{\partial g_1}\right)} + g_1\sum_{r=1}^{d-1}\nu_r(1)\uc^r\left(\ptau-g_1\frac{\vec{\partial}}{\partial g_1}\right)\right] h(\ptau, g_1).
    \label{eq:uerrT_form}
\end{equation}
The function $h(\ptau, g_1)$ originates in the sum over pairings within the braces of Eqs.~\eqref{eq:pairingsum1} and \eqref{eq:pairingsum2}; from the above expression, it is formally related to the persistence amplitude at $\ptau$ by
\begin{equation}
    z(\ptau, t_0) = (1-\vareps_1)^{\ptau/2} h\left(\ptau, \sqrt{\frac{\vareps_1}{1-\vareps_1}}\right).
\end{equation}

\subsection{Gaussian estimate}

The persistence amplitude at $\ptau+1$ can be expressed in terms of the coefficients in $\uerr^{\ptau}$ and $\uerr^1$ as follows:
\begin{equation}
    z(\ptau+1, t_0) = \left\lvert \frac{1}{d}\Tr\left[\uerr^1 \uerr^{\ptau}\right]\right\rvert = \left\lvert \sqrt{1-\vareps_1}\sqrt{1-\vareps_{\ptau}}+\sqrt{\vareps_1\vareps_\ptau} \sum_{r=1}^{d-1}\nu_{r}(1)\nu_{-r}(\ptau)\right\rvert.
\end{equation}
Substituting the appropriate expressions for $\vareps_\ptau$ and $\nu_\ptau$ from Eq.~\eqref{eq:uerrT_form}, we get
\begin{equation}
    z(\ptau+1, t_0) \approx (1-\vareps_1)^{1/2}\left\lvert z(\ptau,t_0) +\left\lbrace (1-\vareps_1)^{\ptau/2} \vphantom{\left(\ptau-g_1\frac{\vec{\partial}}{\partial g_1}\right)} g_1^2\sum_{r=1}^{d-1}\nu_{r}(1)\nu_{-r}(1)\left(\ptau-g_1\frac{\vec{\partial}}{\partial g_1}\right) h(\ptau, g_1) \right\rbrace \right\rvert
    \label{eq:gestimate_intermediate}
\end{equation}

Now, we assume that the second term within the absolute value is smaller than the first, and $\ptau h \gg g_1 \partial h/\partial g_1$; both will be justified retroactively. Further defining
\begin{equation}
    \nu_C = -\sum_{r=1}^{d-1} \nu_r(1) \nu_{-r}(1),
\end{equation}
which happens to measure the goodness of the approximation in Eq.~\eqref{eq:nu_symmetry}, we are led to
\begin{equation}
    z(\ptau+1, t_0) \approx (1-\vareps_1)^{1/2}\left[1-g_1^2\ptau \nu_C \right]z(\ptau, t_0).
\end{equation}
It is now straightforward to multiply over values of $\ptau$ from some given $\overline{\ptau}$ through to $1$. For $\vareps_1 \ll 1$ and setting $\nu_C \approx 1$ as per Eq.~\eqref{eq:nu_symmetry}, we get
\begin{equation}
    z(\overline{\ptau}, t_0) \approx \exp\left[-\frac{\vareps_1}{2}\lvert \ptau\rvert-\frac{g_1^2}{2}\ptau^2\right].
\end{equation}
We see that the smallness of the second term in Eq.~\eqref{eq:gestimate_intermediate} and $\ptau h \gg g_1 \partial h/\partial g_1$  are both satisfied when $\ptau \ll 1/g_1$, i.e. when the persistence amplitude is still close to $1$.

\subsection{Minimum error constraints from the SFF}
\label{app:sfferror}

Substituting the form $K(t) = \lambda t^\gamma$ in Eq.~\eqref{eq:SFFerror_relation} and dropping subleading terms in $\vareps_1 = \vareps_C(1,t_0)$ gives
\begin{equation}
    2\lambda t_0^\gamma\sum_{\ptau=1}^{1/(M\sqrt{\vareps_1})} \ptau^{\gamma-2} \lessapprox \vareps_1.
\end{equation}
For $\gamma \in [0,1)$, the left hand side is dominated by small $\ptau$ and is independent of $M$. Replacing $1/(M\sqrt{\vareps_1}) \to \infty$, we obtain
\begin{equation}
    \vareps_1 \gtrapprox 2\lambda t_0^\gamma\zeta(2-\gamma),
\end{equation}
where $\zeta(x)$ is the Riemann zeta function. In particular, for $\gamma = 0$ and $\lambda = 1/d$ (Poisson statistics), we have $\vareps \gtrapprox \pi^2/(3d) \in O(1/d)$. For $\gamma > 1$, it is instead the terms with larger $\ptau$ that dominate. Using the leading term in Faulhaber's formula for the sum (formula (0.121) in Ref.~\cite{mathGR}; equivalent to replacing the sum with an integral), we have
\begin{equation}
    2\lambda t_0^\gamma\frac{[1/(M\sqrt{\vareps_1})]^{\gamma-1}}{\gamma-1} \lessapprox \vareps_1.
    \label{eq:errsff_gamma>1}
\end{equation}
The presence of $M \in O(1) \geq 1$ in this expression allows us to make only order of magnitude statements. We get
\begin{equation}
    \vareps_1^{(1+\gamma)/2} \gtrapprox 2\lambda t_0^\gamma\frac{ (\gamma-1)}{M^{\gamma-1}},
\end{equation}
which implies $\vareps_1 \gtrsim O(d^{-4/(\gamma+1)})$ when $\lambda \in O(d^{-2})$ and $t_0 \in O(1)$, for any $\gamma \in O(1) > 1$. The most generic case (i.e. typical for Haar random~\cite{Mehta, Haake} systems), $\gamma = 1$, is a bit more subtle. Here, it is again the large-$\ptau$ terms that dominate, so we take the $\gamma \to 1$ limit of Eq.~\eqref{eq:errsff_gamma>1}, which gives
\begin{equation}
    \frac{\vareps_1}{\ln\left(\frac{1}{M\sqrt{\vareps_1}}\right)} \gtrapprox 2\lambda t_0.
\end{equation}
This is a transcendental equation for $\vareps_1$, but we can nevertheless invert it to leading order in $\lambda^{-1}$ (i.e. substituting $\vareps_1 = \mu(\lambda)\lambda$ and solving for $\mu$, neglecting $\ln(\ln \lambda)$), obtaining
\begin{equation}
    \vareps_1 \gtrapprox \lambda t_0 \ln \frac{1}{\lambda}.
\end{equation}
For Wigner-Dyson statistics, $\lambda \in O(d^{-2})$ and $t_0 \in O(1)$ gives $\vareps \geq c d^{-2}\ln d$ for a constant $c$.

\subsection{Numerical evidence for error coefficient pairing}

\begin{figure*}[!hbt]

\subfloat[][$\ptau = 100$.]{\includegraphics[width = 0.28\textwidth]{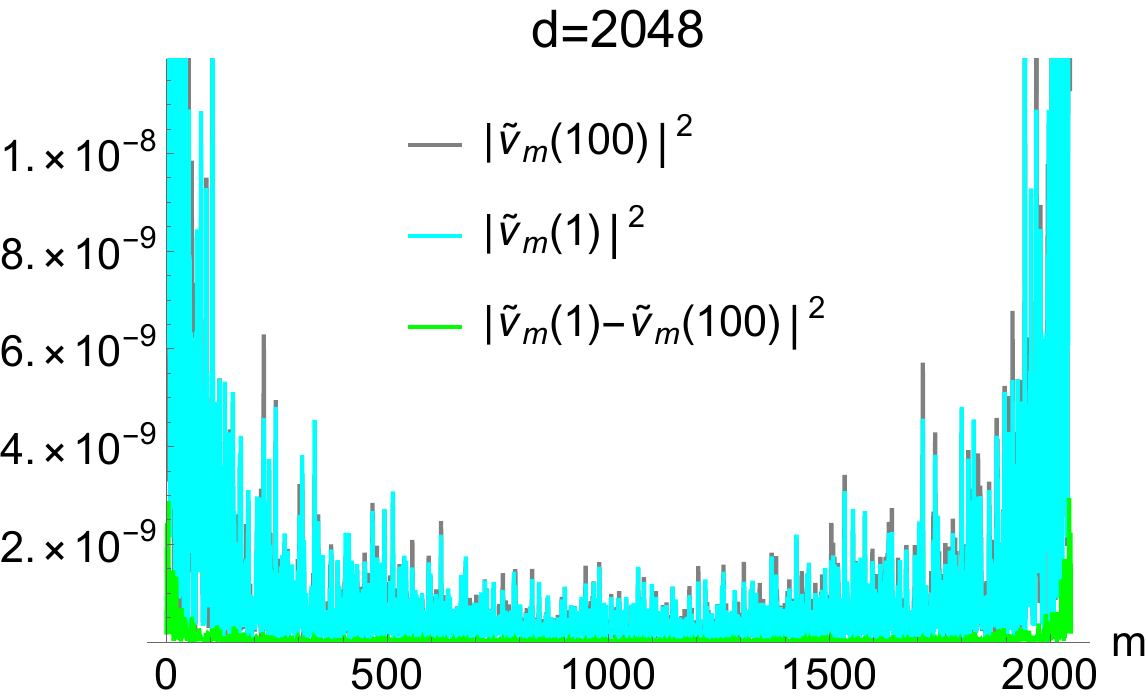}
    \label{fig:errorparing100}}
\hfill
\subfloat[][$\ptau = 100$, zoomed in.]{\includegraphics[width = 0.28\textwidth]{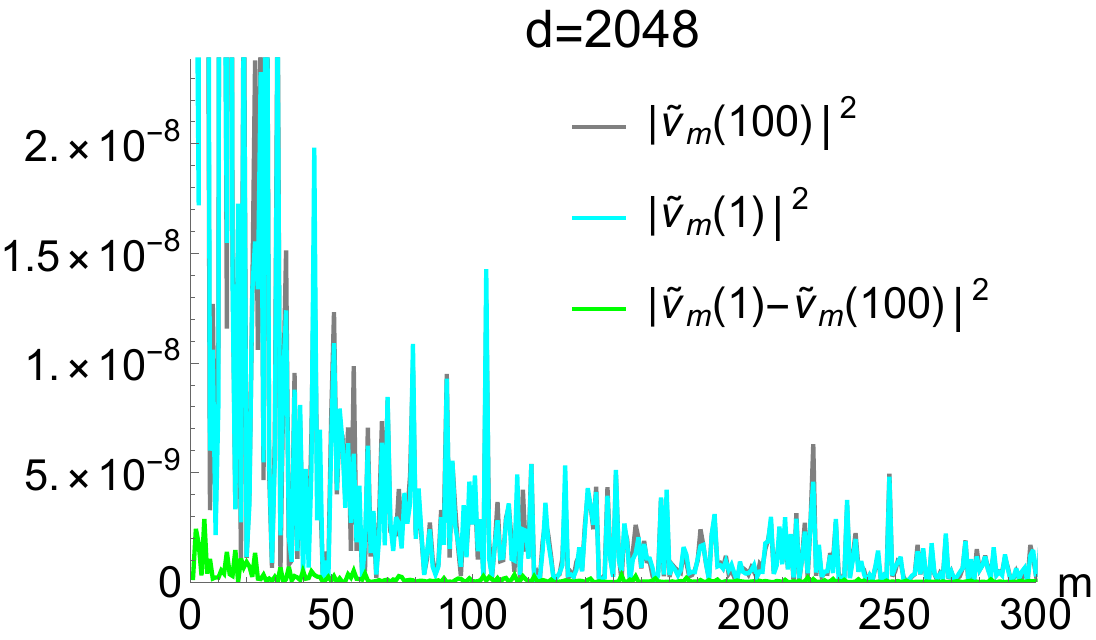}
    \label{fig:errorpairing100zoomed}}
\hfill
\subfloat[][$\ptau = 250$.]{\includegraphics[width = 0.28\textwidth]{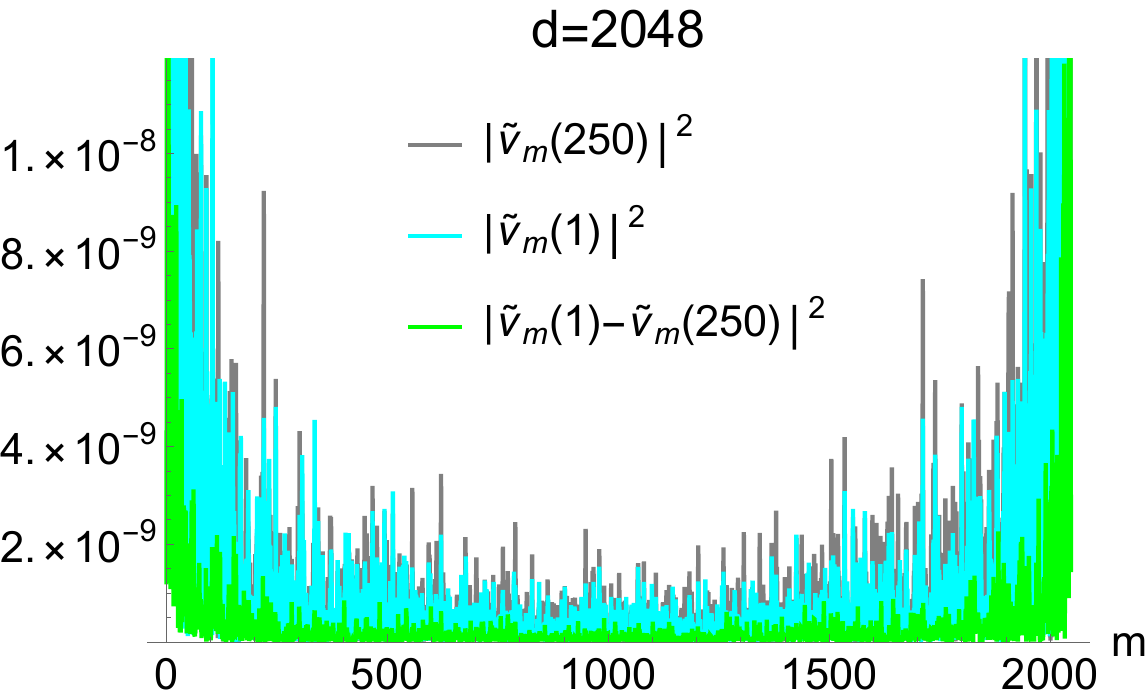}
    \label{fig:errorpairing250}}
    
    \caption{Comparison of $\tilde{\nu}_m(\ptau)$ with $\tilde{\nu}_m(1)$, using magnitudes $\lvert \tilde{\nu}_m(\ptau)\rvert^2$, $\lvert \tilde{\nu}_m(1)\rvert^2$ and residuals $\lvert \tilde{\nu}_m(1)-\tilde{\nu}_m(\ptau)\rvert^2$ for $d=2048$. The residuals are predicted to be negligible compared to the magnitudes at the same $m$ for $\ptau \ll 525$, which these plots are in good agreement with even when $\ptau$ is a considerable fraction of $525$.}
    \label{fig:errorpairing}
\end{figure*}

To provide numerical evidence for the pairing of error coefficients, we test the prediction of Eq.~\eqref{eq:uerrT_form} when $g_1 \partial h/\partial g_1$ is negligible, i.e. Eq.~\eqref{eq:uerrT_approx} in the main text. More directly, we define
\begin{equation}
    \tilde{\nu}_m(\ptau) = \frac{1}{\ptau z(\ptau, t_0)}\nu_m(\ptau).
\end{equation}
Eqs.~\eqref{eq:uerrT_form}, \eqref{eq:uerrT_approx} then imply that $\tilde{\nu}_m(\ptau) = \tilde{\nu}_m(1)$ for any $\ptau \ll 1/\sqrt{\vareps_C(1,t_0)}$. This is verified in Fig.~\ref{fig:errorpairing} for the ($\beta=2$, $d=2048$) CUE dataset of Fig.~\ref{fig:RMT_ergodicity}, for which $1/\sqrt{\vareps_C(1,t_0)} \approx 525$.

\section{Non-ergodicity and spectral fluctuations in rational tori}
\label{app:toruscyclic}
When $\alpha = m/\ell$ is rational with coprime $m$ and $\ell$, all energy eigenstates of the form $(J_x - r m, J_y + r \ell)$, $r\in \mathbb{Z}$ for a given $(J_x,J_y)$ have the same energy~\cite{BerryTabor}. Consequently, taking into account the restriction of $(J_x,J_y)$ to a rectangle of area $d$ in $\Eshell{d}$, the energy spectrum is $O(\sqrt{d})$-fold degenerate with $O(\sqrt{d})$ unique energy values (when $d_x = d_y$). Thus, in the calculation of the mode fluctuations $\Delta_n = [E_{q(n)} t_0 d/(2\pi)] - n$ (see Eq.~\eqref{eq:deltamodefluctuation}) for generic $t_0$, each unique value of $E$ is associated with $O(\sqrt{d})$ different (at best, consecutive) values of the integer $n$ due to the degeneracy, and we have $\Delta_n \gtrsim O(\sqrt{d})$ in general. This argument means that for rational tori, we should expect (based on Eqs.~\eqref{eq:q_dft_persistence} and \eqref{eq:persistence_modefluctuations}) 
\begin{equation}
    \vareps_C(1,t_0)[q] \sim O(d^{-2} \sigma_\Delta^2) \gtrsim O(d^{-1}),
\end{equation}
which is at least of the same order of magnitude of spectral fluctuations as Poisson level statistics. Given that the $1$-step error for any cyclic permutation must be $\gtrsim O(d^{-1})$ in this case by Theorem~\ref{thm:dftoptimal}, it follows that rational tori are unlikely to possess any ergodic cyclic permutation in $\Eshell{d}$. We also note that an analytically solvable illustration of this argument is given by the sorted DFT cyclic permutation for $t_0 = 2\pi/(\omega_x d_x)$ when $d_x = d_y$, in which the (wrapped) energy levels are \textit{exactly} $\sqrt{d}$-fold degenerate.

\end{appendices}


\printbibliography

\end{document}